\documentclass[12pt]{article}
\usepackage{latexsym}
\usepackage{amsthm}
\usepackage{amsmath,amssymb,amsfonts,bm,amsbsy,bbm}
\usepackage{epsfig}
\usepackage{graphicx,rotating,lscape}

\usepackage{hyperref}
\usepackage{verbatim}
\usepackage{natbib}
\usepackage{multirow,color}
\usepackage{geometry}
\usepackage{setspace}
\usepackage{subfigure}
\usepackage{bbm}
\usepackage[flushleft]{threeparttable}
\usepackage{enumitem}

\newtheorem{Pro}{Proposition}
\newtheorem{Lem}{\underline{\bf Lemma}}
\newtheorem{Th}{\underline{\bf Theorem}}

\newtheorem{Rem}{\underline{\bf Remark}}

\newcommand{\indep}{\;\, \rule[0em]{.03em}{.67em} \hspace{-.25em}
\rule[0em]{.65em}{.03em}
\hspace{-.25em}\rule[0em]{.03em}{.67em}\;\,}

\def\tr{{\rm tr}}
\def\wh{\widehat}
\def\wt{\widetilde}
\def\wb{\overline}
\def\n{\nonumber}   
\def\var{\mbox{var}}
\def\cov{\mbox{cov}}

\def\vec{\mbox{vec}}
\def\dvec{\mbox{dvec}}

\def\pr{\mbox{pr}}
\def\log{\mbox{log}}
\def\trans{^{\top}}

\def\sumi{\sum_{i=1}^n}
\def\sumia{\sum_{i=1}^{n_1}}
\def\sumib{\sum_{i=n_1+1}^n}

\def\sumj{{\sum_{j=1}^n}}
\def\sumja{\sum_{j=1}^{n_1}}
\def\sumjb{\sum_{j=n_1+1}^n}

\def\0{{\bf 0}}
\def\1{{\bf 1}}

\def\real{{\mathrm{R}}}

\def\A{{\bf A}}
\def\a{{\bf a}}
\def\B{{\bf B}}

\def\c{{\bf c}}
\def\D{{\bf D}}

\def\I{{\bf I}}

\def\r{{\bf r}}

\def\V{{\bf V}}

\def\r{{\bf r}}

\def\s{{\bf s}}
\def\S{{\bf S}}

\def\u{{\bf u}}

\def\v{{\bf v}}
\def\W{{\bf W}}

\def\w{{\bf w}}
\def\X{{\bf X}}
\def\b0{{\bf 0}}

\def\x{{\bf x}}
\def\Y{{\bf Y}}
\def\y{{\bf y}}
\def\Z{{\bf Z}}
\def\z{{\bf z}}
\def\me{\mathcal{E}}

\def\ba{\boldsymbol\alpha}
\def\bg{\boldsymbol\gamma}

\def\bpsi{{\boldsymbol\psi}}
\def\bg{{\boldsymbol\gamma}}

\def\btheta{{\boldsymbol\theta}}

\def\bSigma{{\boldsymbol \Sigma}}
\def\bmu{{\boldsymbol \mu}}

\def\bse{\begin{eqnarray*}}
\def\ese{\end{eqnarray*}}
\def\be{\begin{eqnarray}}
\def\ee{\end{eqnarray}}
\def\bsq{\begin{equation*}}
\def\esq{\end{equation*}}
\def\bq{\begin{equation}}
\def\eq{\end{equation}}

\def\squarebox#1{\hbox to #1{\hfill\vbox to #1{\vfill}}}
\def\boxit#1{\vbox{\hrule\hbox{\vrule\kern6pt\vbox{\kern6pt#1\kern6pt}\kern6pt\vrule}\hrule}}

\allowdisplaybreaks

\definecolor{mygreen}{RGB}{34, 139, 34} 

\begin{document}
\begin{center}
{\LARGE{\bf A KL-divergence based test for elliptical distribution}}\\
\vspace{0.5cm}
Yin Tang\textsuperscript{$\ast$a}\footnote{The majority of this work was carried out while the author was a graduate student at the Pennsylvania State University.}, Yanyuan Ma\textsuperscript{b}, Bing Li\textsuperscript{b}\\
\textsuperscript{a}\textit{Dr. Bing Zhang Department of Statistics, University of Kentucky, USA}\\
\textsuperscript{b}\textit{Department of Statistics, Pennsylvania State University, USA}\\
\textsuperscript{$\ast$}Corresponding author. Email address: yin.tang@uky.edu
\end{center}

\abstract
We conduct a KL-divergence based procedure for testing elliptical
distributions. The procedure simultaneously takes into account
the two defining properties of an elliptically distributed random vector: 
independence between length and direction, and uniform distribution of
the direction.
The test statistic is constructed based on the $k$ nearest neighbors
($k$NN) method,
and two cases are considered where the mean vector and covariance 
matrix are known and unknown. 
First-order asymptotic properties of the test
statistic are rigorously established by creatively utilizing sample splitting,
truncation and transformation between Euclidean space and unit
sphere, while avoiding assuming Fr\'echet differentiability of any
functionals. Debiasing and variance inflation are further proposed to
treat the degeneration of the influence function. Numerical
implementations suggest better size and power performance than the
state of the art procedures.

{\bf Keywords}: Elliptical distribution, Entropy, KL-divergence,
Influence function, $k$NN method.
\baselineskip=18pt

\section{Introduction}

Elliptical distribution is arguably the most frequently used
semiparametric family of distributions. It includes many familiar
distributions such as normal, student t, Laplace, Cauchy, etc.
See, for example, \cite{fang2013symmetric}.
Often, when we have established certain statistical properties for the 
normal distribution and plan to widen were  the study to a more general
setting, the first thing we would inspect is the elliptical distribution
family. 
Indeed, the elliptical distribution family often retains
some nice properties that originally were only discovered to hold
under normality assumption. Thus, in practice, it is crucial to
examine whether or not the elliptical distribution assumption holds.

Many works have been devoted to  testing the elliptical distribution.  
A significant portion of the literatures, especially in the earlier
years,  are designed for  testing either the spherical distribution, or the
elliptical distribution with known 
mean vector and covariance matrix. These two problems are indeed 
the same under rescaling. For example,
\cite{kariya1977robust} provides several tests on
mean  and/or variance in the spherical
  distribution family, and show that their tests are uniformly most
  powerful (UMP) and/or uniformly most
  powerful and unbiased (UMPU).
  \cite{king1980robust} further  studies these tests and shows that
  the UMP/UMPU properties are retained under linear transformations, 
  and applies them to linear models.
  \cite{romano1989bootstrap} introduces  bootstrap and randomization
  tests and applies them to test a class of nonparametric hypotheses,
  including spherical distributions.
    \cite{baringhaus1991rotationally} proposes a test based on a
    second-order V-statistic involving the ranks of the lengths and
     transformed pairwise angles, which is
    equivalent to the $L^2$ distance between the empirical distribution and
    its nearest centered spherical distribution.
  \cite{fang1993necessary} considers projecting the data to different
  directions and forming the pairwise differences between directions,
  then using a
  Wilcoxon-type test statistic to test for spherical symmetry.  They
  require the projection directions to be sufficient rich and scattered on
  the unit sphere, and otherwise the procedure is not an omnibus test.
  \cite{koltchinskii1998testing} uses a multivariate quantile approach
  to measure the spherical asymmetry, and constructs a test for
  spherical distributions with unknown centers. In this sense, it no
  longer requires to know the mean vector, but still requires to know
the  covariance matrix when used as a test for elliptical distributions.  
  In practice, they propose a
  bootstrap approach or Monte Carlo approach  to approximate
  the null distribution.
  \cite{diks1999test} proposes a Monte Carlo based test for various types
  of multivariate symmetries, which can be used to test the spherical
  distribution as well.
   \cite{liang2008necessary} devises a test for spherical distribution
   via the Rosenblatt transformation on each element, which only tests
   a necessary condition for spherical distributions.
\cite{einmahl2012testing} proposes an 
    omnibus test for spherical distribution 
    using the localized empirical likelihood, but it only applies to
    the bivariate case.
    \cite{liang2012t3plot} gives a test  based on plots
        related to the
      third derivative of the empirical moment generating
      function. This test also requires projecting the data to a
      specific direction, and the confidence interval is computed via
      Monte Carlo. 
  \cite{henze2014testing} forms a test by checking whether the
  characteristic function is constant on the unit sphere, which is
  only a necessary condition for spherical distributions.   
   \cite{guha2021visual} constructs a test statistic derived from
   scale-scale plots, and uses Monte Carlo to get the reference
   distribution. However, the procedure only tests whether
      different directions are uniformly distributed on the unit sphere, which
   is again only a necessary condition hence the test is not omnibus.
  \cite{huang2023multivariate} further introduces a test following the
  same line via optimal transport but is also not omnibus.
  \cite{banerjee2024consistent} proposes a test based on an
  energy distance between the raw data and the augmented data
  formed by the raw length and newly-generated direction
  vectors, and the critical value is determined via
  resampling. 
   
 All the testing procedures discussed above assume a known  covariance
 matrix, and most of them also assume a known mean vector.
When the mean and variance are not known, it is natural to
   estimate them and then insert into the testing procedures. 
   Such a plugin procedure gives rise to significant complications
   and have been studied in the literature as well.
    For example,
\cite{fang1998projection} proposes a test by projecting the
normalized data to different directions and 
computing the skewness and kurtosis for each of them, and further
checking whether their behavior is significantly heterogeneous
while taking into account the estimation errors of the mean and variance.
However, similar to \cite{fang1993necessary}, the test is not omnibus
without considering all possible directions. 
\cite{manzotti2002statistic} uses the spherical harmonics over the
    projections of the scaled residual to perform a test for the
    uniform distribution of the direction vector, 
  \cite{schott2002testing} tests whether the fourth moments have the
  structure indicated by elliptical distributions. 
   \cite{babic2019optimal} proposes a test statistic for elliptical
   distribution which is asymptoticly $\chi^2$ for elliptical distribution.
   But the test  is only optimal against some generalized
   skew-elliptical distributions where their test statistic is
   non-central $\chi^2$ distributed.
   The three tests above take into
  account the estimation of the mean and variance in deriving the
  theoretical properties of the tests, but  only detect certain
  departures from the elliptical distribution.
  \cite{iwashita2020necessary} introduces a test for elliptical
  distribution by only testing the uniform distribution on the Stiefel
  manifold based on modified degenerate V-statistics.
  \cite{soloveychik2023robust} constructs a goodness-of-fit test by
  comparing the distribution of the unnormalized direction vector to
  its true distribution, but also does not test the independence
  between the length and the direction vector. In summary,
    although the above tests take into account the estimated
    mean/covariance, they only test certain aspects of the elliptical
    distribution, hence are not omnibus.
  
  Furthermore, there are many tests for elliptical distributions
    based on transformations and resampling
    or bootstrap approaches, which result in
   additional computational cost. For example, 
  \cite{beran1979testing} gives a test statistic based on
  transformations of the length and the direction, where, to achieve
  the asymptotic results, the number of transformations needs to go to
  infinity, while the convergence rate of the test statistic is slower
  than $O_p(n^{-1/2})$.  
  \cite{koltchinskii2000testing} provides a test which requires
  taking supremum over a function class that characterizes the
  distribution, and the test is conducted using the bootstrap
  distribution.
    \cite{zhu2003conditional} constructs the test statistic via the integral of a squared empirical  process of trigonometric
functions. They use a 
  resampling distribution by keeping the lengths unchanged but
  independently sampling from the uniform distribution on the unit
  sphere.  
  \cite{huffer2007test} constructs a test by slicing the data, which
  results in
  loss of information. Further, except when the null distribution is
  normal, they resort to bootstrap in general elliptical distributions.
\cite{bianco2017conditional} further considers estimating the
  mean and covariance robustly for the test statistic in \cite{zhu2003conditional}, but still applies the resampling technique. 
  \cite{su2012smooth} also derives a smooth test based on Cholesky
  decomposition and spherical harmonics, but also applies the
  resampling technique to conduct the test.
   Finally, \cite{albisetti2020testing} provides a Kolmogorov-Smirnov type test,
  which requires many projecting transformations on the data, and also
  utilizes the bootstrap distribution.
Some of the tests for elliptical distribution mentioned above are
summarized and implemented in R \citep{babic2021rjournal}.
  
    There are also some tests that are related to
    elliptical distributions. Some of them are designed to
  test whether the
  data are symmetric about one point. 
  For example, \cite{heathcote1995testing} introduces a test for
  whether a multivariate distribution is symmetric about a point using
  a stochastic process via the empirical characteristic function.
  \cite{neuhaus1998permutation,henze2003invariant} propose different
  tests for whether the distribution is symmetric about an unspecified
  point based on the vanishing imaginary part of the characteristic
  function, and conduct the tests via a permutation procedure.  
  Based on the similar idea, \cite{ngatchouwandji2009testing}
  conducted a Cram\'er-von Mises test using the asymptotic
  distribution. 
  Also, \cite{cassart2008optimal} proposes a pseudo-Gaussian test
  which is optimal against some violations of Fechner-type symmetry. 
  On the other hand, some tests are specific to some specific type of
  distribution rather than general elliptical distributions.
  For example, \cite{batsidis2013necessary} and
  \cite{ducharme2020goodness} propose different classes of 
  goodness-of-fit tests for several specific types of elliptical
  distributions, with some specific density generators. Their null
  hypothesis is a specific type of distributions rather than the
  general elliptical distribution, and their test statistic
  construction also relies on the specific density generator.

  In addition, there are a few other works
  that cannot be classified into the categories above. Specifically,
\cite{li1997qq} provides a QQ-plot based procedure to check spherical and
elliptical distributions visually.  
  \cite{sakhanenko2008testing} conducts comparison studies between
  three types tests in \cite{koltchinskii2000testing} and three
 tests given respectively by
  \cite{beran1979testing,manzotti2002statistic} and \cite{huffer2007test}.

Recently, \cite{tang2024nonparametric} points out that a consistent
test for a spherical distribution  should take into account both of
its two characterizing properties: 
  (i) the length and the direction of the random vector  
are independent, and (ii) the direction of the random vector is uniformly
distributed on the unit sphere, and that relying only  on one of the
them would lead to false acceptance. Correspondingly, a consistent
test for the elliptical distribution should take into account the
counterparts of these two conditions  in the elliptical setting.
Based on the two conditions, \cite{tang2024nonparametric} proposes a
nonparametric test for elliptical distributions by  embedding   two
distributions -- the empirical joint distribution  and a factorized
distribution derived from the two conditions -- into a tensor product
of reproducing kernel Hilbert spaces (RKHS), and then computing the
Hilbert-Schmidt 
distance between  two operators. 
The distance is close to zero under  elliptical distribution, but is
large if at least one of the two conditions are
violated. However, the procedure critically relies on the
  Fr\'echet differentiability at the null distribution, which may not
  be satisfied by all elliptical distributions.

In this paper, we propose a  Kullback-Leibler (KL)-divergence (also known as relative
entropy) based testing procedure. KL-divergence is first introduced in
\cite{kullback1951information}. KL-divergence is widely
used in many fields to quantify the difference between two
distributions, for example in information theory (see, for 
example, Chapters 2 and 8 of \cite{cover2012elements}), machine
learning (see, for example, Chapter 2 of \cite{eguchi2022minimum} and
Chapters 9 and 10 of \cite{bishop2016pattern}) and statistical
inference (see, for example, Chapter 1 of
\cite{pardo2020statistical}). In particular, the KL-divergence
  is sometimes
used for goodness-of-fit tests. \cite{li2005goodness}
summarizes two classical approaches of goodness-of-fit tests: minimum
cross entropy (MinxEnt) principle (see
\cite{kullback1959information,kullback1951information}) and the
Vasicek-Song test via $M$-spacing method (see
\cite{song2002goodness,vasicek1976test}). Also, KL-divergence is
closely related to many concepts in information theory, such as
entropy (\cite{shannon1948mathematical}) and mutual information. See
Chapter 2 of \cite{cover2012elements} for more details. In particular,
for two random elements, the mutual information is exactly the
KL-divergence between their joint distribution and the product of
their marginal distributions, which characterizes the dependency
between them. Consequently, mutual information can also be used to test
independence (see, for example,
\cite{bs2019,ai2024testing,pethel2014exact}).

The problem of testing for ellipticity is a combination of a
goodness-of-fit test and an independence test. Compared to the
goodness-of-fit test, our null distribution, the elliptical
distribution, has the additional complexity of containing a nonparametric component, i.e., the distribution of the
length.  Compared to the independence test, we have the additional
complexity of testing  the suitability of 
the spherical distribution for the direction vector.  We handle
  both complexities in one step  by
directly constructing KL-divergence based test statistic, and representing the
KL-divergence by entropies.  To this end, \cite{bsy2019} gives a
  generalized version 
of the Kozachenko-Leonenko entropy estimator (\cite{kl1987}) by using 
$k$-nearest neighbor ($k$NN) method, 
and \cite{bsy2019} establishes its first-order asymptotic properties. 

One difficulty in utilizing the  $k$NN  method in our problem
is that we are faced with computing distance and determining neighbors
on a sphere while the original method is designed for Euclidean
distance. To bypass this complexity, we convert the problem 
back to the Euclidean space by engaging mathematical tools to
establish equivalence presentation. 
We further derive the asymptotic property of our resulting test
statistic. Different from \cite{tang2024nonparametric}, 
  where all steps only involve moment estimators, establishing
  theoretical properties of the proposed method is much more
  challenging. First of all,   it is 
challenging to establish the Fr\'echet differentiability for our
$k$NN-based estimator, which prohibits us from using many
high-level functional tools. 
Therefore, we derive the influence function of
the test statistic as well as the order of the remainder terms
directly. In particular, when the mean vector and the covariance
matrix are unknown, in order to bound the remainder term, we need to
conduct sample splitting and involve transformation and
truncation tools that are mathematically challenging and
difficult.
Secondly, we handle the degeneration of the influence
function under the null hypothesis by creatively debiasing and inflating
variance. 

The rest of the paper is organized as follows. We first introduce the
problem setting and our estimating approach in Section
\ref{sec:approach}. We then consider a basic case where the mean
vector and the covariance matrix are known, and give the test
statistic and asymptotic properties in Section
\ref{sec:known}. The general case where the mean
vector and the covariance matrix are unknown is studied in Section
\ref{sec:unknown}.  We further 
propose the debias method in practice due to the degeneration of the
influence function in Section \ref{sec:debias}. 
The regularity conditions and proofs are placed in the supplementary materials.
 We conduct simulation studies in Section
\ref{sec:simu}, where we also  compare our method to 
\cite{tang2024nonparametric}  and five most popular methods in the literature.
We further apply our method to a real
dataset in Section \ref{sec:real}. Finally, we give some
  discussions and conclude 
the paper in
Section \ref{sec:conclusion}.

\section{General Approach}\label{sec:approach}

\subsection{Problem Setting}\label{sec:setting}

We first state our problem and define some notations. Suppose we are
given a sample of $p$-dimensional independent and identically distributed (iid)
random vectors $\X_1, \dots, \X_n$ with an absolutely
continuous distribution and  a  probability density function (pdf) $f_\X(\x)$.
 Throughout this paper, we assume that the covariance matrix $\bSigma = \cov(\X)$ exists.
Our interest is in testing whether or not $\X$ follows an elliptical
distribution. First note that a $p$-dimensional
random vector $\X$ has elliptical distribution if and only if its pdf
can be written as 
\be\label{eq:ellip-def}
f_\X(\x)=\{\det(\W)\}^{-1/2}g\{(\x-\bmu)\trans\W^{-1}(\x-\bmu)\}
\ee
where $\bmu \in \real^p$, $\W \in \real^{p \times p}$, and $g$ is a function satisfying $\int g(\x\trans\x) d\x =1$.
We assume $\W$ is a strictly positive definite matrix.
Under the model \eqref{eq:ellip-def}, we have $E(\X) = \bmu$ and 
$\cov(\X) = \bSigma = c \W$ for some constant $c>0$. 
Note that \eqref{eq:ellip-def} is not identifiable in that 
it always gives the same model if $\W$ is multiplied by 
a constant and $g$ is scaled accordingly. 
For identifiability purpose, a common choice is to replace 
$\W$ directly by $\bSigma$ and use the corresponding $g$.
Let $\Y$ be the normalized version of $\X$, i.e.,
$\Y\equiv\bSigma^{-1/2}(\X-\bmu)$, 
where 
$\bSigma$ is strictly positive definite.  We assume $\bmu$ and
  $\bSigma$ are finite,
 and we can always write
$f_\X(\x)=\{\det(\bSigma)\}^{-1/2}f_\Y\{\bSigma^{-1/2}(\x-\bmu)\}$, where $f_\Y$ is the pdf of $\Y$. 
Let $U\equiv\|\Y\|$, $\V\equiv\Y/U$.
It can be verified that $\X$ has elliptical distribution
if and only if  $U\indep\V$ and $\V$ has a uniform distribution on the $p$-dimensional unit sphere 
 $\mathrm{S}^{p-1}$.
Using this fact, the problem of testing
whether $\X$ has an elliptical distribution is equivalently
written as testing
\be\label{eq:h01}
H_0:  f_{\V\mid U} (u,\v) = f_0 (\v) \mbox{\  versus \ } H_a:  H_0 \text{ is not true, }
\ee
where  $f_{\V\mid U} (u,\v)$ is the conditional pdf of $\V$ given $U$, and $f_0(\v)$ is the pdf of the uniform distribution on the $p$-dimensional unit
sphere. Or  equivalently,
\be\label{eq:h02}
H_0: f_{U, \V}(u,\v)=f_U(u)f_0(\v) \quad \mbox{versus} \quad  H_a: \mbox{$H_0$ is not true, } 
\ee
 where $f_{U, \V}(u,\v)$ is the joint pdf of $(U,\V)$.
Note that the pdf $f_{U, \V}(u,\v)$ is with respect to the product measure $\lambda \times \sigma$, 
where $\lambda$ is the Lebesgue measure on $(0,\infty)$ and $\sigma$ is the surface measure
on the unit sphere $\mathrm{S}^{p-1}$. See Section 2.7 of \cite{folland1999real} or Section 3.2 of \cite{stein2005real} for rigorous derivations. The density $f _0$ has  explicit form 
\begin{align*}
  f_0(\v)=c_p I(\|\v\|=1), \ \text{where} \ c_p\equiv\Gamma(p/2)/(2\pi^{p/2}).   
\end{align*}

\subsection{Our Approach}\label{sec:ourapproach}

Our general approach to the problem is to use the Kullback-Leibler (KL)
divergence, a well known criterion,  to quantify the difference between two
distributions. Given
two probability density functions $f_1$ and $ f_0$, the
KL divergence between them  is defined  as 
\begin{align*}
  d(f_1\|f_0)\equiv
E[\log \{f_1(\X)/f_0(\X)\}] =\int \log\{ f_1(\x)/f_0(\x)\} f_1(\x)d\x.  
\end{align*} 
Note
that the expectation is computed under $f_0$ and the KL divergence is not
symmetric. To use the KL divergence to perform the test, we certainly
need to approximate the expectation. It will be shown that this boils
down to 
estimating the entropies $H(\Y)\equiv-E\{\log f_\Y(\y)\}$ and
$H(U)\equiv-E\{\log f_U(u)\}$. 

Entropy estimation has been studied in the statistical
literature. Here, we adopt the Kozachenko-Leonenko estimator
\citep{kl1987}, as  detailed below.
Let $H(\Z)\equiv-E\{\log f_\Z(\Z)\}$ be the entropy
for an arbitrary $d$-dimensional random vector $\Z$. 
Based on the iid data
$\Z_1, \dots, \Z_n$,  $H(\Z)$ can be
approximated as
\be\label{eq:estimator}
\wh H_n(\Z)=\frac{1}{n}\sumi\sum_{j=1}^kw_j\log\left[
\frac{(n-1)\rho_{(j),i}^d{\cal V}_d}{\exp\{\psi(j)\}}\right], 
\ee
where $k\in\{1, \dots, n-1\}$,  $w_1, \dots w_k$ are weights
that satisfy $\sum_{j=1}^kw_j=1$, ${\cal V}_d=\pi^{d/2}/\Gamma(1+d/2)$,
$\psi(z)\equiv \Gamma'(z)/\Gamma(z)$ is the digamma function, and
$\rho_{(j),i}^d\equiv \|\Z_{(j),i}-\Z_i\|_2^d$, where $\Z_{(1),i},
\dots, \Z_{(n-1),i}$ is a permutation of $\{\Z_1, \dots,
\Z_n\}$ excluding $\Z_i$ such that $ \|\Z_{(1),i}-\Z_i\|\le\dots\le
 \|\Z_{(n-1),i}-\Z_i\|$. We have the freedom to choose $k, \{w_j\}$
and the optimal choices are given in \cite{bsy2019}. This
method first uses a weighted $k$-nearest neighbors to
approximate $f_{\Z}(\z)$ by
\begin{align*}
\wh f_{\Z}(\z_i)=\sum_{j=1}^kw_j\log\left[
\frac{(n-1)\rho_{(j),i}^d{\cal V}_d}{\exp\{\psi(j)\}}\right], 
\end{align*}
and then replaces expectation $E$ by sample average. The nearest
neighbor is in terms of Euclidean distance.

Since $[0,\infty)$ is a subset of $\real$, we can use the
Kozachenko-Leonenko estimator to directly
obtain $\wh E\{\log f_U(U)\}=-\wh H_n(U)$.
However, to construct a similar estimator of $E_{U,\V}\{\log f_{U,\V}(U,\V)\}$, 
it is difficult to define the nearest neighbor on $(0,\infty) \times \mathrm{S}^{p-1}$. 
To solve this problem, we transform the parametrization $(U,\V)$ back to the original $\Y$, 
which is in $\real^p$, and use the Kozachenko-Leonenko estimator to obtain 
$\wh E\{\log f_\Y(\Y)\}=-\wh H_n(\Y)$. We first derive the relationship between 
$f_{U,\V}(u, \v)$ and $f_\Y(u\v)$ in the following lemma.
\begin{Lem}\label{lem:fuv}
Let $\Y$ be a random vector in $\real^p$, and let $U=\|\Y\|$ and $\V=\Y/U$. Suppose that the pdf of $\Y$ is $f_\Y(\y)$ with respect to the Lebesgue measure on $\real^p$, and the joint pdf of $(U,\V)$ is $f_{U,\V}(u,\v)$ with respect to $\lambda \times \sigma$, where $\lambda$ is the Lebesgue measure on $(0,\infty)$ and $\sigma$ is the surface measure on the unit sphere $\mathrm{S}^{p-1}$. Then, $f_{U,\V}(u, \v) = f_\Y(u\v) u^{p-1}$.
\end{Lem}
The proof of Lemma \ref{lem:fuv} is placed in Section
\ref{sec:proof-fuv} of the supplementary materials.
Based on Lemma \ref{lem:fuv}, we can derive the relationship between $E_{U,\V}\{\log f_{U,\V}(U,\V)\}$ and $E_\Y\{\log f_\Y(\Y)\}$ as follows:
\bse
E_{U,\V}\{\log f_{U,\V}(U,\V)\}
&=&E_\Y[\log f_{U,\V}\{U(\Y),\V(\Y)\}]\\
&=&E_\Y\{\log f_\Y(\Y)\|\Y\|^{p-1}\}\\
&=&E_\Y\{\log f_\Y(\Y)\}+(p-1)E_\Y\{\log \|\Y\|\}\\
&=&E_\Y\{\log f_\Y(\Y)\}+(p-1)E_U\{\log (U)\}.
\ese
 As is mentioned before, we can estimate the first term using the Kozachenko-Leonenko
estimator. The second term can be estimated by the sample average,
i.e., 
\bse
(p-1)\wh E_n\{\log (U)\}
=(p-1)n^{-1}\sumi\log (u_i).
\ese
This allows us to calculate
\be\label{eq:uvy}
\wh H_n(U,\V)=-\wh E_n\{\log f_{U,\V}(U,\V)\}
=\wh H_n(\Y)-(p-1)\wh E_n\{\log (U)\}.
\ee

\section{Test with known mean and covariance}\label{sec:known}

To set the stage for the main idea, we start by considering a possibly overly
simplified situation, where
$\bmu$ and $\bSigma$ are both known. In this case, we directly observe
$\Y_i$ and subsequently also directly observe $U_i, \V_i$ for $i=1, \dots, n$.

\subsection{Test Statistic Construction}

We adopt the hypotheses forms in \eqref{eq:h02} to devise our test.  
First, the KL divergence between $f_{U,\V}(u,\v)$ and
$f_U(u)f_0(\v)$  is calculated as
\bse
d(f_{U,\V}\|f_Uf_0)  &=&E\left\{\log
  \frac{f_{U,\V}(U,\V)}{f_U(U)f_0(\V)}\right\}\\
&=& \int_{\mathrm{S}^{p-1}} \int_0^\infty \log\frac{f_{U,\V}(u,\v)}{f_U(u)f_0(\v)} f_{U,\V}(u,\v)dud\sigma(\v)\\
  &=&E\{\log f_{U,\V}(U,\V)\}-E\{\log f_U(U)\}-E\{\log f_0(\V)\}\\
  &=&E\{\log f_{U,\V}(U,\V)\}-E\{\log f_U(U)\}-\log c_p.
  \ese
\begin{Rem}
If we replace $f_0(\v)$ with $f_\V(\v)$ in the definition of
$d(f_{U,\V}\|f_Uf_0) $, then it is the mutual 
information $I(U,\V)$, which is used to measure the dependence between
$U$ and $\V$. 
So we can write
$d(f_{U,\V}\|f_Uf_0)=I(U,\V)+E\{\log f_\V(\V)\}-E\{\log
f_0(\V)\}=I(U,\V)+d(f_\V\|f_0)$. This indicates that
$d(f_{U,\V}\|f_Uf_0)$ consists of two nonnegative components, one measures the
dependence relation, the other measures the deviation of $f_\V$ to 
$f_0$. If $U$ and $\V$ are independent, but $\V$ is not necessarily
uniformly distributed on the unit sphere, then  $I(U,\V)=0$ while
$d(f_\V\|f_0)>0$. On the other hand, if $\V$ is marginally
uniformly distributed on the unit sphere, but $U$ and $\V$ are
dependent, then $d(f_\V\|f_0)=0$ but $I(U,\V)>0$.
Under $H_0$, both are zero and hence $d(f_{U,\V}\|f_Uf_0)=0$. Otherwise,
$d(f_{U,\V}\|f_Uf_0)>0$. 
\end{Rem}

  \begin{Rem}
Our test procedure is very different from both \cite{bsy2019} and
  \cite{bs2019}, and is far from a simple extension.
 In terms of the problem treated,  \cite{bsy2019}
  provides an estimator of a single entropy and derives its asymptotic
  properties. 
  On the other hand,   \cite{bs2019} exclusively considered an
  independence testing problem. 
  We considered a different problem, where we test elliptical
  distribution, which involves simultaneously testing
  whether the marginal distribution of a properly scaled variable set
is uniformly distributed on the unit sphere and whether two sets of variables are
independent. In addition, our methodology is also different.
\cite{bsy2019} developed a nonparametric estimation procedure which
does not involve testing methodology. \cite{bs2019}
relies on resampling to form the testing methodology. 
In contrast, we derive the asymptotic properties and rely on the
theoretical properties to develop the testing methodology, as
established in the following development.
\end{Rem}

Using \eqref{eq:uvy},
we define the test statistic to be
\be\label{eq:T}
 T\equiv -\wh H_n(U,\V)+\wh H_n(U)-
 \log c_p
 =-\wh H_n(\Y)+(p-1)\wh E_n\{\log (U)\}+\wh H_n(U)-
 \log c_p,
  \ee
  where $\wh H_n(\Y)$ and $\wh H_n(U)$ are given by
  \eqref{eq:estimator} with $\Z$ replaced by $\Y$ and $U$.
 Specifically, 
  \bse
\wh H_n(\Y) &=& \frac{1}{n}\sumi\sum_{j=1}^{k_p}w_{pj}\log\left[
\frac{(n-1)\|\Y_{(j),i}-\Y_i\|^p{\cal V}_p}{\exp\{\psi(j)\}}\right], \\
\wh H_n(U) &=& \frac{1}{n}\sumi\sum_{j=1}^{k_1}w_{1j}\log\left[
\frac{(n-1)|U_{(j),i}-U_i|{\cal V}_1}{\exp\{\psi(j)\}}\right],
\ese
where $\Y_{(j),i}$ and $U_{(j),i}$ are the $j$-th nearest neighbor of 
$\Y_i$ and $U_i$, respectively. Note that the numbers ($k_p,k_1$) of
the nearest neighbors 
and the weights ($w_{pj}, w_{1j}$) for $\wh H_n(\Y)$ and $\wh H_n(U)$
can be different.

\subsection{Asymptotic Properties}

Next, we develop the asymptotic distribution of 
 $T$. To this end, we note that \cite{bsy2019} gives the asymptotic expansion
\be\label{eq:hz-asymp}
n^{1/2}\{\wh H_n(\Z)-H(\Z)\}=n^{-1/2}\sumi \{-\log f_\Z(\z_i)
-H(\Z)\}+o_p(1),
\ee 
which allows us to get the asymptotic property of $T$ under $H_0$ or
$H_a$ directly. Specifically, under $H_0$, 
\bse
n^{1/2}T
&=&
n^{-1/2}\sumi[
\log f_\Y(\y_i)
+(p-1)\log (u_i)
-\log f_U(u_i)\\
&&-E\{\log f_\Y(\Y)\}
-(p-1)E\{\log (U)\}
+E\{\log f_U(U)\}]+o_p(1)\\
&=&
n^{-1/2}\sumi [
\log f_{U,\V}(\u_i,\v_i)
-\log f_U(u_i)-E\{\log f_{U,\V}(U,\V)\}
+E\{\log f_U(U)\}]+o_p(1)\\
&=&n^{-1/2}\sumi
\{\log f_0(\v_i)-\log c_p\}+o_p(1)\\
&=&o_p(1).
\ese
By \cite{bsy2019}, there are choices of $k$ in \eqref{eq:estimator} so that
the above $o_p(1)$ term is $O_p(n^{-c})$ for some constant $c>0$.
This is an estimator with convergence rate faster than $n^{-1/2}$,
which is to our benefit. However, in practice, we suggest to approximate
the variance using the formula under alternative derived in \eqref{eq:var}.
Under the alternative,
\bse
&&n^{1/2}\{T-d(f_{U,V}\|f_Uf_0)\}\\
&=&n^{-1/2}\sumi [
\log f_{U,\V}(u_i,\v_i)
-\log f_U(u_i)-E\{\log f_{U,\V}(U,\V)\}
+E\{\log f_U(U)\}]+o_p(1)\\
&=&
n^{-1/2}\sumi
\{(p-1)\log u_i+\log f_\Y(\y_i)-\log f_U(u_i)+H(U,\V)
-H(U)\}+o_p(1)\\
&=&n^{-1/2}\sumi
[(p-1)\log (u_i)+\log f_\Y(\y_i)-\log f_U(u_i)+H(\Y) - (p-1) E \{ \log (U) \}
-H(U)]\\
&&+o_p(1)\\
&\to&N\left(0, E[
(p-1)\log (U) +\log f_\Y(\Y)
-\log f_U(U)+H(\Y) - (p-1) E\{\log (U)\}
-H(U)]^2\right).
\ese

The above analysis allows us to perform test using
\be\label{eq:var}
n\wh\var(T)
&\approx&
n^{-1}\sumi
\left(
- \sum_{j=1}^{k_p}w_{pj}\log\left[
  \frac{(n-1)\|\Y_{(j),i}-\Y_i\|^p{\cal V}_p}{\exp\{\psi(j)\}}\right]
+\sum_{j=1}^{k_1}w_{1j}\log\left[
  \frac{(n-1)|U_{(j),i}-U_i|{\cal V}_1}{\exp\{\psi(j)\}}\right]\right.\n\\
&&\left. +(p-1)\log (u_i)+\wh H_n(\Y) - (p-1) \wh E (\log U ) -\wh H_n(U)
\right)^2+n^{-c}. 
\ee

Under $H_0$, $n^{1/2}T/\sqrt{n\wh\var(T)}=O_p(n^{-c/2})$,
while under the alternative, it has an approximate 
normal
distribution with mean $n^{1/2}d(f_{U,V}\|f_Uf_0)>0$ variance 1.  We
summarize the results in Theorem \ref{th:known} below. 
This result allows us to perform a score type test or a
Wald type test, which are capable of detecting a local alternative with
its KL divergence  of the order $n^{-1/2}$ from the null distribution.
\begin{Th}\label{th:known}
  Let $T$ be given in \eqref{eq:T}, where $k_1$ and $w_{1j}, j=1,
    \dots, k_1$ satisfy the requirements in Theorem 1 of
  \cite{bsy2019} with respect to $f_U$, and $k_p$ and $w_{pj},
  j=1, \dots, k_p$  satisfy these requirements with respect to $f_\Y$.
 See also Section \ref{sec:conditions} in the supplementary materials for
  these specific requirements.
  Then
  \bse
n^{1/2}\{T-d(f_{U,V}\|f_Uf_0)\}
=
n^{-1/2}\sumi \psi(u_i,\v_i)
+o_p(1)
\ese
where
\bse
\psi(u,\v)=\{\log f_{U,\V}(u,\v)-\log f_U(u)+H(U,\V)-H(U)\}.
\ese
Thus, under $H_0$, $n^{1/2}T=o_p(1)$ and under $H_a$,
$n^{1/2}\{T-d(f_{U,V}\|f_Uf_0)\}\to N[0, E\{\psi(U,\V)^2\}]$ in distribution. 
\end{Th}

\begin{Rem}
The proof of Theorem \ref{th:known} is a direct application of Theorem
1 of \cite{bsy2019}, hence is omitted. 
  To ensure the conditions of Theorem 1
 of \cite{bsy2019}, 
we require the orders of $k_1$ and $k_p$ to
depend on the smoothness parameters of the function classes containing
$f_U$ and $f_\Y$. After determining $k_1$ and
$k_p$, we can proceed to set the weights $w_{1j}$'s and
$w_{pj}$'s to satisfy the 
requirements explicitly stated in \cite{bsy2019}.
In other words,  as long as the function classes of
$f_U, f_\Y$ are sufficiently smooth, we can proceed to choose suitable
$k_1,k_p$ and the weights so that the conditions are satisfied.
\end{Rem}

 Let the variance of dominating term
$\psi(u_i,\v_i)$ be $\sigma_1^2$ and the variance contributed from the
  residual $o_p(1)$ term be $\sigma_2^2$, i.e., 
  \bse
  \sigma_1^2 = \var \left\{ n^{-1/2}\sumi \psi(u_i,\v_i)\right\}, \quad 
  \sigma_2^2 = \var \left[ n^{1/2}\{T-d(f_{U,V}\|f_Uf_0)\}
- n^{-1/2}\sumi \psi(u_i,\v_i)\right].
  \ese
    It is natural to approximate the
  variance of $T$ using the estimated version
   $\wh\sigma_1^2$. Under $H_a$, $\wh\sigma_1^2$ will lead to a
   good approximation of $\sigma_1^2$, which dominates $\var(T)$,
   while under $H_0$,
$\wh\sigma_1^2>\sigma_1^2$ 
since $\sigma_1^2=0$, hence hopefully $\wh\sigma_1^2\ge\var(T)$ as
well.
  However, contrary to our expectation, in our implementation, we find
  that this practice often
  leads to an underestimated variance, especially under $H_0$. Our
  suspicion is that the variance $\sigma_2^2$ is not so small under finite
  sample size $n$ although in theory it is ignorable, and in fact
  $\sigma_2^2>\wh\sigma_1^2$.
To ensure that we maintain the test level, we hence intentionally over
estimate the variance of $T$ by overestimating $\sigma_1^2$.

\subsection{Variance inflation}\label{sec:var-inflation}

Note that by Cauchy-Schwarz inequality, 
\bse
\var\{\psi(\X,\bmu,\bSigma)\} 
\le  \sigma^2 \equiv 2 \left( V_1 + V_2    \right),
\ese
where
\bse
V_1 = \var \{ \log f_{\Y} (\Y) \}, \quad V_2 = \var\{(p-1)\log(U) - \log f_{U} (U) \}.
\ese
We can estimate $V_1$ and $V_2$ respectively by
\be\label{eq:v1hat-v2hat-known}
\wh V_1 &=& n^{-1} \sumi  \left( \sum_{j=1}^{k_p}  w_{pj} \log \left[\frac{(n-1)\|\Y_{(j),i}-\Y_i\|^p {\cal V}_p}{\exp\{\psi(j)\}}\right] - \wh H_n(\Y) \right)^2,\n\\
\wh V_2 &=& n^{-1} \sumi  \left((p-1) \left[ \log(U_i) - \wh E \{ \log(U) \} \right] + \sum_{j=1}^{k_1}  w_{1j} \log \left[\frac{(n-1)|U_{(j),i}-U_i| {\cal V}_1}{\exp\{\psi(j)\}}\right] - \wh H_n(U) \right)^2.\quad
\ee

Thus, we can get a $(1-\alpha)$-level confidence interval for $d(f_{U,V}\|f_U,f_0)$ as
\bse
\left[ T - n^{-1/2} z_{\alpha} \wh \sigma, +\infty \right),
\ese
where
\be\label{eq:sigma2hat}
\wh \sigma^2 \equiv 2 (\wh V_1 + \wh V_2).
\ee
Under $H_0$, for a specified size $\alpha$, our decision rule will be rejecting $H_0$ if $T - n^{-1/2} z_{\alpha}  \wh \sigma>0$.
The p-value can be defined as $1-\Phi(n^{1/2}T/ \wh \sigma)$, where $\Phi$ is the cdf of $N(0,1)$.
The next proposition shows that, after the variance inflation
  step, our method can asymptotically control the size while preserve
  the power. The proof of the proposition is in the supplementary materials.
\begin{Pro}\label{prop:level-power-known}
Let $T$ be given in \eqref{eq:T}, and $\wh\sigma^2$ be given in \eqref{eq:sigma2hat} with $\wh V_1, \wh V_2$ defined by \eqref{eq:v1hat-v2hat-known}.
Under the same assumptions as in Theorem \ref{th:known}, for any
$0<\alpha<1/2$, when $n$ is sufficiently large, 
    under $H_0$, 
  $\pr(\mathrm{reject} \, H_0)\le\alpha$,
  under $H_a$, $\pr(\mathrm{reject} \, H_0)\to1$.
\end{Pro}

\section{Test with estimated mean and covariance}\label{sec:unknown}

\subsection{Test Statistic Construction}

 We now consider the realistic situation where $\bmu, \bSigma$
 are unknown and need to be estimated.
 A most natural thing is to estimate $\bmu$ and $\bSigma$ so as to
 obtain the estimated $\Y_i$'s and then proceed with the testing
 procedure described in Section \ref{sec:known}. However, to avoid  the
 dependence among the normalized observations induced by $\wh\bmu, \wh\bSigma$, we split the data into two subsets with sample sizes $n_1$ and $n_2$, respectively. We use the first $n_1$ observations to estimate $\bmu,
\bSigma$, and then form $\wh\Y_i, \wh U_i, \wh\V_i$ in the last $n_2$
observations and perform the test. Specifically,
we form
the sample mean and variance $\wh\bmu=n_1^{-1}\sumja\X_j$ and 
$\wh\bSigma=n_1^{-1}\sumja(\X_j-\wh\bmu)^{\otimes2}$. Subsequently,
we let  
$\wh\Y_i=\wh\bSigma^{-1/2}(\X_i-\wh\bmu)$, $\wh U_i=\|\wh\Y_i\|$ and
$\wh \V_i=\wh\Y_i/\wh U_i$ for $i=n_1+1, \dots, n$.
We now define the test statistic to be
\be\label{eq:T1}
T_1 &\equiv& -\wh H_{n_2}(\wh U,\wh\V)+\wh H_{n_2}(\wh U)-
\log c_p\n\\
&=&-\wh H_{n_2}(\wh \Y)+(p-1)\wh E_{n_2}\{\log (\wh U)\}+\wh
H_{n_2}(\wh U)-\log c_p
\ee
based on the data $(\wh U_i, \wh\V_i), i=n_1+1, \dots, n$.

\subsection{Asymptotic Properties}
We further obtain the asymptotic distribution of  $T_1$. The
asymptotic expansion of $T_1$ is established in Theorem
\ref{th:unknown1}.

\begin{Th}\label{th:unknown1}
Let $T_1$ be as given in \eqref{eq:T1}, where $k_1$ and
  $w_{1j}, j=1, \dots, k_1$ satisfy the requirements in Theorem 1 of \cite{bsy2019}
  with respect to $f_U$, and $k_p$ and $w_{pj},j=1, \dots, k_p$
  also satisfy the 
  requirements with respect to $f_\Y$.  See also Section \ref{sec:conditions} in the supplementary materials for
  these specific requirements.
Under the regularity conditions A, B, C, D, E, which are stated in  Section \ref{sec:assumptions} of the supplementary materials, there
  exists $k \ge 2$, so that
\bse
T_1 -d(f_{U,\V}\|f_Uf_0)&=&
n_1^{-1} \sumja\psi_1(\x_j,\bmu,\bSigma)
+n_2^{-1}\sumib\psi_2(\x_i,\bmu,\bSigma)\\
&& +o_{p}(n_2^{-1/2}+n_1^{-1/2}) +
O_p(n_2n_1^{-k}),
\ese 
where
  \bse
 \psi_1(\x,\bmu,\bSigma)
&=&\tr\left(
    \left[(p-1) E\left(\V\V\trans\right)-\I-E \left\{\frac{d\log f_U(U)}{dU}
U\V\V\trans\right\}\right]
\bpsi_{\bSigma^{-1/2}}(\x,\bmu,\bSigma) \bSigma^{1/2}
\right)\n\\
&&-  \left[ (p-1) E(\V\trans/U)
-E\left\{\frac{d\log f_U(U)}{du}\V\trans\right\}\right]
\y,\n\\
\psi_2(\x,\bmu,\bSigma)&=&  \log f_\Y(\y) + (p-1)\log(u) 
-\log f_U(u)
 +H(\Y) - (p-1) E \{ \log (U )\}-H(U),\\
\bpsi_{\bSigma^{-1/2}}(\x,\bmu,\bSigma) &=&
\dvec\{-(\bSigma^{1/2}\otimes\bSigma+
\bSigma\otimes\bSigma^{1/2})^{-1}\vec(\bpsi_{\bSigma}(\x,\bmu,\bSigma))\},\\
\bpsi_{\bSigma}(\x,\bmu,\bSigma) &=&(\x-\bmu)^{\otimes2}-\bSigma.
\ese
\end{Th}
\begin{Rem}
We can see that the estimation error of $T_1$ consists of two terms,
the first term captures the error caused by estimating
$\bmu$ and $\bSigma$, hence is of the order $n_1^{-1/2}$,
while the second term is the error caused by estimating the
entropies via the Kozachenko-Leonenko estimator, which is performed
using $n_2$ observations hence is of the order $n_2^{-1/2}$. Note that
$\psi_2(\x_i, \bmu,\bSigma)$ is identical to the influence function in
Theorem \ref{th:known}. 
\end{Rem}

  \noindent\underline{\bf Highlight of Proof of Theorem \ref{th:unknown1}:}
While we provide the complete proof rigorously in Section
  \ref{sec:proofth2} in the supplementary materials, we provide a summary of the
  proof and highlight the main difficulties overcome here.
  Our goal is to find the asymptotic expansion of the three terms:
  $\wh E_{n_2}\{\log(\wh U)\} - E\{\log(U)\}$, $-\wh H_{n_2}(\wh \Y) +
  H(\Y)$, and $\wh H_{n_2}(\wh U) - H(U)$. We discuss the three terms
  separately. 

On expanding $\wh E_{n_2}\{\log(\wh U)\} - E\{\log(U)\}$, it is
crucial to find the asymptotic expansion of \\ $n_2^{-1}\sumib \{
\log(\wh U_i) - \log(U_i) \}$. It is then natural to link the term
$\log(\wh U_i) - \log(U_i)$ to $\wh U_i - U_i$ by Taylor expansion,
and further link to $\wh \Y_i - \Y_i$. However, two issues appear in
this procedure: (i) we need to guarantee that the average of the
individual ignorable terms in the expansion of $\wh U_i - U_i$ or $\wh
\Y_i - \Y_i$ are still ignorable; and (ii) the Lagrange form remainder
involves a term $\frac{1}{\wt U_i^2}$ as a coefficient, where $\wt
U_i$ is between $U_i$ and $\wh U_i$, and this term will be
uncontrollable when $\wh U_i$ is too small.  

To deal with issue (i), we retain all residual terms
explicitly in $\wh U_i - U_i$ or $\wh \Y_i - \Y_i$, and directly show
that each term is ignorable after averaging. To solve
issue (ii), instead of the Lagrange form remainder, we utilize the
integral form remainder in the Taylor expansion. Also, we truncate
each $\wh U_i$ at $\epsilon_n U_i$ for some small $\epsilon_n$. On one
hand, under the nice event where none of $\wh U_i$'s are
  less than $\epsilon_n U_i$, we can bound the remainder and
 derive its order. On the other hand, we show that the
complementary event, where some $\wh U_i$ is extremely
small, is sufficiently rare to be ignored. Combining the two parts
gives the final order of the remainder term. 

On expanding $-\wh H_{n_2}(\wh \Y) + H(\Y)$, we mainly need to expand
two parts: $n_2^{-1}\sumib\{\log f_{\wh\Y}(\wh\y_i)-\log
f_\Y(\wh\y_i)\}$ and $n_2^{-1}\sumib\{\log f_\Y(\wh\y_i)-\log
f_\Y(\y_i)\}$. On expanding the first part, we first link $\log f_{\wh
  \Y}(\y)$ to $\log f_\Y(\y)$ through variable transformation and
Taylor expansion, and further bound the remainder terms in $\log
f_{\wh\Y}(\wh\y_i)-\log f_\Y(\wh\y_i)$ using the Lipschitz conditions
on the second derivative. The second part can be treated similarly
through Taylor expansion, and the remainder term in $\log
f_\Y(\wh\y_i)-\log f_\Y(\y_i)$ can be similarly bounded. 

Similar to the previous term, on expanding $\wh H_{n_2}(\wh U) -
H(U)$, we mainly need to expand two parts: $n_2^{-1}\sumib\{\log
f_{\wh U}(\wh u_i)-\log f_U(\wh u_i)\}$ and $n_2^{-1}\sumib\{\log
f_U(\wh u_i)-\log f_U(u_i)\}$. In the first term, linking a single
$f_{\wh U}(u)$ to $f_U(u)$ requires transforming back to integrals
related to $f_\Y(\y)$, which adds to the complexity of the remainder
terms in a single expansion of $\log f_{\wh U}(\wh u_i)-\log f_U(\wh
u_i)$. Afterwards, we again encounter the two issues as
in expanding $\wh E_{n_2}\{\log(\wh U)\} - E\{\log(U)\}$, and similar
techniques are applied to solving them. In particular, for the second
issue, here we do a truncation on each $f_{\wh U}(\wh u_i)$ and show
the similar two parts as before. The second part is also handled by
Taylor expansion, where we use the Lipschitz condition and also
conduct further analysis on the ignorable terms.
Combining the three terms leads to the desired result.\qed

Theorem \ref{th:unknown1} also indicates that the convergence rate of $T_1-
d(f_{U,\V}\|f_Uf_0)$ relies on the sample sizes of both subsets and is
determined by the smaller subsample size. Thus, a
natural choice is to set $n_1=n_2=\lfloor n/2\rfloor$, which leads to
$n_1^{1/2}\{ T_1 -d(f_{U,\V}\|f_Uf_0)\}=
n_1^{-1/2} \sumja\psi_1(\x_j,\bmu,\bSigma)
+n_2^{-1/2}\sumib\psi_2(\x_i,\bmu,\bSigma)+o_p(1)$.
To compensate for the lost power caused by the reduced sample size due
to the data splitting, we reverse
the roles of the two subsets of the data to obtain $T_2$ and form
$T=(T_1+T_2)/2$. Specifically, $T$ can be represented by
\be\label{eq:T-def}
T = -\wb H_n(\wh\Y) + (p-1)\wb E_n\{\log(\wh U)\}  + \wb H_n(\wh U) - \log c_p,
\ee
where
\be\label{eq:hbar-ebar}
\wb H_n(\wh\Y) &=& \{ \wb H_{n_1}(\wh\Y) + \wb H_{n_2}(\wh\Y) \} /2,\n\\
\wb E_n\{\log(\wh U)\} &=& [\wb E_{n_1}\{\log(\wh U)\}+\wb E_{n_2}\{\log(\wh U)\}]/2, \n\\
\wb H_n(\wh U) &=& \{ \wb H_{n_1}(\wh U) + \wb H_{n_2}(\wh U) \} /2.
\ee
The final test statistic has the property established in
Theorem \ref{th:unknown}.

\begin{Th}\label{th:unknown}
Under the same conditions as in Theorem \ref{th:unknown1},
the test statistic $T$ satisfies
$
n^{1/2}\{T -d(f_{U,\V}\|f_Uf_0) \}=n^{-1/2}\sumi\psi(\x_i,\bmu,\bSigma)+o_p(1),
$
where
$\psi(\x,\bmu,\bSigma)=\psi_1(\x,\bmu,\bSigma)+\psi_2(\x,\bmu,\bSigma)$. 
Under $H_0$, $n^{1/2}T = o_p(1)$,  and
under $H_a$,
\bse
n^{1/2}\{T -d(f_{U,\V}\|f_Uf_0) \}
\to N[0, E\{\psi(\X,\bmu,\bSigma)^2\}]
\ese
in distribution when $n\to\infty$.
\end{Th}

In practice, we can perform test using
\bse
n\wh\var(T)
&\approx&
n^{-1}\sumi
\left(
\wh \psi_1 ( \X_i, \wh \bmu, \wh \bSigma) + \wh \psi_2 ( \X_i, \wh \bmu, \wh \bSigma)
\right)^2+n^{-c},
\ese
where
\bse
\wh \psi_1(\X_i,\wh \bmu,\wh \bSigma)
&=&\tr\left(
    \left[(p-1) n^{-1} \sumj \wh \V_j \wh \V_j \trans -\I- n^{-1} \sumj \frac{\wh f_U'(\wh U_j)}{\wh f_U(\wh U_j)}
\wh U_j \wh \V_j \wh \V_j \trans \right]
\wh\bpsi_{\bSigma^{-1/2}}(\X_i,\wh \bmu,\wh \bSigma) \wh \bSigma^{1/2}
\right)\n\\
&&-  \left[ (p-1) n^{-1} \sumj \wh \V_j\trans / \wh U_j
-n^{-1} \sumj \frac{\wh f_U'(\wh U_j)}{\wh f_U(\wh U_j)}\wh\V_j\trans\right]
\wh \bSigma^{-1/2} (\X_i-\wh \bmu),\n\\
\wh \psi_2(\X_i,\wh \bmu,\wh \bSigma)&=&  (p-1)\log(\wh U_i) -\sum_{j=1}^{k_p}w_{pj}\log\left[
  \frac{(n-1)\|\wh \Y_{(j),i}-\wh \Y_i\|^p{\cal V}_p}{\exp\{\psi(j)\}}\right]\\
  &&
+\sum_{j=1}^{k_1}w_{1j}\log\left[
  \frac{(n-1)|\wh U_{(j),i} - \wh U_i| {\cal V}_1}{\exp\{\psi(j)\}}\right] +\wh H_n(\wh\Y) - (p-1) n^{-1} \sumj \log \|\wh \Y_j\|  -\wh H_n (\wh U),\\
\wh \bpsi_{\bSigma^{-1/2}}(\X_i,\wh \bmu,\wh \bSigma) &=&
\dvec\{-(\wh \bSigma^{1/2}\otimes \wh \bSigma+
\wh \bSigma\otimes \wh \bSigma^{1/2})^{-1}\vec(\wh \bpsi_{\bSigma}(\X_i,\wh \bmu,\wh \bSigma))\},\\
\wh \bpsi_{\bSigma}(\X_i,\wh \bmu,\wh \bSigma) &=&(\X_i-\wh \bmu)^{\otimes2}-\wh \bSigma,
\ese
Here, we use the kernel density estimator for $f_U$ and its derivative as
\bse
\wh f_U(u) = \frac{1}{nh} \sum_{k=1}^n K\left(\frac{u-\wh U_k}{h}\right), \quad \text{and} \quad \wh f_U'(u) = \frac{1}{nh^2} \sum_{k=1}^nK'\left(\frac{u-\wh U_k}{h}\right).
\ese
We may set $K$ as the Gaussian kernel, i.e.,
$K(u)=(2\pi)^{-1/2}\exp(-u^2/2)$. Note that the corresponding $K'(u)=-(2\pi)^{-1/2}\exp(-u^2/2)u$.

\begin{Rem}\label{rem:noloss}
  A comparison between Theorems \ref{th:unknown1} and \ref{th:unknown}
 suggests that the sample size loss  
due to sample splitting for $T_1, T_2$ is fully recovered in
 $T$. Similarly, the sample size loss in estimating the entropies is
 also fully recovered in $T$. On the other hand, the potential
 dependence between the estimation of $\bmu, \bSigma$ and the estimation
 of the entropies based on the same data is reflected in the fact that
 $\phi_1(\x_i, \bmu,\bSigma)$ and $\phi_2(\x_i, \bmu,\bSigma)$ are
 evaluated on the same observations in Theorem \ref{th:unknown} in
 contrast to in Theorem \ref{th:unknown1}. We thus conjecture that if
 we had not performed sample splitting, i.e., if we had performed all the
 estimations, including that for $\bmu, \bSigma$ and the entropies
 using the whole sample, the test statistic would have the same
 asymptotic properties. In other words, the sample splitting procedure
 does not incur any power loss.
\end{Rem}

\subsection{Variance inflation}\label{sec:var-inflation-2}

Similar to the known $\bmu$ and $\bSigma$ case in Section \ref{sec:var-inflation}, we consider two terms separately:
\begin{enumerate}
\item We know that 
\bse
n^{1/2} \{ -\wb H_{n}(\wh \Y) + H (\Y) \} 
&=&n^{-1/2}\sumi \psi_{T1}(\x_i,\bmu,\bSigma)+o_p(1)
\ese
where
\bse
\psi_{T1}(\x,\bmu,\bSigma)&=& \log f_\Y(\y) +  H (\Y) -  \tr\left\{\bpsi_{\bSigma^{-1/2}}(\x,\bmu,\bSigma)\bSigma^{1/2}\right\}.
\ese
Let $V_1 = \var \{ \psi_{T1}(\x,\bmu,\bSigma) \}$. We can estimate $V_1$ by
\be\label{eq:v1hat-unknown}
\wh V_1 &=&  n^{-1} \sumi \left(- \sum_{j=1}^{k_p} w_{pj} \log \left[\frac{(n-1)\|\wh \Y_{(j),i}-\wh \Y_i\|^p {\cal V}_p}{\exp\{\psi(j)\}}\right] + \wh H_n(\wh \Y) \right. \n\\
&&\left.- \tr\{\bpsi_{\bSigma^{-1/2}}(\X_i, \wh \bmu, \wh \bSigma) \wh \bSigma^{1/2}\} \right)^2.
\ee
\item We know that 
\bse
n^{1/2}(\wb E_n\{\log(\wh U)\} - E\{\log(U)\}) = n^{-1/2} \sumi  \psi_{T2}(\x_i,\bmu,\bSigma) + o_p(1),
\ese
and
\bse
n^{1/2}\{\wb H_n(\wh U) - H(U)\} = n^{-1/2} \sumi  \psi_{T3}(\x_i,\bmu,\bSigma)  + o_p(1),
\ese
where
\bse
 \psi_{T2}(\x,\bmu,\bSigma) 
&=& \tr\left\{ E\left(\V\V\trans\right)
\bpsi_{\bSigma^{-1/2}}(\x,\bmu,\bSigma) \bSigma^{1/2}\right\}
-E\left(\V\trans/U\right)\y +\log(u) -E\{\log(U)\},\\
\psi_{T3}(\x,\bmu,\bSigma)&=&
- \log f_U(u) -  H (U)
 -\tr\left[\bSigma^{1/2}
E\left\{\frac{d\log f_U(U)}{du}
U\V\V\trans\right\}\bpsi_{\bSigma^{-1/2}}
(\x,\bmu,\bSigma)\right]\\
&&
+ E\left\{\frac{d\log f_U(U)}{du}\V\trans\right\}
\y.
\ese
Let $V_2=\var\{(p-1) \psi_{T2}(\X,\bmu,\bSigma) +
\psi_{T3}(\x,\bmu,\bSigma) \}$, and we can estimate $V_2$ by 
\be\label{eq:v2hat-unknown}
\wh V_2 &=& n^{-1} \sumi \left( (p-1) \left[\tr\left\{\A_1
\wh \bpsi_{\bSigma^{-1/2}}(\X_i,\wh \bmu,\wh \bSigma) \wh \bSigma^{1/2}\right\}
-\A_2 \wh \Y_i +\log(\wh U_i) - \wb E_n \{ \log (\wh U) \} \right] \right.\n\\
&& + \left. \sum_{j=1}^{k_1} w_{1j} \log \left[\frac{(n-1)|U_{(j),i}-U_i| {\cal V}_1}{\exp\{\psi(j)\}}\right] - \wh H_n(U)  \right. \n\\
&&\left.-\tr\left\{\B_1 \wh \bpsi_{\bSigma^{-1/2}}
(\X_i,\wh \bmu,\wh \bSigma)\wh \bSigma^{1/2}\right\}+
\B_2 \wh \Y_i \right)^2.
\ee
where
\bse
&&\A_1= n^{-1} \sumj \wh \V_j \wh \V_j \trans,\quad \A_2=n^{-1} \sumj \wh \V_j\trans / \wh U_j , \\
&&\B_1 = 
n^{-1} \sumj \frac{\wh f_U'(\wh U_j)}{\wh f_U(\wh U_j)}
\wh U_j \wh \V_j \wh \V_j \trans,\quad
\B_2 = n^{-1} \sumj \frac{\wh f_U'(\wh U_j)}{\wh f_U(\wh U_j)}\wh\V_j\trans.
\ese
\end{enumerate}

The confidence interval, decision rule and p-value are the same
  as in Section \ref{sec:var-inflation}, with $\wh V_1$ and $\wh V_2$
   in $\wh \sigma^2$ replaced by the corresponding quantities above. 
Similar to Proposition \ref{prop:level-power-known}, the next proposition shows that our test can still control the size while preserve the power after the above variance inflation steps. The proof is  in parallel with that of Proposition \ref{prop:level-power-known} by directly replacing Theorem \ref{th:known} by Theorem \ref{th:unknown}, and thus is omitted.
\begin{Pro}
Let $T$ be given in \eqref{eq:T-def}, and $\wh\sigma^2$ be given in \eqref{eq:sigma2hat} with $\wh V_1, \wh V_2$ defined by \eqref{eq:v1hat-unknown} and \eqref{eq:v2hat-unknown}, respectively.
    For any $0<\alpha<1/2$, when $n$ is sufficiently large,
    under $H_0$, 
  $\pr(\mathrm{reject} \, H_0)\le\alpha$,
  and under $H_a$, $\pr(\mathrm{reject} \, H_0)\to1$.
\end{Pro}

\section{Debiasing}\label{sec:debias}

Under $H_0$,  the first-order influence function of $\wh T$ is zero, 
while the bias in estimation is usually only guaranteed to be
$o(n^{-1/2})$, and hence the bias may dominate the standard deviation of
$\wh T$. Even under $H_0$, when bias is ignorable, it is always
desirable to correct the potential bias. 
We adopt a resampling technique to perform the debiasing.

Specifically, when $\bmu$ and $\bSigma$ are known, for $b=1, \ldots, B$
and $i=1, \ldots, n$, 
we first independently generate $\V_{bi}^*$ from the uniform
distribution on the unit sphere, by first generating
a standard normal sample $\Z_{bi}^*$ and then setting
$\V_{bi}^*=\Z_{bi}^*/\|\Z_{bi}^*\|$. We then form
$\X_{bi}^* = \bmu + \bSigma^{1/2}U_i \V_{bi}^*$.
Thus, for each $b=1,\ldots,B$, we compute a new
test statistic $T_b^*$ using the data $\X_{b1}^*, \ldots,
\X_{bn}^*$. We then use $\wb T_b = B^{-1}\sum_{b=1}^B T_b^*$ as an estimate
of the bias, and use $T'=T-\wb T_b$ as the debiased test statistic.
When $\bmu$ and $\bSigma$ are unknown, we follow the same procedure
except that we replace the construction of $\X_{bi}^*$ by
$\X_{bi}^* = \wh \bmu + \wh\bSigma^{1/2}\wh U_i \V_{bi}^*$.

\section{Simulations}\label{sec:simu}
We consider $p=2,5,10$ and $n=500,1000$. 
For each $n$ and $p$, we select  the tuning parameters $k_p$ and
$k_1$  following Theorem 1 of \cite{bs2019}.
 Specifically, Theorem 1 of \cite{bsy2019}  gives the requirements on the order of
  $k$.  We thus determine $k_1$ and $k_p$ based on these
  order requirements and the finite sample performance. 
  In terms of the order requirements, we may plug in
$\alpha=\beta=\infty$,  i.e., we 
assume the density is sufficiently smooth and the tail is sufficiently
light.  Specifically, we set $k_p=\lceil p n^{\tau(p)} \rceil$, where  $\tau(p)
\le \min(\frac{2}{5},\frac{4}{4+3p},1-\frac{p/4}{1+\lfloor
    p/4\rfloor})$. In practice, we find $\tau(1) = \frac{1}{4}$, 
    $\tau(2) = \frac{2}{5}$, $\tau(5) = \frac{4}{19}$, and $\tau(10) =
    \frac{2}{17}$ performs well and we suggest to use these values for
    simplicity. When $p \ge 2$, a possibly good choice is
 $\tau(p) = \min(\frac{4}{4+3p},1-\frac{p/4}{1+\lfloor
    p/4\rfloor})$, and the above values of $\tau(2)$, $\tau(5)$ and $\tau(10)$
 are calculated based on this. This is to ensure that we have enough
nearest neighbors to use when estimating the entropies, especially
when the dimension $p$ is high.
If $p>3$, we select the optimal weights $\{w_{p1},\ldots,w_{pk_p}\}$
by the function \texttt{L2OptW} in the R package \texttt{IndepTest}
\citep{IndepTest}, which, as recommended by \cite{bs2019}, minimizes
the Euclidian norm of $(w_{p1},\ldots,w_{pk_p})\trans$ under 
  the constraints 
$\mathcal{W}^{(k_p)}$ in \eqref{eq:wk}. 
If $p\le 3$, we directly set $w_{p1}=\ldots=w_{pk_p}=k_p^{-1}$
  to incorporate the information of
more nearest neighbors. 
Similarly, we set $w_{11}=\ldots=w_{1k_1}=k_1^{-1}$. 
When $\bmu$ and $\bSigma$ are unknown, we set the
bandwidth $h$ to be $n^{-1/5}$. We set the resampling sample size as $B=100$.

\subsection{Cases with known mean and covariance}
We first consider the case when $\bmu$ and $\bSigma$ are known.
We consider the following setting to generate data.
  
\textbf{Setting 1}: For a fixed
skewness parameter
$s \in \{0,1,\ldots,p\}$, we generate $\X=(X_1,\ldots,X_p)\trans$,
where $X_1,\ldots,X_s$ are iid $(\chi^2(2)-2)/2$ and
$X_{s+1},\ldots,X_p$ are iid $N(0,1)$. Thus, when
$s=0$, $X_1,\ldots,X_p$ are iid $N(0,1)$; when $s=p$, $X_1,\ldots,X_p$
are iid $(\chi^2(2)-2)/2$. 

Under Setting 1, $\X$ follows an elliptical distribution
when $s=0$, and deviates from the elliptical distribution as $s$
increases. Thus, $s$ can be considered as a parameter that
measures the departure from the null distribution, with a larger 
  $s$ indicates larger departure from elliptical distribution.
  
\begin{figure}[htbp]
\centering
\includegraphics[width=0.9\textwidth]{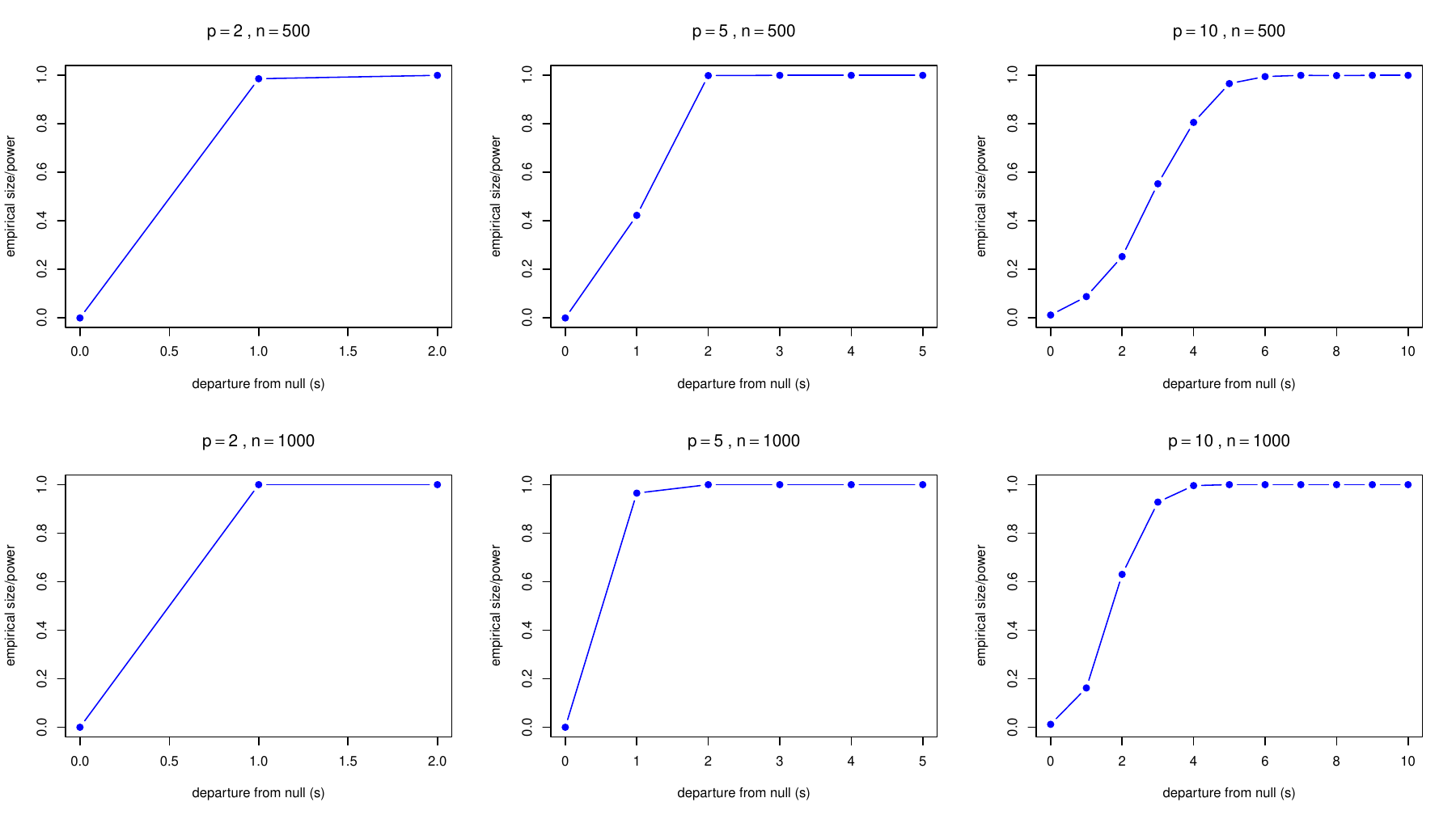}
\caption{Simulation results for known $\bmu$ and $\bSigma$ in Setting 1.}
\label{fig:res-known}
\end{figure}

The preliminary simulation results are presented in Figures
\ref{fig:res-known}. As we can see,
when $s=0$, which corresponds to $H_0$, the size of our test is well
controlled (all the sizes are below the nominal value 0.05 at the 0.05 test level).
As $s$ increases, the empirical power also
increases very sharply.  This suggests that the test is consistent
while achieves good power.

\subsection{Cases with estimated mean and covariance}\label{sec:un}
We now consider the case when $\bmu$ and $\bSigma$ are estimated. In 
addition to Setting 1 above without using the information of $\bmu$
and $\bSigma$, we further consider the following two settings.

\textbf{Setting 2}: We first generate $U \sim \mathrm{Gamma} (4,2)$,
where the second parameter is the rate, and $\V \sim
\mathrm{Uniform}(\mathrm{S}^{p-1})$. For a fixed parameter $s
\in \{0,1,\ldots,p\}$, we generate $W_j \sim \mathrm{Gamma} (
\gamma_j, \alpha_j)$ for $j=1,\ldots,s$, with $\bg$ and $\ba$
pre-specified. We then compute $\X=(X_1,\ldots,X_p)\trans$, where $X_j
= U V_j / \sqrt{W_j}$ for $j=1,\ldots,s$, and $X_j = U V_j$ for
$j=s+1,\ldots,p$. In practice, we truncate each $W_j$ on the left at
$10^{-3}$ to avoid singularity. Regarding the pre-specified
parameters, we set $\bg =(0.5,2,0.4,3,0.3,4,0.2,5,0.1,6)\trans$ and
$\ba =(1,2,3,0.1,0.2,0.3,4,5,0.4,0.5)\trans$.
Note that when $s=0$, $\X$ has an elliptical distribution.

\textbf{Setting 3}: We first generate $\V=(V_1,\ldots,V_p)\trans \sim
\mathrm{Uniform}(\mathrm{S}^{p-1})$. 
Then based on $\V$, for a fixed parameter $s \in \{0,1,\ldots,p\}$, we
further generate $U|\V \sim \mathrm{Uniform}(\sum_{j=1}^s j^2 V_j^2,
\sum_{j=1}^s j^2 V_j^2 +1)$. We finally set $\X = U \V$.
Note that when $s=0$, $U \sim \mathrm{Uniform}(0,1)$,
$\X$ has an elliptical distribution.

\textbf{Setting 4}: We first generate $\Z = (Z_1, \ldots, Z_p) \trans
\sim t(6)$, the multivariate $t$ distribution with mean $\0$, shape
matrix $\I_p$ and 6 degrees of freedom. We also independently generate
$W \sim \mathrm{Bernoulli} (1/2)$.  For a fixed parameter $s
\in \{0,1,\ldots,p\}$, we compute $\X = (X_1, \ldots, X_p) \trans$ where $X_j = 20W + Z_j$ for $j=1,\ldots,s$, and $X_j=Z_j$ for $j=s+1,\ldots,p$.
Note that when $s=0$, $\X$ has a multivariate t distribution.

Note that, in
Setting 2, both the null and alternatives are centrally
  symmetric around $\0$ \citep{babic2019comparison},
while the alternatives are anisotropic due
to the variety of value combinations in $\bg$ and $\ba$. 
In Setting 3, the independence condition between $U$ and $\V$ are
violated under alternative, while $\V$ is still uniformly
distributed on the unit sphere. Setting 4 is the bimodal case, and the
null distribution is similar to Setting 1 but has a heavier tail. 
In all four
settings, the parameter $s$ measures the deviation from the null
hypothesis. 

\begin{figure}[htbp]
\centering
\includegraphics[width=\textwidth]{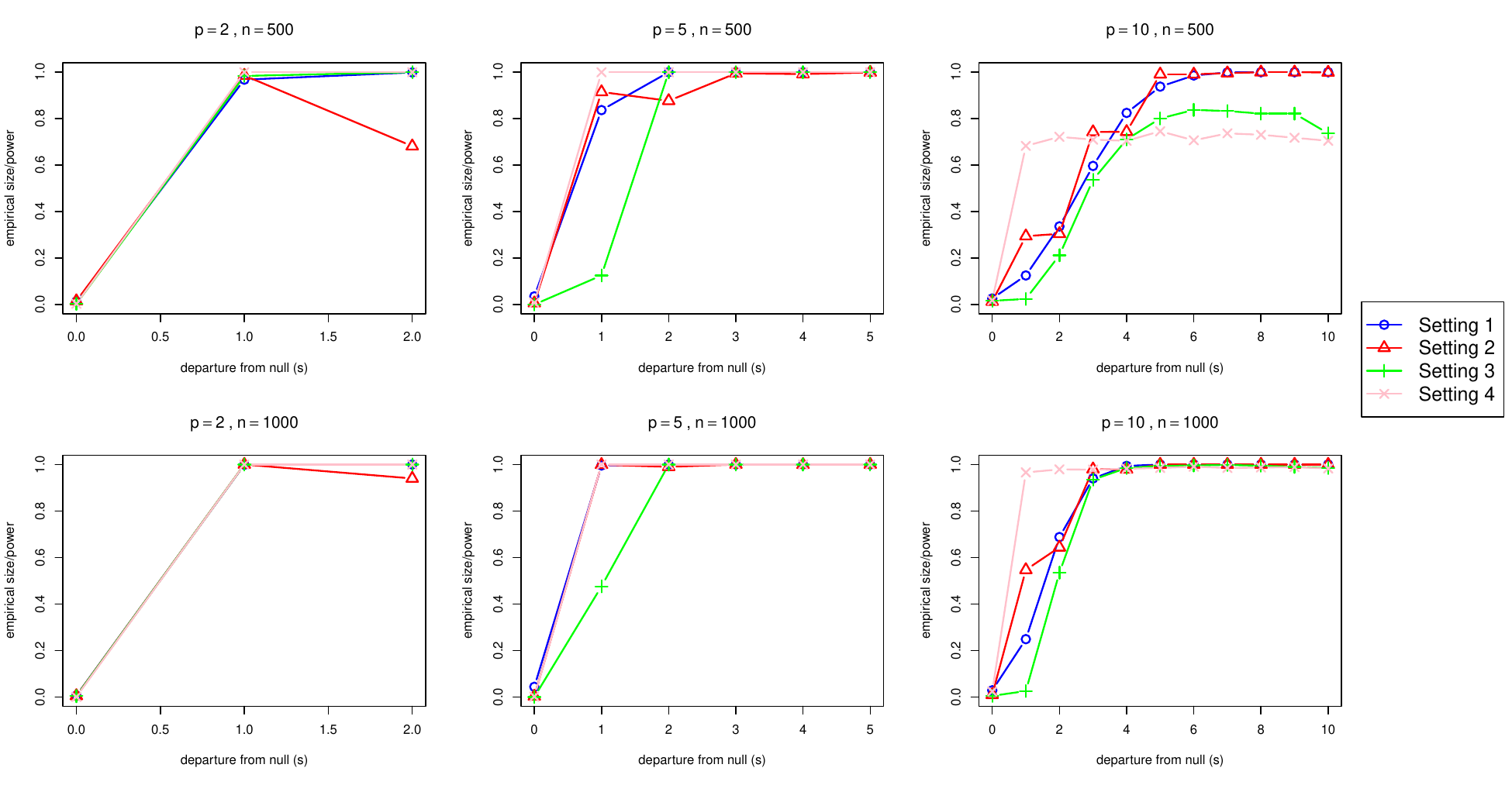}
\caption{Simulation results for unknown $\bmu$ and $\bSigma$ in four settings.}
\label{fig:res-unknown}
\end{figure}

The simulation results in the four settings above are presented in
Figure \ref{fig:res-unknown}. In all cases, when $s=0$, the size of
our test is also well controlled  with the maximum size being 0.044
  at the 0.05 level test. As we can see from Figure
\ref{fig:res-unknown}, the empirical power increases with the
departure from the null hypothesis in most cases. 
 In certain situation, we observe the empirical powers decrease
  slightly when $s=p$,  possibly because the distribution is more isotropic compared 
to the ones when $1 \le s \le p-1$ and hence
hard to distinguish.

\subsection{Comparison with existing methods}\label{sec:comp}
In this section we compare our method (KL) to  the
kernel-embedding-based
test (KE) in \cite{tang2024nonparametric}, as well as five other
tests summarized in \cite{babic2021rjournal}, including the tests
proposed by \cite{huffer2007test} (HP), by \cite{manzotti2002statistic}
(MPQ) , by \cite{cassart2008optimal} (PG), by \cite{schott2002testing} (SW),
and by \cite{babic2019optimal} (SO). The seven abbreviations of the tests will be used below.

We conduct our experiment under the same settings as in
  Section \ref{sec:un} and repeated the experiments 1000 times in
  general. However, one exception is made for the competing method KE,
  where we only repeated the experiments 100 times for sample size
  1000, due to its extremely large computational cost.

\begin{figure}[htbp]
\centering
\includegraphics[width=\textwidth]{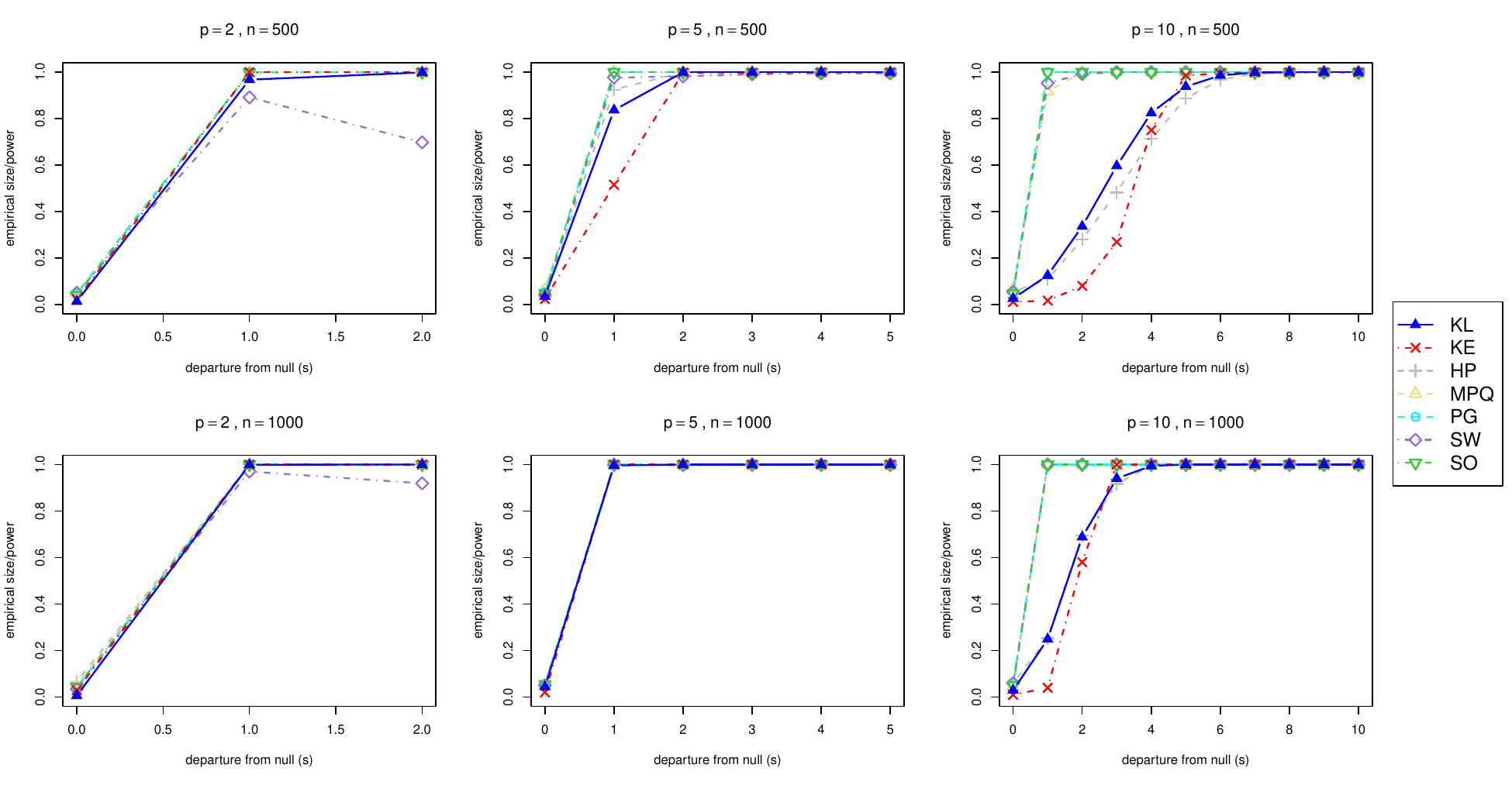}
\caption{Comparisons of seven methods under Setting 1.} 
\label{fig:res-comparison-1}
\end{figure}

\begin{figure}[htbp]
\centering
\includegraphics[width=\textwidth]{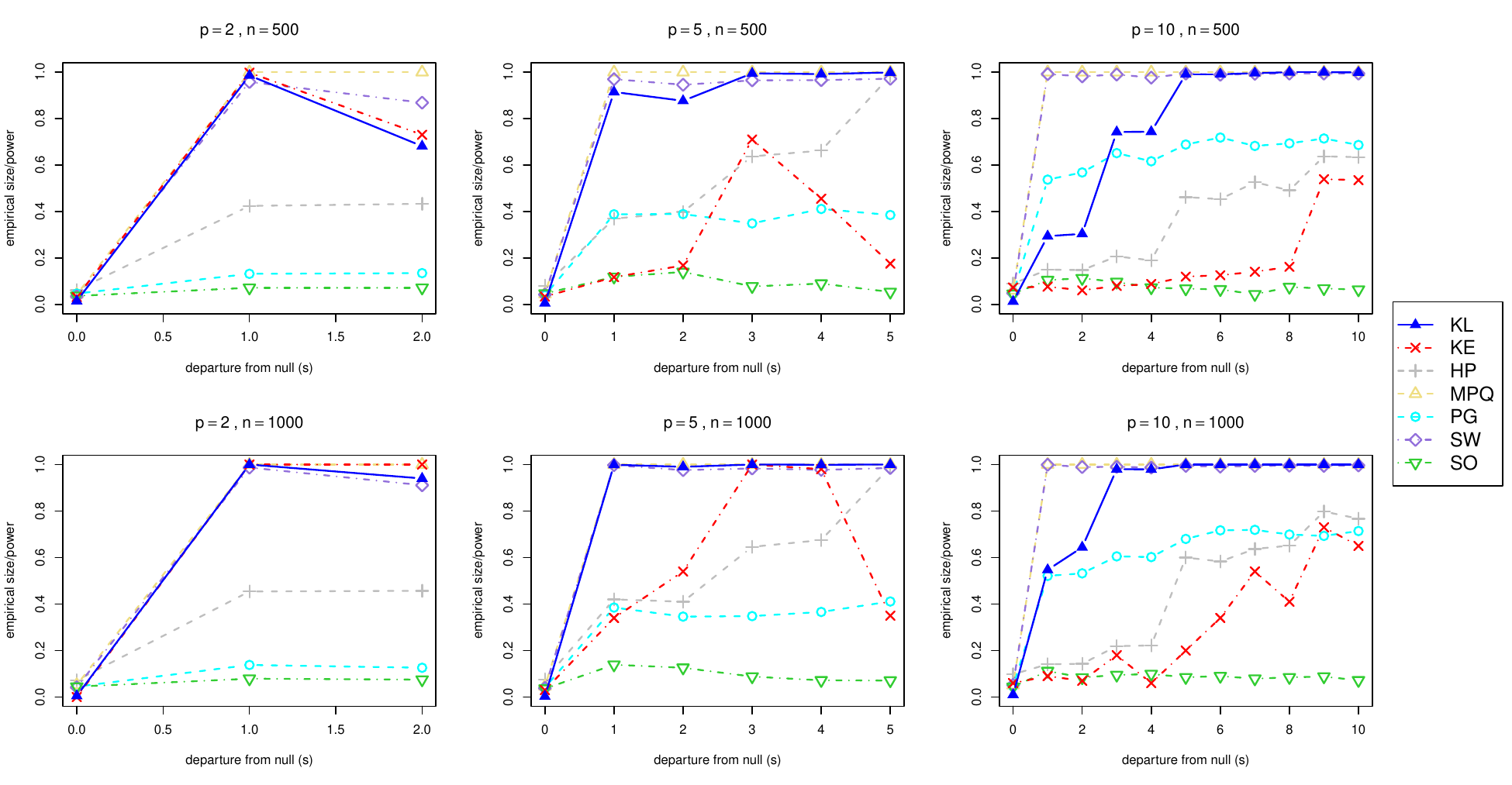}
\caption{Comparisons of seven methods under Setting 2.} 
\label{fig:res-comparison-2}
\end{figure}

\begin{figure}[htbp]
\centering
\includegraphics[width=\textwidth]{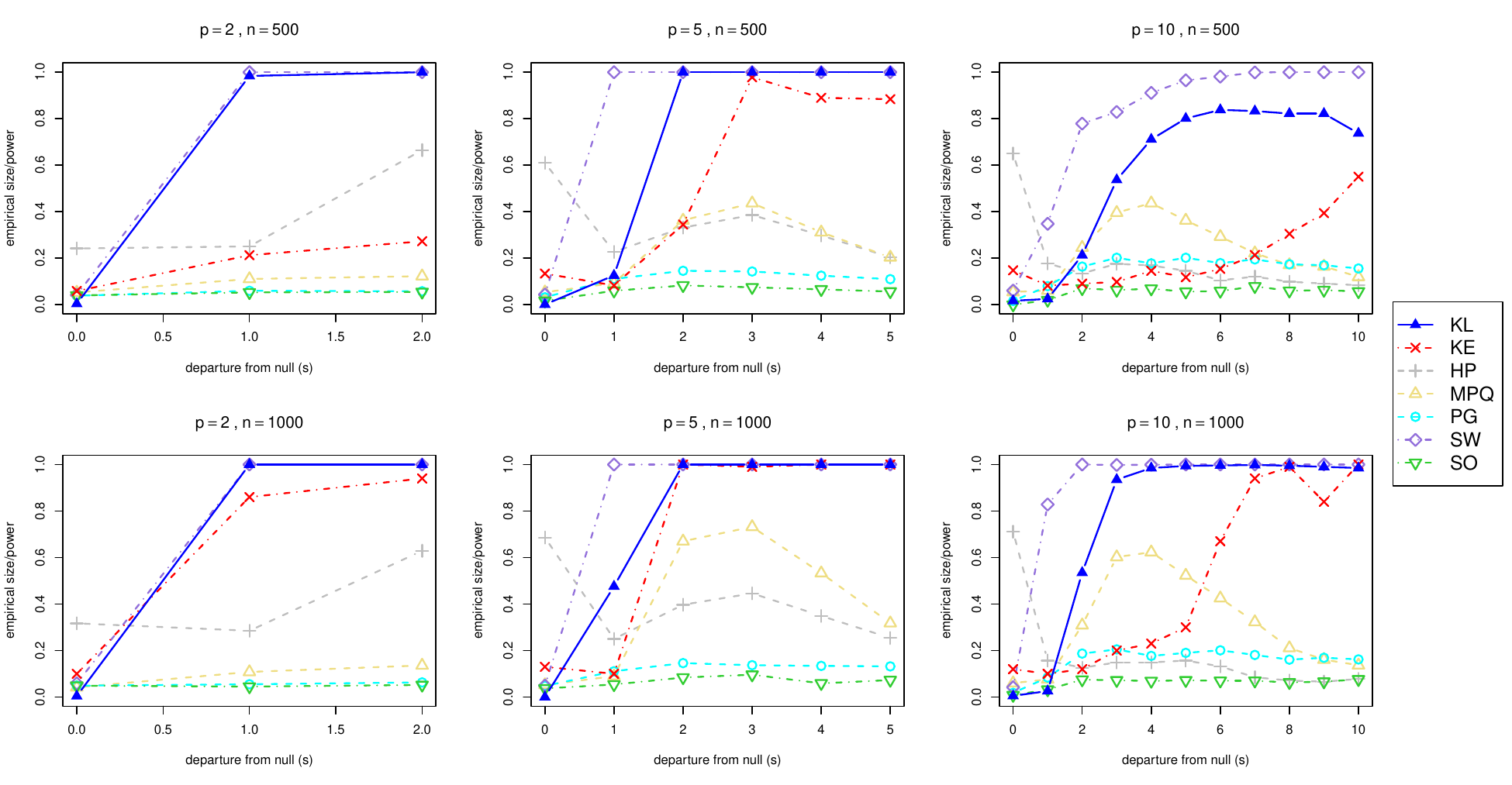}
\caption{Comparisons of seven methods under Setting 3.} 
\label{fig:res-comparison-3}
\end{figure}

The empirical sizes of the seven tests in the four settings
are summarized in Table \ref{tab:size}, while the rejection rate
  as a function of $s$, i.e., the departure level from null, are presented in
Figures \ref{fig:res-comparison-1}, \ref{fig:res-comparison-2},
\ref{fig:res-comparison-3} and \ref{fig:res-comparison-4}. 
The exact rejection values are given
in Tables S.1 -- S.4 in the supplementary materials.

\begin{table}[!htbp]
\centering
\begin{tabular}{crrrrrrrrr}
\hline
Setting & $n$ & $p$ & KL & KE & HP & MPQ & PG & SW & SO\\
\hline
\multirow{6}{*}{1}
 & 500 & 2 & 0.014 & 0.024 & 0.046 & 0.052 & 0.047 & 0.051 & 0.040\\
 & 500 & 5 & 0.036 & 0.024 & 0.034 & 0.065 & 0.054 & 0.043 & 0.049\\
 & 500 & 10 & 0.027 & 0.012 & 0.057 & 0.068 & 0.057 & 0.055 & 0.048\\
 & 1000 & 2 & 0.006 & 0.03* & 0.065 & 0.049 & 0.039 & 0.035 & 0.043\\
 & 1000 & 5 & 0.044 & 0.02* & 0.050 & 0.048 & 0.056 & 0.051 & 0.055\\
 & 1000 & 10 & 0.029 & 0.01* & 0.059 & 0.043 & 0.058 & 0.060 & 0.050\\
\hline
\multirow{6}{*}{2}
 & 500 & 2 & 0.016 & 0.033 & 0.063 & 0.044 & 0.048 & 0.036 & 0.038\\
 & 500 & 5 & 0.007 & 0.035 & 0.082 & 0.046 & 0.041 & 0.044 & 0.048\\
 & 500 & 10 & 0.014 & 0.074 & 0.091 & 0.062 & 0.047 & 0.050 & 0.047\\
 & 1000 & 2 & 0.005 & 0.00* & 0.072 & 0.052 & 0.045 & 0.042 & 0.045\\
 & 1000 & 5 & 0.003 & 0.03* & 0.076 & 0.046 & 0.041 & 0.039 & 0.036\\
 & 1000 & 10 & 0.010 & 0.06* & 0.098 & 0.051 & 0.046 & 0.052 & 0.044\\
\hline
\multirow{6}{*}{3}
 & 500 & 2 & 0.004 & 0.059 & 0.242 & 0.053 & 0.039 & 0.043 & 0.040\\
 & 500 & 5 & 0.001 & 0.134 & 0.610 & 0.054 & 0.032 & 0.044 & 0.017\\
 & 500 & 10 & 0.017 & 0.148 & 0.650 & 0.055 & 0.016 & 0.061 & 0.000\\
 & 1000 & 2 & 0.004 & 0.10* & 0.317 & 0.042 & 0.049 & 0.059 & 0.049\\
 & 1000 & 5 & 0.000 & 0.13* & 0.686 & 0.052 & 0.048 & 0.049 & 0.037\\
 & 1000 & 10 & 0.005 & 0.12* & 0.711 & 0.060 & 0.018 & 0.042 & 0.008\\
\hline
\multirow{6}{*}{4}
 & 500 & 2 & 0.004 & 0.040 & 0.068 & 0.057 & 0.042 & 0.029 & 0.049\\
 & 500 & 5 & 0.009 & 0.051 & 0.097 & 0.049 & 0.046 & 0.031 & 0.051\\
 & 500 & 10 & 0.019 & 0.042 & 0.085 & 0.055 & 0.064 & 0.043 & 0.044\\
 & 1000 & 2 & 0.001 & 0.07* & 0.081 & 0.049 & 0.041 & 0.021 & 0.053\\
 & 1000 & 5 & 0.002 & 0.02* & 0.108 & 0.052 & 0.027 & 0.039 & 0.047\\
 & 1000 & 10 & 0.024 & 0.00* & 0.073 & 0.047 & 0.046 & 0.030 & 0.033\\
\hline
\end{tabular}
\caption{Empirical sizes of the seven tests under four settings at significance level $\alpha=0.05$. The entries with * are based on 100 experiments, while others are based on 1000 experiments.} 
\label{tab:size}
\end{table}

\begin{figure}[htbp]
\centering
\includegraphics[width=\textwidth]{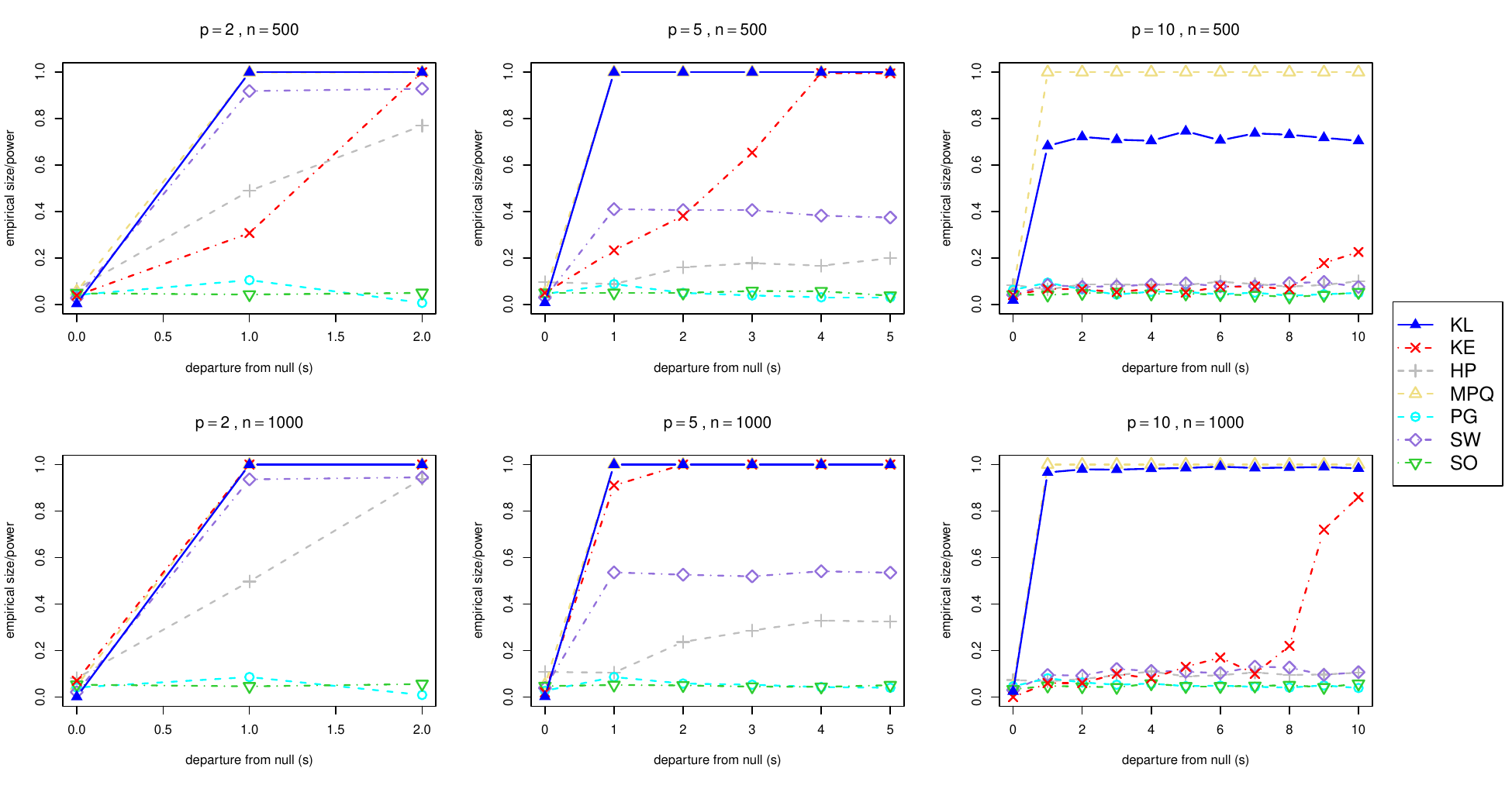}
\caption{Comparisons of seven methods under Setting 4.} 
\label{fig:res-comparison-4}
\end{figure}

The results in
 Table \ref{tab:size} indicate that the sizes of all methods except HP and KE are
 controlled well in all cases. HP has difficulties in
   maintaining the size under 0.05 in Settings 2, 3 and 4,  likely because the
 asymptotic null distribution of HP is based on the multivariate
 normal  distribution, but our null distributions in these three settings are
 elliptical distributions very
 different from normal. 
 The size of KE is not well controlled in Setting 3, and occasionally in Setting 2,
   probably because the distribution of $\Y$ is very concentrated
   around zero under the two circumstances (especially in Setting 3),
which leads to inaccurate estimation of the direction vector
  $\V$. This is the case where the Fr\'echet differentiability condition in
   \cite{tang2024nonparametric} is likely violated, leading KE to
   mistakenly reject the null hypothesis. In theory, our method
   KL should also encounter the similar issue in Setting 3, where
   the requirement $E(1/\|\Y\|^3)<\infty$ is violated
   when $U=\|\Y\| \sim \mathrm{Uniform}(0,1)$. However, KL exhibits
   certain robustness in that it still controls the size well after the debiasing
   and variance inflation step.

  Because a test has to maintain its size in order to
   control the type I error and deliver trustworthy decision, hence
   we will direct our attention more towards the power comparison
     of the remaining five testing procedures: KL, MPQ, PG, SW and SO.
   To this end, it
   is difficult to 
   claim an obvious winner in all settings. Figure
    \ref{fig:res-comparison-1} indicates that in Setting 1,  MPQ, PG, SW and SO are
   more powerful in general than KL.
   This is not a surprise since SO is shown
   to be the most  powerful test in detecting skew-elliptical distributions
   \citep{babic2019optimal}, while it is the skewness that causes the
   departure from null in this setting. 
  Also, as is shown in \cite{cassart2008optimal}, PG is optimal
     against a type of multivariate Fechner asymmetry, which includes
     our alternatives in Setting 1. MPQ is based on the uniform
     distribution of $\V$ on the unit sphere. When some
       components of $\X$  are  changed to $\chi^2$
     distributions, the resulting distribution of $\V$ is very far
     from uniform.
   SW is based on the structure of the fourth moments of the
     distributions. Under the skewness introduced by the $\chi^2$
     distributions, the moment structure deviates very much from
       the null.  Note that as the departure from null
       increases, all five methods are very powerful.
   
However, Figures
\ref{fig:res-comparison-2}, \ref{fig:res-comparison-3} and
\ref{fig:res-comparison-4} tell very different stories, 
in that SO is the least powerful in Settings 2, 3 and 4. This is because
in these settings, the departure from null is no longer due to the
skewness.  
Also, PG does not have good powers in Settings 3 and 4,
possibly due to the similar reason that the Fechner symmetry is not
strongly violated. MPQ does not perform well in Setting 3, because it
only tests the uniform distribution of $\V$ but does not consider the
dependency between $U$ and $\V$. On the other hand, SW does not
perform well in Setting 4, possibly because under the bimodal
setting, the fourth moment structure is not affected much.

Lastly, we comment briefly on the power performance of HP and
KE, the two tests that 
do not control the size well in all settings. The powers of HP in Settings 2, 3
and 4 are not good, possibly because the slicing step in HP leads to
much information loss. The powers of KE are generally not good when 
the sample size $n=500$, but have some improvement 
when the sample size increases from 
$n=1000$.  In addition, KE does not perform well  when
the dimension is high. This is probably because the deviations from the
null are in the direction orthogonal to the chosen basis functions.
 In contrast, KL controls size well, and retains good power in all settings,
even though it may not be the most powerful in each scenario.

\section{Application}\label{sec:real}
In this section, we apply our test to a real dataset on Travel Reviews
\citep{travel-reviews-dataset}. The dataset can be found from
{\text url {https://archive.ics.uci.edu/dataset/484/travel+reviews}.} In this
dataset, there are $n=980$ records, each of them including  $p=9$
features. Note that we exclude category 7 because it is discrete.
Thus, $X_1,\ldots,X_9$ correspond to categories 1,2,3,4,5,6,8,9,10.
The scatterplot of the original dataset is given in Figure
\ref{fig:original-dataset} of the supplementary materials.
Applying our test to the original dataset, the p-value is
approximately $2.0994 \times 10^{-6}$, which indicates that the
original dataset is very unlikely to be elliptically distributed.

We further perform a Box-Cox transformation on the dataset, using the method
given  in Section 7.2 of \cite{li2018sufficient}. The optimal Box-Cox
transformation parameter 
$\lambda$'s are: 
\bse
0.1053,  0.3158,  0.1053, -0.5263,  0.3158,  0.7368,  0.5263,  0.1053,  0.3158.
\ese
The scatterplot of the Box-Cox transformed dataset is given in Figure
\ref{fig:bctrans-dataset} of the supplementary materials.

We subsequently apply our test to the Box-Cox transformed dataset, and
obtain the 
p-value $6.0511 \times 10^{-6}$. This suggests that
the Box-Cox transformed dataset is also unlikely to be elliptically
distributed. This is reasonable because by Section 7.3 of
\cite{li2018sufficient}, the Box-Cox transformation can only improve
elliptical shapes to pairwise distributions of a data set, but highly
non-elliptical shapes may still appear in the joint distribution.

To further investigate the effect of the Box-Cox transformation, 
we conduct analysis on the ellipticity of pairwise distributions
$(X_i,X_j)$ for $1\le i<j \le p$, both on the original data and the
transformed data.  The p-values for the pairwise
comparisons on the original data are shown in the upper triangular part
of Table
\ref{tab:pairwise-comp}, and those on the Box-Cox transformed
data are shown in the lower triangular part
of Table
\ref{tab:pairwise-comp}.

\begin{table}[htbp]
\centering
\begin{tabular}{rrrrrrrrrrr}
\hline
\multicolumn{2}{c}{$j$} & 1 & 2 & 3 & 4 & 5 & 6 & 7 & 8 & 9\\
\hline
\multirow{9}{*}{$i$}
& 1& NA & 0.9984 & 0.0005 & 0.7355 & 0.9376 & 0.8552 & 0.9951 & 0.9960 & 0.9163\\
& 2& 0.9994 & NA & 0.0056 & 0.6425 & 0.9726 & 0.8955 & 0.9971 & 0.9886 & 0.9788\\
& 3& 0.2510 & 0.2164 & NA & 0.0000 & 0.0000 & 0.0000 & 0.0005 & 0.0015 & 0.0002\\
& 4& 0.9995 & 0.9934 & 0.0132 & NA & 0.2817 & 0.1073 & 0.6811 & 0.4973 & 0.1361\\
& 5& 0.9968 & 0.9985 & 0.0566 & 0.9316 & NA & 0.9034 & 0.9958 & 0.9846 & 0.9506\\
& 6& 0.9841 & 0.9693 & 0.0074 & 0.7380 & 0.9167 & NA & 0.9537 & 0.9473 & 0.7517\\
& 7& 0.9999 & 0.9961 & 0.1670 & 0.9632 & 0.9999 & 0.9524 & NA & 0.9976 & 0.9882\\
& 8& 1.0000 & 0.9991 & 0.1530 & 0.9882 & 0.9989 & 0.9682 & 0.9997 & NA & 0.9632\\
& 9& 0.9965 & 0.9670 & 0.0408 & 0.9648 & 0.9781 & 0.9548 & 0.9970 & 0.9975 & NA\\
\hline
\end{tabular}
\caption{P-values for pairwise comparisons on the original data (upper
  triangle) and the transformed data (lower triangle).}
\label{tab:pairwise-comp}
\end{table}

At the overall level $\alpha=0.05$, we test
$H_0$: $(X_i,X_j)$ is elliptically distributed for all the $(i,j)$
pairs. 
Using Bonferroni correction, we set the individual significant level 
$\alpha'=\alpha/36=0.0014$ since there are in total $p(p-1)/2=36$ comparisons. 
For the original data, the pairs $(X_i,X_j)$ that violates the
elliptical distribution requirement are
\bse
(X_1,X_3), (X_3,X_4),  (X_3,X_5),  (X_3,X_6), (X_3,X_7),  (X_3,X_9).
 \ese
 In contrast, all the pairs pass the test after the Box-Cox
 transformation. This result verifies empirically that the Box-Cox
 transformation indeed improves the elliptical shape
on the pairwise distributions of the data.

\section{Conclusion and Discussion}\label{sec:conclusion}
We have proposed a KL-divergence based test for elliptical
distribution. In assessing the KL-divergence, we convert it to
entropies in the Euclidean space, and estimate the entropies via $k$
nearest neighbor methods. We derive the asymptotic properties of the
test statistic when both the mean and covariance of the data are known
and unknown. Due to the
degeneration of the influence function, we have further conducted
debiasing and variance inflating,
 which leads to a more conservative test procedure in theory.
Achieving a test at the target level using this method will require
higher order influence function and is worth future studying. However,
based on the simulation
studies, our test achieves better size and power performance than the
state of the art procedures.

In this paper, we considered the fixed 
dimension $p$. Things can be very different when $p$ is allowed
to increase with $n$. Even when $p$ increases very slowly,
all the results need to be reinvestigated, including the theoretical and
numerical properties of the $k$NN-based estimator for the entropy.
When $p$ increases sufficiently fast, even the basic procedures such as
estimating the covariance matrix will encounter issues and additional
sparsity assumption will be needed and very different analysis will be
required. Indeed, testing for elliptical distribution in high
dimensional data is challenging, and we are aware of the work
\cite{banerjee2024consistent}, which discussed this issue and allowed
a rate of departure from the null slower than parametric rate under an
increasing $p$ without giving concrete results.
Recently, \cite{wang2025testing} proposed an asymptotic goodness-of-fit test
in high dimensions, where $p$ and $n$ are allowed to diverge proportionally.
Much more in depth analysis is need to investigate the
high dimension testing for elliptical distributions.

\clearpage
\pagenumbering{arabic}
\section*{Supplementary Materials}\label{sec:proofs}
\pagenumbering{arabic}
\setcounter{page}{1} 
\setcounter{equation}{0}\renewcommand{\theequation}{S.\arabic{equation}}
\setcounter{section}{0}\renewcommand{\thesection}{S.\arabic{section}}
\setcounter{subsection}{0}\renewcommand{\thesubsection}{S.\arabic{section}.\arabic{subsection}}
\setcounter{subsubsection}{0}\renewcommand{\thesubsubsection}{S.\arabic{section}.\arabic{subsection}.\arabic{subsubsection}}
\setcounter{Th}{0}\renewcommand{\theTh}{S.\arabic{Th}}
\setcounter{Lem}{0}\renewcommand{\theLem}{S.\arabic{Lem}}
\setcounter{Rem}{0}\renewcommand{\theRem}{S.\arabic{Rem}}
\setcounter{Cor}{0}\renewcommand{\theCor}{S.\arabic{Cor}}
\setcounter{table}{0}\renewcommand{\thetable}{S.\arabic{table}}
\setcounter{figure}{0}\renewcommand{\thefigure}{S.\arabic{figure}}

The supplementary materials contain regularity
  conditions, proofs of the main results, and additional tables and
  plots for 
 simulation results and the real data analysis. Because the proof Theorem \ref{th:known} is much
  simpler than that of Theorems \ref{th:unknown1} and
  \ref{th:unknown}, we omit its proof. 
  
  In the following sections, we 
first list the requirements of Theorem 1 of
\cite{bsy2019}
in Section \ref{sec:conditions}.
We   then provide the  proofs of Lemma \ref{lem:fuv} and Proposition \ref{prop:level-power-known} in  Sections   \ref{sec:proof-fuv} and \ref{sec:proof-known-prop}. 
We also
present the proofs of the Theorems \ref{th:unknown1} and \ref{th:unknown} as well as the conditions involved.
The proof of Theorem \ref{th:unknown1} is lengthy. We first list the
regularity assumptions in Section \ref{sec:assumptions}.
We classify these assumptions in five groups
named A, B, C, D,  and E, based on where these Assumptions are
used. We then prove some basic lemmas in Section \ref{sec:lemmas},
and prove Theorems \ref{th:unknown1} and \ref{th:unknown} in Sections \ref{sec:proofth2} and \ref{sec:proofth3}, respectively. Sections \ref{sec:values} and \ref{sec:plots} include some additional tables and
  plots for 
 numerical results. 

\section{List of Requirements in Theorem 1 of
  \cite{bsy2019}}\label{sec:conditions}

 Before stating the requirements for $k_1$, $k_p$, $\w_1$ and $\w_p$, we start with some concepts given in \cite{bsy2019}. We first give a general statement for dimension $d$. Suppose that $\Z$ is a random vector in $\real^d$ with density function $f$. 

Let $\mathcal{F}_d$ be the class of density functions in $\real^d$ with respect to the Lebesgue measure, 
and let $\mathcal{A}$ be the class of decreasing functions $a:(0,\infty) \to [1,\infty)$ such that $a(\delta) = o(\delta^{-\epsilon})$, as $\delta\searrow 0$ for all $\epsilon>0$. 
Let $\Theta = (0,\infty)^4 \times \mathcal{A}$, and define the parameter $\btheta = (\alpha, \beta, \nu, \gamma, a) \in \Theta$. 

Denote the target function class
\bse
\mathcal{F}_{d,\btheta} &=& \left\{ f \in \mathcal{F}_d: \mu_\alpha(f) \le \nu, \|f\|_\infty \le \gamma, 
  \sup_{\x: f(\x) \ge \delta} M_{f,\alpha,\beta}(\x) \le a(\delta) \mbox{ for } \forall \delta>0 \right\},
\ese
where
\bse
\mu_\alpha(f) = \int_{\real^d} \|\x\|^\alpha f(\x) d\x,
\ese
for $f \in \mathcal{F}_d$ and $\alpha>0$, and
\bse
M_{f,\alpha,\beta} (\x) = \max \left\{ \max_{t=1,\dots,m} \frac{\|f^{(t)}(\x)\|}{f(\x)}, \, \sup_{\y \in B_\x^\circ\{r_a(\x)\} }\frac{\|f^{(m)}(\y) - f^{(m)}(\x) \|}{f(\x) \|\y-\x\|^{\beta - m}}\right\},
\ese
where $m = \lceil \beta \rceil -1$, and $r_a(\x) = [8 d^{1/2} a\{ f(\x) \} ]^{-1/(\beta \wedge 1)}$. Note that when we mention $M_{f,\alpha,\beta} (\x)$, we assume that $f$ is $m$-times differentiable at $\x$. 

We then introduce the weight constraints. For a fixed $k \in \natural$, the class of weights $\w = (w_1,\dots,w_k)\trans$ is defined by
\be\label{eq:wk}
\mathcal{W}^{(k)} 
&=& \left\{ \w \in \real^k: \sum_{j=1}^k w_j \frac{\Gamma(j+2l/d)}{\Gamma(j)} = 0 \, \mbox{ for } \, l = 1, \dots, \lfloor d/4 \rfloor, \right. \n\\
&&\left. \sum_{j=1}^k w_j =1, \, \mbox{ and } \, w_j=0 \,\mbox{ if }\, j \notin \{ \lfloor k/d \rfloor, \lfloor 2k/d \rfloor, \dots, k \} \right\}.
\ee

Based on the above concepts, we now state the requirements for $k$ and $\w$ as in Theorem 1 of \cite{bsy2019}. We set $\alpha>d$ and $\beta>d/2$. 
We require $k \in \{k_0^*, \dots, k_1^*\}$,
where $k_0^* = k_{0,n}^*$ and $k_1^* = k_{1,n}^*$ are defined as two deterministic sequences of positive integers satisfying $k_0^* \le k_1^*$, $k_0^*/ \log^5 n \to \infty$, $k_1^* = O(n^{\tau_1})$ and $k_1^* = o(n^{\tau_2})$, where
\bse
\tau_1< \min \left( \frac{2\alpha}{5\alpha+3d}, \frac{\alpha-d}{2\alpha}, \frac{4\beta^*}{4\beta^*+3d}\right), \quad \tau_2 =  \min \left( 1 - \frac{d/4}{1+\lfloor d/4 \rfloor}, 1- \frac{d}{2\beta} \right), 
\ese
where $\beta^* = \beta \wedge 1$.  
In addition, we require $k \ge k_{\min}$, where $k_{\min} \in \natural$ satisfies that for each $k' \ge k_{\min}$, there exists $\w = \w^{(k')} \in \mathcal{W}^{(k')}$ with $\sup_{k' \ge k_{\min}} \|\w^{(k')}\| < \infty$. 
Note that $k_{\min}$ depends only on $d$, and its existence is guaranteed by Theorem 1 of \cite{bsy2019}.
When the sample size $n$ is sufficiently large, $k_{0,n}^* \ge k_{\min}$ is satisfied.
Also, the asymptotic result in \eqref{eq:hz-asymp} holds given that $k \in \{k_0^*, \dots, k_1^*\}$ and $\w = \w^{(k)} \in \mathcal{W}^{(k)}$ (see Theorems 1 and 8 of \cite{bsy2019}). 

The requirements for of $k_p$, $k_1$, $\w_p$ and $\w_1$ are based on the above statements, where $d$ is replaced by $p$ and $1$, respectively. To be specific, suppose that $f_U \in \mathcal{F}_{1,\btheta_1}$, where $\btheta_1 = (\alpha_1,\beta_1,\gamma_1,\nu_1,a_1)$, then we calculate $\tau_{11}$ and $\tau_{12}$ based on the form of $\tau_1$ and $\tau_2$ with $\alpha_1$, $\beta_1$ and $d=1$, and the order of the lower and upper bounds for $k_1$, namely $k_{10}^*$ and $k_{11}^*$, can be calculated accordingly. Similarly, when $f_\Y \in \mathcal{F}_{p,\btheta_p}$ where $\btheta_p = (\alpha_p,\beta_p,\gamma_p,\nu_p,a_p)$, then we calculate $\tau_{p1}$ and $\tau_{p2}$ using $\alpha_p$, $\beta_p$ and $d=p$, and the order of the lower and upper bounds for $k_p$, namely $k_{p0}^*$ and $k_{p1}^*$, can be calculated similarly. Based on the $k_1$ and $k_p$, we can select $\w_1 \in \mathcal{W}^{(k_1)}$ and $\w_p \in \mathcal{W}^{(k_p)}$.

\section{Proof of Lemma \ref{lem:fuv}}\label{sec:proof-fuv}
Define  $T: (0,\infty) \times \mathrm{S}^{p-1} \to \real^p \backslash \{\0\}, (u,\v) \mapsto u\v$. Clearly, $T$ is a continuous bijection, and $T^{-1}$ is also continuous.
For any measurable set $A \in (0,\infty) \times \mathrm{S}^{p-1}$, we have
\bse
\pr \{ (U,\V) \in A \} =  \int_{\mathrm{S}^{p-1}} \int_0^\infty I_{A} (u,\v) f_{U,\V}(u,\v) du d\sigma(\v).
\ese
Also, using $\Y=U\V=T(U,\V)$, we have another representation of $\pr \{ (U,\V) \in A \}$ as follows: 
\bse
\pr \{ (U,\V) \in A \} &=& \pr \{ \Y \in T(A) \} = \int_{\real^p} I_{T(A)} (\y) f_\Y(\y) d\y \\
&=& \int_{\mathrm{S}^{p-1}} \int_0^\infty I_{T(A)}(u\v) f_\Y(u\v) u^{p-1} du d\sigma(\v)  \\
&=& \int_{\mathrm{S}^{p-1}} \int_0^\infty I_{A} (u,\v) f_\Y(u\v) u^{p-1} du d\sigma(\v),
\ese
where we use Theorem 3.4 of \cite{stein2005real} in the second last equality.
Therefore,
\bse
\int_{\mathrm{S}^{p-1}} \int_0^\infty I_{A} (u,\v) f_{U,\V}(u,\v) du d\sigma(\v) = \int_{\mathrm{S}^{p-1}} \int_0^\infty I_{A} (u,\v) f_\Y(u\v) u^{p-1} du d\sigma(\v).
\ese
\qed

\section{Proof of Proposition \ref{prop:level-power-known}} \label{sec:proof-known-prop}
Note that our test rejects $H_0$ if and only if
$T-n^{-1/2}z_\alpha \wh\sigma>0$.
By Theorem \ref{th:known},
under $H_0$, we have $n^{1/2} T \to 0$ and $\wh \sigma^2 \to \sigma^2$
in probability. Slutsky's theorem implies that $n^{1/2} T / \wh
\sigma^2 \to 0$ in probability, which implies that  
\bse
\pr (T-n^{-1/2}z_\alpha \wh\sigma>0) = \pr (n^{1/2} T / \wh\sigma >
z_\alpha ) \to 0 
\ese
for any $z_\alpha>0$.
Hence $\pr(\mathrm{reject} \, H_0)=\pr (T-n^{-1/2}z_\alpha
  \wh\sigma>0) <\alpha$ for sufficiently large $n$.
Under $H_a$, however, we have $T \to d(f_{U,V}\|f_U,f_0) >0$ and $\wh
\sigma^2 \to \sigma^2$ in probability. Slutsky's theorem implies that
$T / \wh \sigma^2 \to d(f_{U,V}\|f_U,f_0) / \sigma^2>0$ in
probability, so $n^{1/2} T / \wh \sigma^2 \to \infty$ in
probability. Therefore, 
\bse
\pr(\mathrm{reject} \, H_0)=\pr (T-n^{-1/2}z_\alpha \wh\sigma>0) = \pr (n^{1/2} T / \wh\sigma > z_\alpha ) \to 1
\ese
for any $z_\alpha>0$. 
\qed

\section{Assumptions}\label{sec:assumptions}
For notational convenience, we define
\bse
\S_1(\y)\equiv \frac{\partial\log
  f_\Y(\y)}{\partial\y},\quad
    \S_2(\y)\equiv \frac{\partial^2\log
      f_\Y(\y)}{\partial\y\partial\y\trans},\quad
\S_3(\y)\equiv \S_2(\y)
+\left\{\S_1(\y)\right\}^{\otimes2},
\ese
\bse
Q_1(u) \equiv \frac{d\log
  f_U(u)}{d u} ,\quad
    Q_2(u)\equiv \frac{d^2\log
      f_U(u)}{d u^2}.
\ese
Furthermore, for any real, vector or matrix valued function
  $\a(\y)$, we define
\bse
E_{\wt p}\{\a(\Y)\mid\wt\bmu,\wt\bSigma\}\equiv\int \a(\y) f_{\wt\bSigma^{-1/2}(\X-\wt\bmu)}(\y)d\y,\quad E_{\wh p}\{\a(\Y)\mid\wh\bmu,\wh\bSigma\}\equiv\int \a(\y) f_{\wh\Y}(\y)d\y.
\ese
We state the regularity conditions as follows.
\begin{enumerate}[label=\Alph*,ref=\Alph*,start=1]
    \item\label{cond:A} Finite moment conditions on $\Y$:
    \begin{enumerate}[label=\arabic*.,ref=A\arabic*,start=1]
        \item\label{cond:A1} $E(\|\Y\|^8)<\infty$;
        \item\label{cond:A2} $E(1/\|\Y\|^3)<\infty$;
        \item\label{cond:A3} $E(\log \|\Y\|)<\infty$;
    \end{enumerate}
    \item\label{cond:B} Finiteness conditions on expectations with respect to $f_\Y$:
    \begin{enumerate}[label=\arabic*.,ref=B\arabic*,start=1]
        \item\label{cond:B1} $E\{\|\S_1(\Y)\|^4\}<\infty$;
        \item\label{cond:B2} $E\{\| \S_3(\Y) \|_F^4\}<\infty$;
    \end{enumerate}
    \item\label{cond:C} Finiteness conditions on expectations with respect to $f_U$:
    \begin{enumerate}[label=\arabic*.,ref=C\arabic*,start=1]
        \item\label{cond:C1} $E\{|Q_1(U)|^3\}<\infty$;
        \item\label{cond:C2} $E\{|Q_2(U)|^2\}<\infty$;
    \end{enumerate}
    \item Continuity assumptions with respect to $\wt\bmu$ and $\wt\bSigma$:
    \begin{enumerate}[label=\arabic*.,ref=D\arabic*,start=1]
    \item\label{cond:D1} At any $\x$, $f_\X(\x,\wt\bmu,\wt\bSigma)$
     as a function of $(\wt\bmu, \wt\bSigma)$
         is continuous at  $(\bmu, \bSigma)$.
   \item\label{cond:D2} 
        The functions of the form $E_{\wt p}(\cdot|\wt\bmu,\wt\bSigma)$ that
        are given in \eqref{eq:cont-1}, \eqref{eq:cont-2},
        \eqref{eq:cont-3}, \eqref{eq:cont-4}, \eqref{eq:cont-5},
        \eqref{eq:cont-6}, \eqref{eq:cont-7-8}, \eqref{eq:cont-9},
        \eqref{eq:cont-10}, \eqref{eq:cont-11}, \eqref{eq:cont-12},
        \eqref{eq:cont-13}, \eqref{eq:cont-14} and \eqref{eq:cont-15-16}
        are continuous functions of $(\wt\bmu,\wt\bSigma)$ at the
          point $(\bmu,\bSigma)$.
    \end{enumerate}
\item\label{cond:F} Lipschitz conditions for $\S_2$ and $Q_2$: there
  exists a finite constant $L$ such that 
\begin{enumerate}[label=\arabic*.,ref=E\arabic*,start=1]
    \item\label{cond:F1} $\|\S_2(\y_1)-\S_2(\y_2)\| \le L \| \y_1 -
      \y_2 \|$ for all $\y_1, \y_2 \in \mathrm{supp}(\Y)$; 
    \item\label{cond:F2} $\|Q_2(u_1)-Q_2(u_2)\| \le L | u_1 - u_2 |$
      for all $u_1, u_2 \in \mathrm{supp}(U)$ .
\end{enumerate}
\end{enumerate}

The assumptions \ref{cond:A}-\ref{cond:F} are all related to
boundedness of various expectations
and smoothness or Lipschitz condition of
various functions, hence are very mild overall. In particular, we do not
impose compact support requirement on the variable $\X$, and we do not
require $\X$ to be bounded away from zero either. 
Thus, the Lipschitz condition in Assumption \ref{cond:F1} need to hold
for all $\y_1,\y_2 \in \mathbb{R}^p$, and that in Assumption
\ref{cond:F2} need to hold for all $u_1,u_2 \ge 0$ if the support
  of $\Y$
  is the entire $p$-dimensional real space.
Assumptions in \ref{cond:B} and \ref{cond:C} are intrinsically
  linked because $U=\|\Y\|$. However, we find that Assumption \ref{cond:B}
does not directly imply Assumption
  \ref{cond:C} unless various other additional conditions are imposed. In view of clarity
  and simplicity, we thus retain Assumption
  \ref{cond:C}. We also point out that Assumptions \ref{cond:B} and
  \ref{cond:C} are not results from other Assumptions listed. 
For example, Assumption \ref{cond:F} is not sufficient to ensure
Assumptions \ref{cond:B} and \ref{cond:C} due to the appearance of
higher moments in the latter Assumptions. 
  In Lemma
    \ref{lem:contiguity}, we show that $f_\X(\x,\bmu,\bSigma)$ is
    contiguity with respect to $f_\X(\x,\bmu_n,\bSigma_n)$ under
    Assumption \ref{cond:D1}, a property that will be used in various
    places of the proof. Similar to Assumption \ref{cond:D1},
      Assumption \ref{cond:D2} only assumes continuity at a single
      point hence these are very mild assumptions.

Interestingly, Assumption \ref{cond:A2} requires $f_U(u)$ to go
to zero faster than $u^2$ as $u \to 0$, thus it excludes
distributions such as normal under 
$p\le3$. Technically, this requirement arises because of the need to
control the average power of the relative errors caused by estimating
the length $U_i$'s. Intuitively, Assumption \ref{cond:A2} requires
that $\Y$ is not concentrated near zero, or equivalently, $\X$ should
not be too concentrated near $\bmu$. This is reasonable because
one aspect of the test is about the directions. The closer a point is to
the center, the harder it is to estimate its direction given that the
center is estimated itself, while at the
center the notion of direction itself degenerates. Thus, Assumption
\ref{cond:A2} limits such cases. In this sense, the exclusion of
difficult cases such as 2 and 3 dimensional normal is likely not due to the
particular test statistic that we construct.

\section{Basic Lemmas}\label{sec:lemmas}
We now prove two lemmas that will be used throughout the proof of
the theorems.
\begin{Lem}\label{lem:a}
$\left(\sum_{i=1}^m x_i\right)^p \leq m^{p-1}\left(\sum_{i=1}^m x_i^p\right).$
\end{Lem}
Proof:
By H\"older's inequality,
\bse
\sum_{i=1}^m x_i \leq \left(\sum_{i=1}^m x_i^p\right)^{1/p} \left(\sum_{i=1}^m 1^q\right)^{1/q} = \left(\sum_{i=1}^m x_i^p\right)^{1/p} m^{1/q},
\ese
where $\frac{1}{p}+\frac{1}{q}=1$. Therefore
\bse
\left(\sum_{i=1}^m x_i\right)^p \leq \left(\sum_{i=1}^m x_i^p\right) m^{p/q} =  \left(\sum_{i=1}^m x_i^p\right) m^{p-1}.
\ese
\qed

\begin{Lem}\label{lem:contiguity}
Under Assumption \ref{cond:D1}, $f_\X(\x,\bmu,\bSigma)$ is contiguous with
respect to $f_\X(\x,\bmu_n,\bSigma_n)$.
\end{Lem}
Proof:
Let the likelihood ratio be
 $L_n(\x,\bmu,\bSigma,\bmu_n,
\bSigma_n)\equiv f_\X(\x,\bmu,\bSigma)/f_\X(\x,\bmu_n,\bSigma_n)$.
Because \\
$1/ L_n(\x,\bmu,\bSigma,\bmu_n,
\bSigma_n)$ is a continuous function of $(\bmu_n, \bSigma_n)$
at $(\bmu,\bSigma)$
and because $\bmu_n\to\bmu, \bSigma_n\to\bSigma$ in probability under
$f_\X(\x,\bmu,\bSigma)$,
when $n\to\infty$, $1/L_n(\x,\bmu,\bSigma,\bmu_n,
\bSigma_n)$ converges to 1 in probability under
$f_\X(\x,\bmu,\bSigma)$.
Thus, by Le Cam's first lemma (see, for example, Lemma 6.4 of
\cite{vaart1998asymptotic}), 
$f_\X(\x,\bmu,\bSigma)$ is  contiguous with
respect to $f_\X(\x,\bmu_n,\bSigma_n)$.
\qed

\section{Proof of Theorem \ref{th:unknown1}}\label{sec:proofth2}
We will divide the proof of Theorem \ref{th:unknown1} into
  several parts. We expand the three centralized non-constant terms in
  \eqref{eq:T1} separately: in Section \ref{sec:part1}, we will expand
  the term $\wh E_{n_2}\{\log(\wh U)\} - E\{\log(U)\}$; in Section
  \ref{sec:part2}, we will expand the term $-\wh H_{n_2}(\wh \Y) +
  H(\Y)$; in Section \ref{sec:part3}, we will expand the term $\wh
  H_{n_2}(\wh U) - H(U)$. For preparation, before exploring their
  expansions, we first provide the asymptotic expansions of some basic
  functions in Section \ref{sec:basic-func}.
In addition to $\S_1,\S_2,\S_3$ and $R_1,R_2$, we further define
$\A\equiv\bSigma^{-1/2}\wh\bSigma^{1/2}$ and
$\c\equiv\wh\bSigma^{-1/2}(\bmu-\wh\bmu)$. Also note that
$E \{\S_1(\Y)\}=\0$ and $E \{\Y\S_1(\Y)\trans\}=-\I$, which will be
used in the proof.

\subsection{Basic functions}\label{sec:basic-func}

\noindent\underline{\bf Asymptotic expansion of $\wh\bmu-\bmu,
  \wh\bSigma^{-1/2}-\bSigma^{-1/2}$}

Define 
\bse
\bpsi_\bmu(\x,\bmu)&\equiv&\x-\bmu, \\
\bpsi_{\bSigma}(\x,\bmu,\bSigma)&\equiv&(\x-\bmu)^{\otimes2}-\bSigma,\\
\bpsi_{\bSigma^{-1/2}}(\x,\bmu,\bSigma) &\equiv&
\dvec\{-(\bSigma^{1/2}\otimes\bSigma+
\bSigma\otimes\bSigma^{1/2})^{-1}\vec(\bpsi_{\bSigma}(\x,\bmu,\bSigma)\},
\ese
where $\dvec$ is the inverse operation of $\vec$.
Then,  since $E(\|\X\|^2)<\infty$ due to $\bmu$ and $\bSigma$ being
finite, by Lemma 9.1 of \cite{li2018sufficient},
we have
\be
n_1^{1/2}(\wh\bmu-\bmu)=n_1^{-1/2}\sumja\bpsi_\bmu(\X_j,\bmu)=
n_1^{-1/2}\sumja(\X_j-\bmu),\label{eq:mu}
\ee
\be
n_1^{1/2}(\wh\bSigma-\bSigma)=
n_1^{-1/2}\sumja\bpsi_{\bSigma}(\X_j,\bmu,\bSigma) + \r_1 
=
n_1^{-1/2}\sumja\{(\X_j-\bmu)^{\otimes2}-\bSigma\}+ \r_1,\label{eq:sigma}
\ee
where $\r_1=O_p(n_1^{-1/2})$, and 
\be
&&n_1^{1/2}(\wh\bSigma^{-1/2}-\bSigma^{-1/2})
=n_1^{-1/2}\sumja\bpsi_{\bSigma^{-1/2}}(\X_j,\bmu,\bSigma) + \r_2\n\\
&=&
n_1^{-1/2}\sumja \dvec\{-(\bSigma^{1/2}\otimes\bSigma+
\bSigma\otimes\bSigma^{1/2})^{-1}\vec(\bpsi_{\bSigma}(\X_j,\bmu,\bSigma)\}+ \r_2,\label{eq:sigmaneghalf}
\ee
where $\r_2=O_p(n_1^{-1/2})$.

\noindent\underline{\bf Asymptotic expansion of $\log | \det ( \A) |$}

For convenience, we also expand $\log | \det ( \A) |$ as follows:
\be\label{eq:logdetA}
n_1^{1/2}\log | \det ( \A) | 
&=& n_1^{1/2}\log | \det ( \bSigma^{-1/2} \wh \bSigma^{1/2}) | \n\\
&=& - n_1^{1/2}\log | \det ( \wh \bSigma^{-1/2}  \bSigma^{1/2}) | \n\\
&=& -n_1^{-1/2} \sumja \tr\{\bpsi_{\bSigma^{-1/2}}(\X_j,\bmu,\bSigma)\bSigma^{1/2}\} + O_p(n_1^{-1/2}).
\ee

\subsection{Expanding $\wh E_{n_2}\{\log(\wh U)\} - E\{\log(U)\}$}\label{sec:part1}

\subsubsection{Asymptotic expansion of $\wh\Y_i-\Y_i$}

We first give the asymptotic expansion of $\wh\Y_i-\Y_i$ in terms of
$\bpsi_{\bmu}$ and $\bpsi_{\bSigma^{-1/2}}$. Under Assumption
\ref{cond:A1},
$E(\|\bpsi_\bmu\|^2)<\infty$,
$E(\|\bpsi_\bSigma\|^2)<\infty$ and
$E(\|\bpsi_{\bSigma^{-1/2}}\|^2)<\infty$.
Further,
\bse
\wh\Y_i-\Y_i&=&(\wh\bSigma^{-1/2}-\bSigma^{-1/2})\X_i-(\wh\bSigma^{-1/2}
-\bSigma^{-1/2})\wh\bmu
-\bSigma^{-1/2}(\wh\bmu-\bmu)\\
&=&(\wh\bSigma^{-1/2}-\bSigma^{-1/2})\X_i-(\wh\bSigma^{-1/2}
-\bSigma^{-1/2})\bmu
-\bSigma^{-1/2}(\wh\bmu-\bmu)+ \r_3,
\ese
 where 
 \bse
 \r_3=(\wh\bSigma^{-1/2}
-\bSigma^{-1/2})(\wh\bmu-\bmu) =O_p(n_1^{-1}).
\ese
So, incorporating \eqref{eq:mu} and \eqref{eq:sigmaneghalf},  we have the expansion
\be\label{eq:yihat}
&&\wh\Y_i-\Y_i \n\\
&=& n_1^{-1}\sumja\{\bpsi_{\bSigma^{-1/2}}(\X_j,\bmu,\bSigma)(\X_i-\bmu)
-\bSigma^{-1/2}\bpsi_\bmu(\X_j,\bmu)\}+ n_1^{-1/2}\r_2(\X_i-\bmu)  +\r_3\n\\
&=&n_1^{-1}\sumja\{\bpsi_{\bSigma^{-1/2}}(\X_j,\bmu,\bSigma)\bSigma^{1/2}\Y_i
-\bSigma^{-1/2}\bpsi_\bmu(\X_j,\bmu)\}+\r_4\Y_i +\r_3\n\\
&=&n_1^{-1}\sumja\bpsi_{\Y_i}(\X_j,\bmu,\bSigma)
+\r_4\Y_i +\r_3,
\ee
where  $\r_4=n_1^{-1/2}\r_2=O_p(n_1^{-1})$,  and 
\bse
\bpsi_{\Y_i}(\X_j,\bmu,\bSigma)=\bpsi_{\bSigma^{-1/2}}(\X_j,\bmu,\bSigma)\bSigma^{1/2}\Y_i
-\bSigma^{-1/2}\bpsi_\bmu(\X_j,\bmu).
\ese

\subsubsection{Two technical lemmas}

Based on \eqref{eq:yihat}, we prove two technical lemmas that will be used several times afterwards.

\begin{Lem}\label{lem:b}
  For any integer $k$,
\bse
n_2^{-1}\sumib\|\wh\Y_i-\Y_i\|^k=O_p(n_1^{-k/2})
\ese 
when
 $E(\|\Y\|^k)<\infty$.
\end{Lem}
Proof: Using Lemma \ref{lem:a} for $m=4, p=k$ in \eqref{eq:yihat} leads to
  \bse
  &&n_2^{-1}\sumib\|\wh\Y_i-\Y_i\|^k\n\\
&\le &4^{k-1}\left[n_2^{-1}\sumib\|n_1^{-1}\sumja\bpsi_{\bSigma^{-1/2}}(\X_j,\bmu,\bSigma)\bSigma^{1/2}\Y_i\|^k
+n_2^{-1}\sumib\|n_1^{-1}\sumja\bSigma^{-1/2}\bpsi_\bmu(\X_j,\bmu)\|^k\right.\n\\
&&\left.+n_2^{-1}\sumib\|\Y_i\|^k\|\r_4\|^k
+n_2^{-1}\sumib\|\r_3\|^k\right]\n\\
&\le & 4^{k-1}\left[\|n_1^{-1}\sumja\bpsi_{\bSigma^{-1/2}}(\X_j,\bmu,\bSigma)\|^k\|\bSigma^{1/2}\|^k\left( n_2^{-1}\sumib\|\Y_i\|^k\right)
+\|n_1^{-1}\sumja\bSigma^{-1/2}\bpsi_\bmu(\X_j,\bmu)\|^k\right.\n\\
&&\left.+\|\r_4\|^k n_2^{-1}\sumib\|\Y_i\|^k 
+\|\r_3\|^k\right]\n\\
&\le&O_p(n_1^{-k/2}) O_p(1) + O_p(n_1^{-k/2})
+O_p(n_1^{-k}) O_p(1)
+O_p(n_1^{-k})\n\\
&=&O_p(n_1^{-k/2})
\ese
under the Assumption that $E(\|\Y\|^k)<\infty$.
\qed

\begin{Lem}\label{lem:c}
  For general $k,l\in\mathbb{N}$, and a general function $a(\y)$,
\bse
n_2^{-1}\sumib |a(\y_i)|\frac{\|\wh\Y_i-\Y_i\|^k}{\|\Y_i\|^l}
=O_p(n_1^{-k/2})
\ese
when $E(|a(\Y)|\|\Y\|^{k-l})<\infty$ and $E(|a(\Y)|\|\Y\|^{-l})<\infty$.
\end{Lem}

Proof:
Using Lemma \ref{lem:a} for $m=4, p=k$ in \eqref{eq:yihat} leads to
\bse
&&n_2^{-1}\sumib |a(\y_i)|\frac{\|\wh\Y_i-\Y_i\|^k}{\|\Y_i\|^l}\n\\
&\le & 4^{k-1} n_2^{-1}\sumib |a(\y_i)| \frac{\|n_1^{-1}\sumja\bpsi_{\bSigma^{-1/2}}(\X_j,\bmu,\bSigma)\bSigma^{1/2}\Y_i\|^k}{\|\Y_i\|^l}\n\\
&&+4^{k-1}n_2^{-1}\sumib |a(\y_i)| \frac{\|n_1^{-1}\sumja\bSigma^{-1/2}\bpsi_\bmu(\X_j,\bmu)\|^k}{\|\Y_i\|^l}\n\\
&&+ 4^{k-1}n_2^{-1}\sumib |a(\y_i)| \|\Y_i\|^{k-l} \|\r_4\|^k
+ 4^{k-1}n_2^{-1}\sumib |a(\y_i)| \frac{1}{\|\Y_i\|^l} \|\r_3\|^k\n\\
&\le& 4^{k-1}
\|n_1^{-1}\sumja\bpsi_{\bSigma^{-1/2}}(\X_j,\bmu,\bSigma)\|^k
\|\bSigma^{1/2}\|^k  n_2^{-1}\sumib |a(\y_i)| \|\Y_i\|^{k-l}\n\\
&&+4^{k-1} \|n_1^{-1}\sumja\bSigma^{-1/2}\bpsi_\bmu(\X_j,\bmu)\|^k n_2^{-1}\sumib |a(\y_i)| \frac{1}{\|\Y_i\|^l}\n\\
&&+4^{k-1} \|\r_4\|^k n_2^{-1}\sumib |a(\y_i)| \|\Y_i\|^{k-l} 
+ 4^{k-1}\|\r_3\|^k n_2^{-1}\sumib |a(\y_i)| \frac{1}{\|\Y_i\|^l} \n\\
&=& O_p(n_1^{-k/2}) O_p(1) +O_p(n_1^{-k/2}) O_p(1) +O_p(n_1^{-k}) O_p(1) +O_p(n_1^{-k}) O_p(1)\n\\
&=&O_p(n_1^{-k/2})
\ese
under the assumption that $E(|a(\Y)|\|\Y\|^{k-l})<\infty$,
$E(|a(\Y)|\|\Y\|^{-l})<\infty$.\qed

\subsubsection{Asymptotic expansion of $\wh U_i-U_i$}

\noindent\underline{\bf Main expansion of $\wh U_i-U_i$}

Next, we give the asymptotic expansion of $\wh U_i-U_i$ based on
\eqref{eq:yihat}. We expand $\wh U_i-U_i$ as follows:
\be
&& \wh U_i-U_i
=\|\wh\Y_i\|-\|\Y_i\|
=\frac{\|\wh\Y_i\|^2-\|\Y_i\|^2}{\|\wh\Y_i\|+\|\Y_i\|}\n\\
&=&\frac{\|\wh\Y_i\|^2-\|\Y_i\|^2}{2\|\Y_i\|\{1+(\|\wh\Y_i\|-\|\Y_i\|)/(2\|\Y_i\|)\}
}\n\\
&=&\frac{\|\wh\Y_i\|^2-\|\Y_i\|^2}{2\|\Y_i\|}
\{1+ C_i (\|\wh\Y_i\|-\|\Y_i\|)/(2\|\Y_i\|)\}\n\\
&=&\frac{\|\wh\Y_i\|^2-\|\Y_i\|^2}{2\|\Y_i\|}
  +C_i
\frac{(\|\wh\Y_i\|+\|\Y_i\|)}{2\|\Y_i\|}(\|\wh\Y_i\|-\|\Y_i\|)
\frac{\|\wh\Y_i\|-\|\Y_i\|}{2\|\Y_i\|}\n\\
&=&\frac{\|\wh\Y_i\|^2-\|\Y_i\|^2}{2\|\Y_i\|}
+\frac{C_i}{2}\frac{(\|\wh\Y_i\|-\|\Y_i\|)^2}{\|\Y_i\|}
+\frac{C_i}{4}\frac{(\|\wh\Y_i\|-\|\Y_i\|)^3}{\|\Y_i\|^2}
\n\\
&=&\frac{\wh\Y_i\trans(\wh\Y_i-\Y_i)+(\wh\Y_i-\Y_i)\trans\Y_i}{2U_i}+ \frac{C_i}{2}\frac{(\|\wh\Y_i\|-\|\Y_i\|)^2}{\|\Y_i\|}
+\frac{C_i}{4}\frac{(\|\wh\Y_i\|-\|\Y_i\|)^3}{\|\Y_i\|^2}
\n\\
&=&\frac{(\wh\Y_i-\Y_i)\trans\Y_i}{U_i}+ \frac{1}{2}\frac{\|\wh\Y_i-\Y_i\|^2}{\|\Y_i\|}
+ \frac{C_i}{2}\frac{(\|\wh\Y_i\|-\|\Y_i\|)^2}{\|\Y_i\|}
+\frac{C_i}{4}\frac{(\|\wh\Y_i\|-\|\Y_i\|)^3}{\|\Y_i\|^2}\n\\
&=&\frac{(\wh\Y_i-\Y_i)\trans\Y_i}{U_i}+ S_{i1}+S_{i2},\label{eq:uihat-1}
\ee
where 
\bse
S_{i1}=\frac{1}{2}\frac{\|\wh\Y_i-\Y_i\|^2}{\|\Y_i\|}
+ \frac{C_i}{2}\frac{(\|\wh\Y_i\|-\|\Y_i\|)^2}{\|\Y_i\|},\quad S_{i2}=\frac{C_i}{4}\frac{(\|\wh\Y_i\|-\|\Y_i\|)^3}{\|\Y_i\|^2},
\ese
and $C_i$ is a random quantity between 0 and $-4$. This is because
$\frac{1}{1+x}=1-\frac{1}{(1+x^*)^2}x$ where $x^*$ is between 0 and
$x$, then plug in $x=(\|\wh\Y_i\|-\|\Y_i\|)/(2\|\Y_i\|)\geq -1/2$.

\noindent\underline{\bf Dominating term of $\wh U_i-U_i$}

We first expand the first term in $\wh U_i-U_i$. Plugging
\eqref{eq:yihat} into \eqref{eq:uihat-1}, we have
\be\label{eq:uihat-ui}
\wh U_i-U_i
&=&n_1^{-1}\sumja\frac{\Y_i\trans\bpsi_{\Y_i}(\X_j,\bmu,\bSigma)}{\|\Y_i\|}
+\frac{\Y_i\trans  \r_4\Y_i}{U_i}+\V_i\trans \r_3 +S_{i1}+S_{i2}\n\\
&=&n_1^{-1}\sumja\frac{\Y_i\trans\bpsi_{\Y_i}(\X_j,\bmu,\bSigma)}{U_i}
+ S_{i4}+S_{i3}  +S_{i1}+S_{i2}\n\\
&=&n_1^{-1}\sumja\psi_{U_i}(\X_j,\bmu,\bSigma)
+ S_{i4}+S_{i3} 
+ S_{i1}+S_{i2},
\ee
where
\bse
S_{i4}=\frac{\Y_i\trans  \r_4\Y_i}{U_i}=\Y_i\trans \r_4 \V_i,\quad S_{i3}=\V_i\trans \r_3,
\ese
and 
\be\label{eq:psiu}
\psi_{U_i}(\X_j,\bmu,\bSigma)
=\frac{\Y_i\trans\bpsi_{\Y_i}(\X_j,\bmu,\bSigma)}{U_i}
=\frac{\Y_i\trans
\{\bpsi_{\bSigma^{-1/2}}(\X_j,\bmu,\bSigma)\bSigma^{1/2}\Y_i
-\bSigma^{-1/2}\bpsi_\bmu(\X_j,\bmu)\}}{U_i}.
\ee

\noindent\underline{\bf Upper bound of $|S_{ik}|$ in $\wh U_i-U_i$ for
$k=1, \dots, 4$.}

For the remainder terms $ S_{i3}$ and $S_{i4}$
in \eqref{eq:uihat-ui}, we have
$|S_{i4}|\le \|\Y_i\| \|\r_4\|$ , $|S_{i3}|\le \|\r_3\|$.
Further, for $S_{i1}$ and $S_{i2}$, we have
\be\label{eq:si1}
|S_{i1}|
\le  \frac{1}{2}\frac{\|\wh\Y_i-\Y_i\|^2}{\|\Y_i\|}
+ \frac{|C_i|}{2}\frac{(\|\wh\Y_i\|-\|\Y_i\|)^2}{\|\Y_i\|}
\le  \frac{1+|C_i|}{2}\frac{\|\wh\Y_i-\Y_i\|^2}{\|\Y_i\|}
\le  \frac{5}{2}\frac{\|\wh\Y_i-\Y_i\|^2}{\|\Y_i\|},
\ee
and
\be\label{eq:si2}
|S_{i2}| 
\le  \frac{|C_i|}{4}\frac{|\|\wh\Y_i\|-\|\Y_i\||^3}{\|\Y_i\|^2}
\le \frac{\|\wh\Y_i-\Y_i\|^3}{\|\Y_i\|^2}.
\ee

\subsubsection{Asymptotic expansion of $\sumib\{\log(\wh U_i)-\log(U_i)\}$}

A Taylor expansion yields
\bse
\log(\wh U_i) - \log(U_i) = \frac{\wh U_i-U_i}{U_i} - R_i,\quad R_i=\int_{U_i}^{\wh U_i} \frac{\wh U_i-t}{t^2} dt.
\ese
In the second term, we let $t=(1-s)\wh U_i + s U_i$, then $dt=(U_i-\wh
U_i)ds$. Thus, 
\bse
R_i=  \int_{U_i}^{\wh U_i} \frac{\wh U_i-t}{t^2} dt 
&=&    \int_{1}^{0} \frac{\wh U_i - (1-s)\wh U_i - s U_i}{\{(1-s)\wh U_i + s U_i\}^2} (U_i-\wh U_i)ds\\
&=&  (U_i-\wh U_i)^2  \int_{0}^{1} \frac{s}{\{(1-s)\wh U_i + s U_i\}^2} ds.
\ese

\noindent\underline{\bf Truncation for the remainder term}

To overcome the difficulties in 
 controlling the remainder term $R_i$ when $\wh U_i$ is 
close to 0, we propose to handle the term $R_i$ using
  truncation on $\wh U_i$. 
  Let the event $\me_1$ be $\me_1=\{\wh U_i\ge \epsilon_n U_i, \forall
  i\in\{n_1+1,\ldots,n\}\} $ and denote $I(\cdot)$ the indicator function.

Note that $\me_1$ contains the nice cases where
  all the $\wh U_i$'s are reasonably large compared to
  their corresponding $U_i$'s. As we will see in the following
  derivation, the remainder term can be controlled under $\me_1$, as
  the extreme values of $\wh U_i$'s are ruled out. Regarding the
  extreme cases under $\me_1^C$, we will further show that they are
  sufficiently rare
   so they are ignorable in probability.

We first discuss the expansion of $\sumib\{\log(\wh U_i)-\log(U_i)\}$
under $\me_1$, that is,
\bse
n_2^{-1/2}\sumib\{\log(\wh U_i) -\log(U_i)\}I(\me_1)
=n_2^{-1/2}\sumib \frac{\wh U_i-U_i}{U_i}I(\me_1) - n_2^{-1/2}\sumib R_i I(\me_1).
\ese

\noindent\underline{\bf Remainder term under $\me_1$}

Note that for $a>0$,
\bse
\int \frac{x}{(ax+b)^2}dx 
= \frac{1}{a}\int \frac{1}{ax+b}dx - \frac{b}{a} \int \frac{1}{(ax+b)^2}dx
= \frac{1}{a^2}\log|ax+b| + \frac{b}{a^2}\frac{1}{ax+b}+C.
\ese

Noting that $R_i$'s are non-negative, we have
\bse
R_i I(\me_1) &=& I(\me_1) (U_i-\wh U_i)^2   \int_{0}^{1} \frac{s}{\{(1-s)\wh U_i + s U_i\}^2} ds \\
& \le &   (U_i-\wh U_i)^2\int_{0}^{1} \frac{s}{\{(1-s)\epsilon_n U_i + s U_i\}^2} ds\\
& = &  (U_i-\wh U_i)^2\int_{0}^{1} \frac{s}{\{s(1-\epsilon_n) U_i + \epsilon_n U_i\}^2} ds \\
& = &  (U_i-\wh U_i)^2\left[\frac{1}{\{(1-\epsilon_n)U_i\}^2}\log|s(1-\epsilon_n) U_i + \epsilon_n U_i| + \frac{\epsilon_n U_i}{\{(1-\epsilon_n)U_i\}^2} \frac{1}{s(1-\epsilon_n) U_i + \epsilon_n U_i}\right]_0^1 \\
& = &  (U_i-\wh U_i)^2\left[\frac{1}{\{(1-\epsilon_n)U_i\}^2} \left\{\log|U_i | - \log|\epsilon_n U_i|\right\} + \frac{\epsilon_n U_i}{\{(1-\epsilon_n)U_i\}^2} \left\{\frac{1}{U_i}-\frac{1}{\epsilon_n U_i}\right\}\right] \\
& = & (U_i-\wh U_i)^2 \left[-\frac{\log(\epsilon_n)}{\{(1-\epsilon_n)U_i\}^2}+ \frac{\epsilon_n -1}{\{(1-\epsilon_n)U_i\}^2}\right]\\
& = & \frac{(U_i-\wh U_i)^2}{U_i^2} \left\{\frac{-\log(\epsilon_n)+\epsilon_n -1}{(1-\epsilon_n)^2}\right\}.
\ese
Let $\epsilon_n=n_1^{-k}$ where $k\ge2$,
then we further have
\bse
R_i I(\me_1) &\le& \frac{(U_i-\wh U_i)^2}{U_i^2}\left\{\frac{-\log(\epsilon_n)+\epsilon_n -1}{(1-\epsilon_n)^2}\right\} 
=  \frac{(U_i-\wh U_i)^2}{U_i^2}\left\{\frac{k\log(n_1)+n_1^{-k} -1}{(1-n_1^{-k})^2}\right\}\\
&\le&2k\frac{(U_i-\wh U_i)^2}{U_i^2}\log(n_1)
\ese
for sufficiently large $n$. 
By Lemma \ref{lem:c},
\bse
  n_2^{-1/2} \sumib \frac{(\wh U_i-U_i)^2}{U_i^2}
  =n_2^{-1/2} \sumib \frac{(\|\wh \Y_i\|-\|\Y_i\|)^2}{\|\Y_i\|^2}
\le n_2^{-1/2} \sumib \frac{\|\wh\Y_i-\Y_i\|^2}{\|\Y_i\|^2}
=O_p(n_2^{1/2}n_1^{-1})
\ese
under $E(1/\|\Y\|^2)<\infty$, which is implied by Assumption \ref{cond:A2}. Thus, we have
\bse
 n_2^{-1/2}\sumib R_i I(\me_1)
\le n_2^{-1/2}\sumib 2k\frac{(U_i-\wh U_i)^2}{U_i^2}\log(n_1)
= O_p(n_2^{1/2}n_1^{-1}\log(n_1)).
\ese

\noindent\underline{\bf Dominating term under $\me_1$}

Note that 
\bse
&&n_2^{-1/2}\sumib\{\log(\wh U_i) -\log(U_i)\} I(\me_1)\n\\
&=&n_2^{-1/2}\sumib \frac{\wh U_i-U_i}{U_i} I(\me_1)
+  O_p(n_2^{1/2}n_1^{-1}\log(n_1))\n\\
&=&n_2^{-1/2}\sumib
\frac{n_1^{-1}\sumja\psi_{U_i}(\X_j,\bmu,\bSigma)
+S_{i4}+S_{i3}
+ S_{i1}+S_{i2}}{U_i} I(\me_1) +  O_p(n_2^{1/2}n_1^{-1}\log(n_1))\n\\
&=&n_1^{-1}n_2^{-1/2}\sumib\sumja
\frac{1}{U_i}\psi_{U_i}(\X_j,\bmu,\bSigma) I(\me_1) + n_2^{-1/2}\sumib
\frac{S_{i4}}{U_i} I(\me_1) \n\\
&&+ n_2^{-1/2}\sumib
\frac{S_{i3}}{U_i} I(\me_1)+ n_2^{-1/2}\sumib
\frac{S_{i1}}{U_i} I(\me_1)+ n_2^{-1/2}\sumib
\frac{S_{i2}}{U_i} I(\me_1)+ O_p(n_2^{1/2}n_1^{-1}\log(n_1)).
\ese

For the first summation, we have
\be
&&n_1^{-1}n_2^{-1/2}\sumib \sumja\frac{1}{U_i}\psi_{U_i}(\X_j,\bmu,\bSigma)\n\\
&=&n_1^{-1}n_2^{-1/2}\sumib \sumja
\frac{1}{U_i^2}\Y_i\trans\bpsi_{\Y_i}(\X_j,\bmu,\bSigma)\n\\
&=&n_1^{-1}n_2^{-1/2}\sumib \sumja 
\frac{1}{U_i^2}\Y_i\trans
\{\bpsi_{\bSigma^{-1/2}}(\X_j,\bmu,\bSigma)(\X_i-\bmu)
-\bSigma^{-1/2}\bpsi_\bmu(\X_j,\bmu)\}\n\\
&=&n_1^{-1}n_2^{-1/2}\sumib \sumja 
\frac{1}{U_i^2}\Y_i\trans
\{\bpsi_{\bSigma^{-1/2}}(\X_j,\bmu,\bSigma)\bSigma^{1/2}\Y_i
-\bSigma^{-1/2}\bpsi_\bmu(\X_j,\bmu)\}\n\\
&=&n_2^{1/2}  \tr\left[\left\{ n_1^{-1}\sumja
\bpsi_{\bSigma^{-1/2}}(\X_j,\bmu,\bSigma)\right\} \bSigma^{1/2}  \left\{n_2^{-1}\sumib 
\frac{\Y_i\Y_i\trans}{U_i^2}\right\}\right]\n\\
&&
-n_2^{1/2}\tr\left[\bSigma^{-1/2}\left\{ n_1^{-1}\sumja\bpsi_\bmu(\X_j,\bmu)\right\}\left\{ n_2^{-1}\sumib
\frac{1}{U_i^2}\Y_i\trans\right\}\right].\label{eq:psiu-u}
\ee

Note that, under Assumption \ref{cond:A2}, the first term of \eqref{eq:psiu-u} is
\bse
&&n_2^{1/2}  \tr\left[\left\{ n_1^{-1}\sumja
\bpsi_{\bSigma^{-1/2}}(\X_j,\bmu,\bSigma)\right\} \bSigma^{1/2}  \left\{n_2^{-1}\sumib 
\frac{\Y_i\Y_i\trans}{U_i^2}\right\}\right]\\
&=&n_2^{1/2}  \tr\left[\left\{n_1^{-1}\sumja
\bpsi_{\bSigma^{-1/2}}(\X_j,\bmu,\bSigma)\right\} \bSigma^{1/2} \left\{E\left( 
\frac{\Y\Y\trans}{\|\Y\|^2}\right)+O_p(n_2^{-1/2})\right\}\right]\\
&=&n_2^{1/2}  \tr\left[\left\{ n_1^{-1}\sumja
\bpsi_{\bSigma^{-1/2}}(\X_j,\bmu,\bSigma)\right\} \bSigma^{1/2} E\left( \frac{\Y\Y\trans}{\|\Y\|^2}\right)\right]+
n_2^{1/2}  \tr\left\{O_p(n_1^{-1/2}) \bSigma^{1/2} O_p(n_2^{-1/2})\right\}\\
&=&n_2^{1/2}  \tr\left[\left\{ n_1^{-1}\sumja
\bpsi_{\bSigma^{-1/2}}(\X_j,\bmu,\bSigma)\right\} \bSigma^{1/2} E\left( \frac{\Y\Y\trans}{\|\Y\|^2}\right)\right]+
 O_p(n_1^{-1/2}),
\ese
and the second term of \eqref{eq:psiu-u} is
\bse
&&n_2^{1/2}\tr\left[\bSigma^{-1/2}\left\{ n_1^{-1}\sumja\bpsi_\bmu(\X_j,\bmu)\right\}\left\{ n_2^{-1}\sumib
\frac{1}{U_i^2}\Y_i\trans\right\}\right]\\
&=&n_2^{1/2}\tr\left[\bSigma^{-1/2}\left\{n_1^{-1}\sumja\bpsi_\bmu(\X_j,\bmu)\right\} \left\{E\left( 
\frac{\Y\trans}{\|\Y\|^2}\right)+O_p(n_2^{-1/2})\right\}\right]\\
&=&n_2^{1/2}\tr\left[\bSigma^{-1/2}\left\{n_1^{-1}\sumja\bpsi_\bmu(\X_j,\bmu)\right\} E\left( 
\frac{\Y\trans}{\|\Y\|^2}\right)\right]+
n_2^{1/2}\tr\left\{\bSigma^{-1/2}O_p(n_1^{-1/2}) O_p(n_2^{-1/2})\right\}\\
&=&n_2^{1/2}\tr\left[\bSigma^{-1/2}\left\{n_1^{-1}\sumja\bpsi_\bmu(\X_j,\bmu)\right\} E\left( 
\frac{\Y\trans}{\|\Y\|^2}\right)\right]+
O_p(n_1^{-1/2}).
\ese
Plugging back into \eqref{eq:psiu-u}, we have
\bse
&&n_1^{-1}n_2^{-1/2}\sumib \sumja\frac{1}{U_i}\psi_{U_i}(\X_j,\bmu,\bSigma)\\
&=&n_1^{-1}n_2^{1/2} \sumja  \left[\tr\left\{ E\left(
\frac{\Y\Y\trans}{\|\Y\|^2}\right)
\bpsi_{\bSigma^{-1/2}}(\X_j,\bmu,\bSigma) \bSigma^{1/2}\right\}
-E\left(
  \frac{\Y\trans}{\|\Y\|^2}\right)\bSigma^{-1/2}\bpsi_\bmu(\X_j,\bmu) \right]\\
  &&+O_p(n_1^{-1/2}).
\ese

\noindent\underline{\bf Other ignorable terms under $\me_1$}

We now analyze the four terms related to $S_{i1}$, $S_{i2}$, $S_{i3}$,
$S_{i4}$. Note that, under $E(1/\|\Y\|^2)<\infty$, which is implied by
Assumption \ref{cond:A2}, using \eqref{eq:si1} and Lemma \ref{lem:c}, we have
\bse
\left|n_2^{-1}\sumib \frac{S_{i1}}{U_i}\right| \le n_2^{-1}\sumib \frac{\left|S_{i1}\right|}{U_i} \le n_2^{-1}\sumib \frac{5}{2}\frac{\|\wh\Y_i-\Y_i\|^2}{\|\Y_i\|^2} =O_p(n_1^{-1}),
\ese
Similarly, under $E(1/\|\Y\|^3)<\infty$, which is also implied by
Assumption \ref{cond:A2}, using \eqref{eq:si2} and Lemma \ref{lem:c}, we have
\bse
\left|n_2^{-1}\sumib \frac{S_{i2}}{U_i}\right| \le n_2^{-1}\sumib \frac{\left|S_{i2}\right|}{U_i} \le n_2^{-1}\sumib \frac{\|\wh\Y_i-\Y_i\|^3}{\|\Y_i\|^3} =O_p(n_1^{-3/2}).
\ese
Furthermore, under $E(1/\|\Y\|)<\infty$, which is also implied by
Assumption \ref{cond:A2}, Lemma \ref{lem:c} implies that
\bse
\left|n_2^{-1}\sumib \frac{S_{i3}}{U_i}\right| \le n_2^{-1}\sumib \frac{\left|S_{i3}\right|}{U_i} \le n_2^{-1}\sumib \frac{\|\r_3\|}{U_i} = \|\r_3\| n_2^{-1}\sumib \frac{1}{\|\Y_i\|}= O_p(n_1^{-1})
\ese
The remaining term can also be bounded through the following:
\bse
\left|n_2^{-1}\sumib \frac{S_{i4}}{U_i}\right| \le n_2^{-1}\sumib \frac{\left|S_{i4}\right|}{U_i} \le n_2^{-1}\sumib \frac{\|\Y_i\| \|\r_4\|}{U_i} = \|\r_4\| = O_p(n_1^{-1}).
\ese

\noindent\underline{\bf Analysis of $\me_1^C$}

We now show that 
the probability for 
the event $\me_1^C$ to happen is ignorable
by estimating the rate of $I(\me_1^C)$ as follows.
\bse
I(\me_1^C)&=&I(\exists i\in\{n_1+1,\ldots,n\} \mbox{ s.t. } \wh U_i\le\epsilon_n U_i)
\le \sumib I(\wh U_i\le\epsilon_n U_i)
=\sumib I(\epsilon_n U_i/\wh U_i\ge 1)\\
&\le&\sumib \epsilon_n \frac{U_i}{\wh U_i}
= \epsilon_n \sumib \frac{\|\Y_i\|}{\|\wh\Y_i\|}
= \epsilon_n \sumib \frac{\|\A(\wh\Y_i-\c)\|}{\|\wh\Y_i\|}
\le  \epsilon_n \sumib \frac{\|\A\|(\|\wh\Y_i\|+\|\c\|)}{\|\wh\Y_i\|}\\
&=& \epsilon_n \|\A\| \sumib \left\{1+\frac{\|\c\|}{\|\wh\Y_i\|}\right\}
= \epsilon_n \|\A\| \left\{n_2+\|\c\|\sumib \frac{1}{\|\wh\Y_i\|}\right\}\\
&=& n_2\epsilon_n \|\A\|+n_2\epsilon_n
\|\A\|\|\c\|E_{\wh p}\left(1/\|\wh\Y\| \mid
  \wh\bmu,\wh\bSigma\right)+n_2\epsilon_n \|\A\| \|\c\|o_{\wh p}(1)\\
&=&n_2\epsilon_n \|\A\|+n_2\epsilon_n
\|\A\|\|\c\|E_{\wh p}\left(1/\|\wh\Y\| \mid
  \wh\bmu,\wh\bSigma\right)+n_2\epsilon_n \|\A\| \|\c\|o_p(1)\\
&=&  O_p(n_2\epsilon_n) +n_2 \epsilon_n o_{p}(1)\\
&=&  O_p( n_2\epsilon_n),
\ese
 where the fourth last equality uses large number theory, the
third last equality uses contiguity under Assumption \ref{cond:D1}, the
second last equality uses Assumption \ref{cond:D2} with respect to
\be\label{eq:cont-1}
E_{\wt p} \left( 1 / \|\wt \Y\| \mid \wt\bmu, \wt\bSigma \right)
\ee
 and
$E(1/\|\Y\|)<\infty$, which is implied by Assumption \ref{cond:A2}.
Specifically, let $m_1(\wt\bmu,\wt\bSigma)\equiv E_{\wt
  p}\{\|\wt\bSigma^{-1/2}(\bSigma^{1/2}\Y+\bmu-\wt\bmu)\|^{-1}\mid  
      \wt\bmu,\wt\bSigma\}$.  As $n_1\to\infty$, we
  have $\wh\bmu\to\bmu$ and $\wh\bSigma\to\bSigma$ in probability, so
  by Slutsky's theorem and continuous mapping theorem,
  $m_1(\wh\bSigma,\wh\bmu)\to m_1(\bSigma,\bmu)$ in probability,
  indicating that $m_1(\wh\bSigma,\wh\bmu)=O_p(1)$.

  Hence,
  when $\epsilon_n=n_1^{-k}$, we have 
\be\label{eq:i-e1-c}
I(\me_1^C) &=& O_p( n_2 n_1^{-k}).
\ee

\noindent\underline{\bf Summary of results under $\me_1$}

Based on the discussions under $\me_1$, we have
\be
&&n_2^{-1/2}\sumib\{\log(\wh U_i) -\log(U_i)\} I(\me_1)\n\\
&=&n_1^{-1}n_2^{1/2} \sumja  \left[\tr\left\{ E\left(
\frac{\Y\Y\trans}{\|\Y\|^2}\right)
\bpsi_{\bSigma^{-1/2}}(\X_j,\bmu,\bSigma) \bSigma^{1/2}\right\}
-E\left(
  \frac{\Y\trans}{\|\Y\|^2}\right)\bSigma^{-1/2}\bpsi_\bmu(\X_j,\bmu) \right]I(\me_1)\n\\
&& +O_p(n_1^{-1/2})I(\me_1) +O_p(n_2^{1/2}n_1^{-1})I(\me_1) +  O_p(n_2^{1/2}n_1^{-1}\log(n_1))\n\\
&=& n_1^{-1} n_2^{1/2} \sumja  \left[\tr\left\{ E\left(\V\V\trans\right)
\bpsi_{\bSigma^{-1/2}}(\X_j,\bmu,\bSigma) \bSigma^{1/2}\right\}
-E\left(\V\trans/U\right)\bSigma^{-1/2}\bpsi_\bmu(\X_j,\bmu) \right]I(\me_1)\n\\
&& +O_p(n_1^{-1/2} + n_2^{1/2}n_1^{-1}\log(n_1)).\label{eq:logu-e1}
\ee

Inserting \eqref{eq:i-e1-c} into \eqref{eq:logu-e1}, we can obtain the
final expansion under $\me_1$ as follows: 
\bse
&&n_2^{-1/2}\sumib\{\log(\wh U_i) -\log(U_i)\} I(\me_1)\n\\
&=& n_1^{-1} n_2^{1/2} \sumja  \left[\tr\left\{ E\left(\V\V\trans\right)
\bpsi_{\bSigma^{-1/2}}(\X_j,\bmu,\bSigma) \bSigma^{1/2}\right\}
-E\left(\V\trans/U\right)\bSigma^{-1/2}\bpsi_\bmu(\X_j,\bmu) \right]I(\me_1)\n\\
&& +O_p(n_1^{-1/2} + n_2^{1/2}n_1^{-1}\log(n_1))\n\\
&=& n_1^{-1} n_2^{1/2} \sumja  \left[\tr\left\{ E\left(\V\V\trans\right)
\bpsi_{\bSigma^{-1/2}}(\X_j,\bmu,\bSigma) \bSigma^{1/2}\right\}
-E\left(\V\trans/U\right)\bSigma^{-1/2}\bpsi_\bmu(\X_j,\bmu) \right]\{1-I(\me_1^C)\}\n\\
&& +O_p(n_1^{-1/2} + n_2^{1/2}n_1^{-1}\log(n_1))\n\\
&=& n_1^{-1} n_2^{1/2} \sumja  \left[\tr\left\{ E\left(\V\V\trans\right)
\bpsi_{\bSigma^{-1/2}}(\X_j,\bmu,\bSigma) \bSigma^{1/2}\right\}
-E\left(\V\trans/U\right)\bSigma^{-1/2}\bpsi_\bmu(\X_j,\bmu) \right]\n\\
&& +n_2^{1/2} O_p(n_1^{-1/2})O_p( n_2 n_1^{-k}) +O_p(n_1^{-1/2} + n_2^{1/2}n_1^{-1}\log(n_1))\n\\
&=& n_1^{-1} n_2^{1/2} \sumja  \left[\tr\left\{ E\left(\V\V\trans\right)
\bpsi_{\bSigma^{-1/2}}(\X_j,\bmu,\bSigma) \bSigma^{1/2}\right\}
-E\left(\V\trans/U\right)\bSigma^{-1/2}\bpsi_\bmu(\X_j,\bmu) \right]\n\\
&& +  O_p(n_2^{3/2}n_1^{-k-1/2})+O_p(n_1^{-1/2} + n_2^{1/2}n_1^{-1}\log(n_1)).
\ese

\noindent\underline{\bf Result under $\me_1^C$}

In contrast, under $\me_1^C$,
\bse
&&n_2^{-1/2}\sumib\{\log(\wh U_i) -\log(U_i)\} I(\me_1^C)\\
&=& n_2^{1/2}\left[E_{\wh p}\{\log(\wh U)|\wh\bmu,\wh\bSigma\} -  E\{\log(U)\}+o_{\wh p}(1)+o_{p}(1)\right]I(\me_1^C)\\
&=& n_2^{1/2} O_p(1) O_p( n_2 n_1^{-k})\\
&=& O_p( n_2^{3/2} n_1^{-k}).
\ese
under Assumptions \ref{cond:A3}, \ref{cond:D1}, \ref{cond:D2} with respect to
\be\label{eq:cont-2}
E_{\wt p}\{\log(\wt U)|\wt\bmu,\wt\bSigma\}.
\ee

\noindent\underline{\bf Final expansion of $\sumib\{\log(\wh U_i) -\log(U_i)\}$}

Combining the above results, we have
\be\label{eq:logu-1}
&&n_2^{-1/2}\sumib\{\log(\wh U_i) -\log(U_i)\} \n\\
&=&n_2^{-1/2}\sumib\{\log(\wh U_i) -\log(U_i)\} I(\me_1) + n_2^{-1/2}\sumib\{\log(\wh U_i) -\log(U_i)\} I(\me_1^C)\n \\
&=& n_1^{-1} n_2^{1/2} \sumja  \left[\tr\left\{ E\left(\V\V\trans\right)
\bpsi_{\bSigma^{-1/2}}(\X_j,\bmu,\bSigma) \bSigma^{1/2}\right\}
-E\left(\V\trans/U\right)\bSigma^{-1/2}\bpsi_\bmu(\X_j,\bmu) \right]\n\\
&& +O_p(n_1^{-1/2} + n_2^{1/2}n_1^{-1}\log(n_1)+ n_2^{3/2} n_1^{-k})\n\\
&=& n_1^{-1} n_2^{1/2} \sumja  \left[\tr\left\{ E\left(\V\V\trans\right)
\bpsi_{\bSigma^{-1/2}}(\X_j,\bmu,\bSigma) \bSigma^{1/2}\right\}
-E\left(\V\trans/U\right) \y_j \right]\n\\
&& +O_p(n_1^{-1/2} + n_2^{1/2}n_1^{-1}\log(n_1)+ n_2^{3/2} n_1^{-k}).
\ee

\subsubsection{Final result for $\wh E_{n_2}\{\log(\wh U)\} - E\{\log(U)\}$}

By \eqref{eq:logu-1},
\be\label{eq:logu}
&&\wh E_{n_2}\{\log(\wh U)\} - E\{\log(U)\}
=n_2^{-1/2}\sumib\{\log(\wh U_i) -E\{\log(U)\}\}\n \\ 
&=&n_2^{-1/2}\sumib\{\log(\wh U_i) -\log(U_i)\} + n_2^{-1/2}\sumib\{\log(U_i) -E\{\log(U)\}\}\n \\
&=& n_1^{-1} n_2^{1/2} \sumja  \left[\tr\left\{ E\left(\V\V\trans\right)
\bpsi_{\bSigma^{-1/2}}(\X_j,\bmu,\bSigma) \bSigma^{1/2}\right\}
-E\left(\V\trans/U\right)\y_j \right]\n\\
&& + n_2^{-1/2}\sumib\{\log(U_i) -E\{\log(U)\}\} +O_p(n_1^{-1/2} + n_2^{1/2}n_1^{-1}\log(n_1)+ n_2^{3/2} n_1^{-k}).
\ee

\subsection{Expanding $-\wh H_{n_2}(\wh \Y) + H(\Y)$}\label{sec:part2}

\subsubsection{Preparations}

We first give two equivalent representations of $\y_i - \wh \y_i$,
which will be used later. Note that
$\wh\Y_i=\wh\bSigma^{-1/2}(\X_i-\wh\bmu)
=\wh\bSigma^{-1/2}(\bSigma^{1/2}\Y_i+\bmu-\wh\bmu)
=\A^{-1}\Y_i+\c$.
We have
\bse
(\A-\I)\wh \y_i-\A\c 
&=& (\bSigma^{-1/2}\wh\bSigma^{1/2}-\I)\wh\bSigma^{-1/2}(\x_i-\wh \bmu)-\bSigma^{-1/2}\wh\bSigma^{1/2}\wh\bSigma^{-1/2}(\bmu-\wh\bmu)\\
&=& \bSigma^{-1/2}(\x_i-\wh \bmu)-\wh \bSigma^{-1/2}(\x_i-\wh \bmu)-\bSigma^{-1/2}(\bmu-\wh\bmu)\\
&=& \bSigma^{-1/2}(\x_i - \bmu)-\wh \bSigma^{-1/2}(\x_i-\wh \bmu)\\
&=& \y_i - \wh \y_i,
\ese
and 
\bse
(\I-\A^{-1})\y_i-\c 
&=& (\I-\wh\bSigma^{-1/2}\bSigma^{1/2})\bSigma^{-1/2}(\x_i-\bmu)-\wh\bSigma^{-1/2}(\bmu-\wh\bmu)\\
&=& \bSigma^{-1/2}(\x_i-\bmu) - \wh\bSigma^{-1/2}(\x_i-\bmu)-\wh\bSigma^{-1/2}(\bmu-\wh\bmu)\\
&=& \bSigma^{-1/2}(\x_i-\bmu) - \wh\bSigma^{-1/2}(\x_i-\wh \bmu)\\
&=& \y_i-\wh \y_i.
\ese

We also link $f_{\wh\Y}(\y)$ to $f_\Y(\y)$ by Taylor expansion as follows.
\bse
f_{\wh\Y}(\y)
&=&f_\Y\{\A(\y-\c)\}|\det(\A)|\\
&=&f_\Y\{\y+(\A-\I)\y-\A\c\}|\det(\A)|\\
&=&f_\Y(\y) |\det(\A)|+\frac{\partial
  f_\Y(\y)}{\partial\y\trans}\{(\A-\I)\y-\A\c\}|\det(\A)|\\
&&+\frac{1}{2}\{(\A-\I)\y-\A\c\}\trans
\frac{\partial^2
  f_\Y(\y^*)}{\partial\y\partial\y\trans}\{(\A-\I)\y-\A\c\}|\det(\A)|,
\ese
where $\y^*$ is between $\y$ and $\wh\y$. Similar operations can be
done on $\log f_{\wh\Y}(\y)$ as follows.
\bse
\log f_{\wh\Y}(\y)
&=&\log f_\Y\{\y+(\A-\I)\y-\A\c\}+\log |\det(\A)|\\
&=&\log f_\Y(\y)+\frac{\partial
  \log f_\Y(\y)}{\partial\y\trans}\{(\A-\I)\y-\A\c\}\\
&&+\frac{1}{2}\{(\A-\I)\y-\A\c\}\trans
\frac{\partial^2
 \log f_\Y(\y^*)}{\partial\y\partial\y\trans}\{(\A-\I)\y-\A\c\}+\log |\det(\A)|\\
&=&\log f_\Y(\y)+\S_1(\y)\{(\A-\I)\y-\A\c\}\\
&&+\frac{1}{2}\{(\A-\I)\y-\A\c\}\trans
\S_2(\y^*)\{(\A-\I)\y-\A\c\}+\log |\det(\A)|,
\ese
where $\y^*$ is also between $\y$ and $\wh\y$.

\subsubsection{Asymptotic expansion of $\sumib\{\log f_{\wh\Y}(\wh\y_i)-\log f_\Y(\wh\y_i)\}$}

\noindent\underline{\bf Main expansion of $\sumib\{\log f_{\wh\Y}(\wh\y_i)-\log f_\Y(\wh\y_i)\}$}

Under Assumptions \ref{cond:D2} with respect to \eqref{eq:cont-3}
 (given near the end of this subsection), \ref{cond:F1},
$E(\|\S_2(\Y)\|_2^2)<\infty$ (implied by Assumptions \ref{cond:B1} and
\ref{cond:B2}) and $E(\|\Y\|^4)<\infty$ (implied by Assumption
\ref{cond:A1}), we have the following expansion:
\be
&&n_2^{-1/2}\sumib\{\log f_{\wh\Y}(\wh\y_i)-\log f_\Y(\wh\y_i)\}\n\\
&=&
n_2^{-1/2}\sumib
\S_1(\wh\y_i)\trans\{(\A-\I)\wh\y_i-\A\c\}+n_2^{1/2}\log |\det(\A)|\n\\
&&+{ n_2^{-1/2}\sumib\frac{1}{2}\{(\A-\I)\wh\y_i-\A\c\}\trans
\S_2(\wh\y_i^*)\{(\A-\I)\wh\y_i-\A\c\}}\n\\
&=&
n_2^{-1/2}\sumib
\S_1(\wh\y_i)\trans\{(\I-\A^{-1})\y_i
-\c\}+n_2^{1/2}\log |\det(\A)|+{ O_p(n_1^{-1}n_2^{1/2})}\n\\
&=&
n_2^{-1/2}\sumib
\S_1(\y_i)\trans\{(\I-\A^{-1})\y_i
-\c\}+n_2^{1/2}\log |\det(\A)|+O_p(n_1^{-1}n_2^{1/2})\n\\
&&-{ n_2^{-1/2}\sumib\{
(\I-\A^{-1})\y_i-\c\}\trans
\S_2(\y_i^*)\{(\I-\A^{-1})\y_i
-\c\}}\n\\
&=&
n_2^{-1/2}\sumib
\S_1(\y_i)\trans\{(\I-\A^{-1})\y_i
-\c\}+n_2^{1/2}\log |\det(\A)|+{ O_p(n_1^{-1}n_2^{1/2})}.\label{eq:fyhat-fy}
\ee
In the above equation, both $\wh \y_i^*$ and $\y_i^*$ are random
between $\y_i$ and $\wh \y_i$. Note that we absorb two ignorable terms
into $O_p(n_1^{-1}n_2^{1/2})$ in the derivation of \eqref{eq:fyhat-fy}
under Assumptions \ref{cond:D2}, \ref{cond:F1},
$E(\|\S_2(\Y)\|_2^2)<\infty$ and $E(\|\Y\|^4)<\infty$, 
  respectively in obtaining the second equality and the last equality, which will be
verified later. 

Continuing from the results in \eqref{eq:fyhat-fy}, we insert the form
of $\A$ and $\c$ and get
\be\label{eq:fy1}
&&n_2^{-1/2}\sumib\{\log f_{\wh\Y}(\wh\y_i)-\log f_\Y(\wh\y_i)\}\n\\
&=&n_2^{-1/2}\sumib
\S_1(\y_i)\trans\{(\I-
\wh\bSigma^{-1/2}\bSigma^{1/2})\y_i-\wh\bSigma^{-1/2}(\bmu-\wh\bmu)
\}+n_2^{1/2}\log |\det(\A)|+O_p(n_1^{-1}n_2^{1/2})\n\\
&=&n_2^{-1/2}\sumib
\S_1(\y_i)\trans\left\{
-n_1^{-1}\sumja\bpsi_{\bSigma^{-1/2}}(\x_j,\bmu,\bSigma)\bSigma^{1/2}\y_i+n_1^{-1}\sumja \y_j
\right\}+n_2^{1/2}\log |\det(\A)|\n\\
&&+O_p(n_1^{-1}n_2^{1/2})\n\\
&=&-n_2^{1/2}n_1^{-1}\sumja\tr  \left[\bSigma^{1/2} E\{\Y
\S_1(\Y)\trans\}
\bpsi_{\bSigma^{-1/2}}(\x_j,\bmu,\bSigma) \right]
+n_2^{1/2}\log |\det(\A)|\n\\
&&+O_p(n_1^{-1}n_2^{1/2}+n_1^{-1/2})\n\\
&=&n_2^{1/2}n_1^{-1}\sumja\tr  \left\{\bSigma^{1/2}
\bpsi_{\bSigma^{-1/2}}(\x_j,\bmu,\bSigma) \right\}
+n_2^{1/2}\log |\det(\A)|+O_p(n_1^{-1}n_2^{1/2}+n_1^{-1/2})
\ee
under $E(\|\S_1(\Y)\|^2)<\infty$ and $E(\|\Y\|^2)<\infty$, which are
implied by Assumptions \ref{cond:A1} and \ref{cond:B1}.

\noindent\underline{\bf Verification of two ignorable terms in \eqref{eq:fyhat-fy}}

To see that the results in \eqref{eq:fyhat-fy} hold under Assumptions \ref{cond:D2}, \ref{cond:F1}, $E(\|\S_2(\Y)\|_2^2)<\infty$ and $E(\|\Y\|^4)<\infty$,
let
\bse
\D_i=\S_2(\y_i^*)-\S_2(\y_i)=\frac{\partial^2 \log f_\Y(\y_i^*)}{\partial\y\partial\y\trans}-\frac{\partial^2 \log f_\Y(\y_i)}{\partial\y\partial\y\trans}.
\ese
We note that $\|\D_i\|_2\le L\|\y_i^*-\y_i\|\le
L\|\wh\y_i-\y_i\|$. Thus,
\be\label{eq:quad-1}
&&\left|n_2^{-1/2}\sumib\{
(\I-\A^{-1})\y_i-\c\}\trans
\S_2(\y_i^*)\{(\I-\A^{-1})\y_i
-\c\}\right|\n\\
&\le&n_2^{-1/2}\sumib\left|(\y_i-\wh\y_i)\trans
\S_2(\y_i)(\y_i-\wh\y_i)\right|+
n_2^{-1/2}\sumib\left|(\y_i-\wh\y_i)\trans\D_i(\y_i-\wh\y_i)\right|\n\\
&\le&n_2^{-1/2}\sumib\|\y_i-\wh\y_i\|^2
\|\S_2(\y_i)\|_2+
n_2^{-1/2}\sumib\|\y_i-\wh\y_i\|^2\|\D_i\|_2
\ee 
The first term in \eqref{eq:quad-1} can be further bounded as follows:
\bse
n_2^{-1/2}\sumib\|\y_i-\wh\y_i\|^2
\|\S_2(\y_i)\|_2
&\le& n_2^{-1/2}\left\{\sumib\|\y_i-\wh\y_i\|^4\right\}^{1/2}
\left\{\sumib\|\S_2(\y_i)\|_2^2\right\}^{1/2}\\
&=&n_2^{-1/2}\left\{O_p(n_1^{-2}n_2 )\right\}^{1/2}
\left\{O_p(n_2)\right\}^{1/2}\\
&=&O_p(n_1^{-1}n_2^{1/2}),
\ese
where the second last equality follows from  Lemma \ref{lem:b} and 
\bse
n_2^{-1}\sumib\|\S_2(\y_i)\|_2^2 = E\|\S_2(\Y)\|_2^2 + o_p(1) = O_p(1).
\ese 
The second term in \eqref{eq:quad-1} is straightforward using  Lemma \ref{lem:b}:
\bse
n_2^{-1/2}\sumib\|\y_i-\wh\y_i\|^2\|\D_i\|_2 
\le n_2^{-1/2}L\sumib\|\y_i-\wh\y_i\|^3 = O_p(n_1^{-3/2}n_2^{1/2}).
\ese
Plugging back into \eqref{eq:quad-1}, we have
\be\label{eq:quad-2}
&&n_2^{-1/2}\sumib\{
(\I-\A^{-1})\y_i-\c\}\trans
\S_2(\y_i^*)\{(\I-\A^{-1})\y_i
-\c\}\n\\
&=&O_p(n_1^{-1}n_2^{1/2})+O_p(n_1^{-3/2}n_2^{1/2})\n\\
&=&O_p(n_1^{-1}n_2^{1/2}).
\ee
Similarly, 
  let
\bse
\wh\D_i=\S_2(\wh\y_i^*)-\S_2(\wh\y_i)=\frac{\partial^2 \log f_\Y(\wh\y_i^*)}{\partial\y\partial\y\trans}-\frac{\partial^2 \log f_\Y(\wh\y_i)}{\partial\y\partial\y\trans}.
\ese
Also note that $\|\wh\D_i\|_2\le L\|\wh\y_i^*-\wh\y_i\|\le
L\|\wh\y_i-\y_i\|$, so we get
 \be\label{eq:quad-3}
&& \left|n_2^{-1/2}\sumib\{(\A-\I)\wh\y_i-\A\c\}\trans
\S_2(\wh\y_i^*)\{(\A-\I)\wh\y_i-\A\c\}\right|\n\\
&\le&n_2^{-1/2}\sumib \left|(\y_i-\wh\y_i)\trans
\S_2(\wh\y_i)(\y_i-\wh\y_i)\right|
+n_2^{-1/2}\sumib \left|(\y_i-\wh\y_i)\trans\wh\D_i (\y_i-\wh\y_i)\right|\n\\
&\le&n_2^{-1/2}\sumib \|\y_i-\wh\y_i\|^2
\|\S_2(\wh\y_i)\|_2
+n_2^{-1/2}\sumib \|\y_i-\wh\y_i\|^2\|\wh\D_i\|_2
\ee
Similarly, the first term in \eqref{eq:quad-3} can be further bounded as follows:
\bse
n_2^{-1/2}\sumib\|\y_i-\wh\y_i\|^2
\|\S_2(\wh\y_i)\|_2
&\le& n_2^{-1/2}\left\{\sumib\|\y_i-\wh\y_i\|^4\right\}^{1/2}
\left\{\sumib\|\S_2(\wh\y_i)\|_2^2\right\}^{1/2}\\
&=&n_2^{-1/2}\left\{O_p(n_1^{-2}n_2 )\right\}^{1/2}
\left\{O_p(n_2)\right\}^{1/2}\\
&=&O_p(n_1^{-1}n_2^{1/2}),
\ese
where the second last equality follows from  Lemma \ref{lem:b} and 
\bse
n_2^{-1}\sumib\|\S_2(\wh\y_i)\|_2^2 = E_{\wh p}\{\|\S_2(\wh\Y)\|_2^2|\wh\bmu,\wh\bSigma\} + o_p(1) 
= E\|\S_2(\Y)\|_2^2 + o_p(1)= O_p(1),
\ese
where we use Assumption \ref{cond:D2} with respect to
\be\label{eq:cont-3}
E_{\wt p}\{\|\S_2(\wt\Y)\|_2^2|\wt\bmu,\wt\bSigma\}
\ee
 and continuous mapping
theorem. Same as the previous term,
the second term in \eqref{eq:quad-3} is straightforward using  Lemma \ref{lem:b}:
\bse
n_2^{-1/2}\sumib\|\y_i-\wh\y_i\|^2\|\wh \D_i\|_2 
\le n_2^{-1/2}L\sumib\|\y_i-\wh\y_i\|^3 = O_p(n_1^{-3/2}n_2^{1/2}).
\ese
Plugging back into \eqref{eq:quad-3}, we have
\be\label{eq:quad-4}
&&n_2^{-1/2}\sumib\{(\A-\I)\wh\y_i-\A\c\}\trans
\S_2(\wh\y_i^*)\{(\A-\I)\wh\y_i-\A\c\}\n\\
&=&O_p(n_1^{-1}n_2^{1/2})+O_p(n_1^{-3/2}n_2^{1/2})\n\\
&=&O_p(n_1^{-1}n_2^{1/2}).
\ee
Therefore, \eqref{eq:quad-2} and \eqref{eq:quad-4} verify why we can
absorb the two terms in \eqref{eq:fyhat-fy} into
$O_p(n_1^{-1}n_2^{1/2})$.

\subsubsection{Asymptotic expansion of $\sumib\{\log f_\Y(\wh\y_i)-\log f_\Y(\y_i)\}$}

On the other hand, we expand the term with respect to
  $f_\Y$ as follows:
\be\label{eq:fy2}
&&n_2^{-1/2}\sumib\{\log f_\Y(\wh\y_i)-\log f_\Y(\y_i)\}\n\\
&=&n_2^{-1/2}\sumib\left\{
\S_1(\y_i)\trans(\wh\y_i-\y_i)
+\frac{1}{2}(\wh\y_i-\y_i)\trans\S_2(\y_i^*)(\wh\y_i-\y_i)\right\}\n\\
&=&n_2^{-1/2}\sumib\left[
\S_1(\y_i)\trans\left\{n_1^{-1}\sumja\bpsi_{\Y_i}(\x_j,\bmu,\bSigma)+\r_4\y_i+\r_3\right\}\right]+O_p(n_1^{-1}n_2^{1/2})\n\\
&=&n_2^{-1/2}\sumib\left[
\S_1(\y_i)\trans n_1^{-1}\sumja\{\bpsi_{\bSigma^{-1/2}}(\x_j,\bmu,\bSigma)
\bSigma^{1/2}\y_i -\y_j
\}\right]+O_p(n_1^{-1}n_2^{1/2})\n\\
&=&n_2^{1/2}n_1^{-1}\sumja\tr\left[\bSigma^{1/2}E\left\{\Y
\S_1(\Y)\trans\right\}\bpsi_{\bSigma^{-1/2}}(\x_j,\bmu,\bSigma)
\right]\n\\
&& -n_2^{1/2}n_1^{-1}\sumja E\left\{
\S_1(\Y)\trans\right\}\y_j
+O_p(n_1^{-1}n_2^{1/2}+n_1^{-1/2})\n\\
&=& -n_2^{1/2}n_1^{-1}\sumja\tr\left\{\bSigma^{1/2}\bpsi_{\bSigma^{-1/2}}(\x_j,\bmu,\bSigma)
\right\}
+O_p(n_1^{-1}n_2^{1/2}+n_1^{-1/2}).
\ee
In the  second
equality in \eqref{eq:fy2}, we use
$(\I-\A^{-1})\y_i-\c=\y_i-\wh\y_i$ and the result in
\eqref{eq:quad-2}. In the fourth and fifth equalities in
\eqref{eq:fy2}, we use $E(|\S_1(\Y)\trans\Y|^2)\le
\{E(\|\S_1(\Y)\|^4)\}^{1/2}\{E(\|\Y|^4)\}^{1/2}<\infty$,
$E(\|\Y\|^4)<\infty$ and $E(\|\S_1(\Y)\|^4)<\infty$, which are implied
by Assumptions \ref{cond:A1} and \ref{cond:B1}. In the last equality,
we use $E \{\S_1(\Y)\}=\0$ and $E \{\Y\S_1(\Y)\trans\}=-\I$.

\subsubsection{Asymptotic expansion of $\sumib\{\log f_{\wh\Y}(\wh\y_i)-\log f_\Y(\y_i)\}$}

Combining \eqref{eq:fy1} and \eqref{eq:fy2}, we obtain
\be\label{eq:fy-1}
&&n_2^{-1/2}\sumib\{\log f_{\wh\Y}(\wh\y_i)-\log f_\Y(\y_i)\}\n\\
&=&n_2^{-1/2}\sumib\{\log f_{\wh\Y}(\wh\y_i)-\log f_\Y(\wh\y_i)\}
+n_2^{-1/2}\sumib\{\log f_{\Y}(\wh\y_i)-\log f_\Y(\y_i)\}\n\\
&=&n_2^{1/2}n_1^{-1}\sumja\tr  \left[\bSigma^{1/2} 
\bpsi_{\bSigma^{-1/2}}(\x_j,\bmu,\bSigma) \right]
+n_2^{1/2}\log |\det(\A)|+O_p(n_1^{-1}n_2^{1/2}+n_1^{-1/2})\n\\
&& -n_2^{1/2}n_1^{-1}\sumja\tr\left[\bSigma^{1/2}\bpsi_{\bSigma^{-1/2}}(\x_j,\bmu,\bSigma)
\right]
+O_p(n_1^{-1}n_2^{1/2}+n_1^{-1/2})\n\\
&=&n_2^{1/2}\log |\det(\A)|
+O_p(n_1^{-1}n_2^{1/2}+n_1^{-1/2}).
\ee

\subsubsection{Final result for $-\wh H_{n_2}(\wh \Y) + H (\Y)$}

By \eqref{eq:fy-1} and Section 4 of \cite{bsy2019}, we give the final
asymptotic expansion of $-\wh H_{n_2}(\wh \Y) + H (\Y)$ as follows:
\be\label{eq:fy-2}
&&n_2^{1/2} \{ -\wh H_{n_2}(\wh \Y)+ H (\Y) \}\n \\
&=& - n_2^{-1/2}\sumib\left\{\wh H_{n_2}(\wh \Y) +\log f_{\wh\Y}(\wh\y_i) \right\} + n_2^{-1/2}\sumib \left\{\log f_{\wh\Y}(\wh\y_i)-\log f_\Y(\y_i)\right\} \n\\
&&+n_2^{-1/2}\sumib \left\{\log f_\Y(\y_i) +   H (\Y)\right\} \n\\
&=& o_{\wh p}(1) + n_2^{1/2}\log |\det(\A)|
+O_p(n_1^{-1}n_2^{1/2}+n_1^{-1/2}) +n_2^{-1/2}\sumib \{\log f_\Y(\y_i) +   H (\Y)\}\n\\
&=&n_2^{1/2}\log |\det(\A)|
 + n_2^{-1/2}\sumib \{\log f_\Y(\y_i) +  H (\Y)\}+O_p(n_1^{-1}n_2^{1/2}+n_1^{-1/2}) +o_p(1),
\ee
where the last equality also uses contiguity, which is implied by
Assumption \ref{cond:D1}.
Plugging \eqref{eq:logdetA} into \eqref{eq:fy-2}, we have
\be\label{eq:fy}
&&n_2^{1/2} \{ -\wh H_{n_2}(\wh \Y)+ H (\Y) \} \n\\
&=&-n_2^{1/2} n_1^{-1}\sumja \tr\{\bpsi_{\bSigma^{-1/2}}(\X_j,\bmu,\bSigma)\bSigma^{1/2}\} +O_p(n_1^{-1}n_2^{1/2}) \n\\
&& + n_2^{-1/2}\sumib \{\log f_\Y(\y_i) +  H (\Y)\}+O_p(n_1^{-1}n_2^{1/2}+n_1^{-1/2}) +o_p(1)\n\\
&=&-n_2^{1/2} n_1^{-1}\sumja \tr\{\bpsi_{\bSigma^{-1/2}}(\X_j,\bmu,\bSigma)\bSigma^{1/2}\}   + n_2^{-1/2}\sumib \{\log f_\Y(\y_i) +  H (\Y)\}\n\\
&&+O_p(n_1^{-1}n_2^{1/2}+n_1^{-1/2}) +o_p(1).
\ee

\subsection{Expanding $\wh H_{n_2}(\wh U) - H(U)$}\label{sec:part3}

\subsubsection{Asymptotic expansion of $\sumib\{\log f_{\wh U}(\wh U_i)-\log f_U(\wh U_i)\}$}

\noindent\underline{\bf Expansion of a single $f_{\wh U}(u)$ and $\log f_{\wh U}(u)$}

We now proceed to analyze $f_{\wh U}(u)$. 
Note that
  \bse
  \frac{\partial^2
    f_\Y(\y)}{\partial\y\partial\y\trans}
  =\left[
    \frac{\partial^2\log
      f_\Y(\y)}{\partial\y\partial\y\trans}
+\left\{\frac{\partial\log
    f_\Y(\y)}{\partial\y}\right\}^{\otimes2}\right]f_{\Y}(\y).
\ese
Recall the notation of $\S_1(\y), \S_3(\y)$, then we have
\bse
\frac{\partial
  f_\Y(\y)}{\partial\y} = \S_1(\y) f_\Y(\y),\quad
\frac{\partial^2
    f_\Y(\y)}{\partial\y\partial\y\trans} = \S_3(\y) f_\Y(\y).
\ese

We first link $f_{\wh U}(u)$ to $f_\Y(\y)$ as follows.
Denote $\y_t=\y+t\{(\A-\I)\y-\A\c\}$, which is a function of $\y$ and $t$.
\be\label{eq:fuhat-1}
f_{\wh U}(u)
&=&\int_{\|\y\|=u}f_{\wh\Y}(\y)d\y\n\\
&=&|\det(\A)|\int_{\|\y\|=u}f_{\Y}\{\y+(\A-\I)\y-\A\c\}d\y \n\\
&=&|\det(\A)|\left[\int_{\|\y\|=u} f_\Y(\y)d\y+\int_{\|\y\|=u}\frac{\partial
  f_\Y(\y)}{\partial\y\trans}\{(\A-\I)\y-\A\c\}d\y\right.\n\\
&&\left.+\int_{\|\y\|=u}
  \int_0^1
\{(\A-\I)\y-\A\c\}\trans \frac{\partial^2 
  f_\Y(\s)}{\partial\s\partial\s\trans}\big|_{\s=\y_t}\{(\A-\I)\y-\A\c\}(1-t)dt
d\y\right].
\ee

Note that the first term in the bracket is exactly $f_U(u)$. The
second term can be written as
\bse
\int_{\|\y\|=u}\frac{\partial
  f_\Y(\y)}{\partial\y\trans}\{(\A-\I)\y-\A\c\}d\y 
  = \int_{\|\y\|=u}\S_1(\Y)\trans\{(\A-\I)\y-\A\c\}f_\Y(\y)d\y.
\ese
Also, we expand the third term into three parts as follows:
\bse
&& \int_{\|\y\|=u}
  \int_0^1
\{(\A-\I)\y-\A\c\}\trans \frac{\partial^2 
  f_\Y(\s)}{\partial\s\partial\s\trans}\big|_{\s=\y_t}\{(\A-\I)\y-\A\c\}(1-t)dt
d\y \\
&=& \int_{\|\y\|=u}
  \int_0^1
\y\trans(\A-\I)\trans \S_3(\y_t)f_\Y(\y_t)(\A-\I)\y(1-t)dt
d\y\\
&&+\int_{\|\y\|=u}
  \int_0^1
(\A\c)\trans \S_3(\y_t)f_\Y(\y_t) (\A\c)(1-t)dt
d\y\\
&&-2\int_{\|\y\|=u}
  \int_0^1
(\A\c)\trans \S_3(\y_t)f_\Y(\y_t) (\A-\I)\y(1-t)dt
d\y.
\ese
We consider the three parts one by one. For the first part, we have
\bse
&& \int_{\|\y\|=u}
  \int_0^1
\y\trans(\A-\I)\trans \S_3(\y_t)f_\Y(\y_t)(\A-\I)\y(1-t)dt
d\y\\
&=&\int_{\|\y\|=u}
  \int_0^1
(\y\trans\otimes\y\trans)\{(\A-\I)\trans\otimes(\A-\I)\trans\} \vec\left\{\S_3(\y_t)\right\}f_\Y(\y_t)(1-t)dt
d\y\\
&=&\tr\left[\{(\A-\I)\trans\otimes(\A-\I)\trans\} \int_{\|\y\|=u}
  \int_0^1
 \vec\left\{\S_3(\y_t)\right\}(\y\trans\otimes\y\trans)f_\Y(\y_t)(1-t)dt
d\y\right]\\
&=&\tr[\{(\A-\I)\trans\otimes(\A-\I)\trans\}\B_1(u)],
\ese
where
\bse
\B_1(u) = \int_{\|\y\|=u}
  \int_0^1
 \vec\left\{\S_3(\y_t)\right\}(\y\trans\otimes\y\trans)f_\Y(\y_t)(1-t)dt
d\y.
\ese
For the second part, we have
\bse
\int_{\|\y\|=u}
  \int_0^1
(\A\c)\trans \S_3(\y_t)f_\Y(\y_t) (\A\c)(1-t)dt
d\y
&=&(\A\c)\trans\int_{\|\y\|=u}
  \int_0^1
 \S_3(\y_t)  f_\Y(\y_t)(1-t)dt 
d\y (\A\c)\\
&=&(\A\c)\trans\B_2(u)(\A\c),
\ese
where
\bse
\B_2(u) = \int_{\|\y\|=u}
  \int_0^1
 \S_3(\y_t)  f_\Y(\y_t)(1-t)dt 
d\y.
\ese
For the third part, we have
\bse
&& \int_{\|\y\|=u}
  \int_0^1
(\A\c)\trans \S_3(\y_t)f_\Y(\y_t) (\A-\I)\y(1-t)dt
d\y \\
&=&\int_{\|\y\|=u}
  \int_0^1
(\A\c)\trans \left\{\y\trans \otimes \S_3(\y_t)\right\} f_\Y(\y_t)\vec(\A-\I)(1-t)dt
d\y \\
&=& (\A\c)\trans\int_{\|\y\|=u}
  \int_0^1
 \left\{\y\trans \otimes \S_3(\y_t)\right\} f_\Y(\y_t)(1-t)dt
d\y\{\vec(\A-\I)\}\\
&=& (\A\c)\trans\B_3(u)\vec(\A-\I), 
\ese
where
\bse
\B_3(u) = \int_{\|\y\|=u}
  \int_0^1
 \left\{\y\trans \otimes \S_3(\y_t)\right\} f_\Y(\y_t)(1-t)dt
d\y.
\ese

Inserting all terms back into \eqref{eq:fuhat-1}, we have
\bse
f_{\wh U}(u)
&=&|\det(\A)|\left(f_U(u)+\int_{\|\y\|=u}\S_1(\Y)\trans
    \{(\A-\I)\y-\A\c\}f_\Y(\y) d\y\right.\\
&&+\left.\tr[\{(\A-\I)\trans\otimes(\A-\I)\trans\}\B_1(u)]
+(\A\c)\trans\B_2(u)(\A\c)
   -2(\A\c)\trans\B_3(u)\vec(\A-\I) \right).
   \ese

Let
\bse
B(u) = \frac{1}{f_U(u)} \left[\tr\{[(\A-\I)\trans\otimes(\A-\I)\trans]\B_1(u)\}
+(\A\c)\trans\B_2(u)(\A\c)
   -2(\A\c)\trans\B_3(u)\vec(\A-\I)\right]
\ese
and 
\bse
D(u) =\frac{1}{f_U(u)} \int_{\|\y\|=u}\S_1(\Y)\trans \{(\A-\I)\y-\A\c\}f_\Y(\y) d\y.
\ese
Further let 
\bse
\Delta(u)=D(u)+B(u),
\ese
and then   
\bse
f_{\wh U}(u)= |\det(\A)|f_U(u)
  \{1+\Delta(u)\}.
  \ese
  Hence, $1+\Delta(u)\ge 0$.

\noindent\underline{\bf Main expansion of $\sumib\{\log f_{\wh U}(\wh
  u_i) - \log f_U(\wh u_i)\}$} 

Using Taylor expansion, we have
\bse
&&\log f_{\wh U}(u)
=\log |\det(\A)| + \log \{f_U(u)\} + \log \{1+\Delta(u)\} \n\\
&=&\log |\det(\A)| + \log \{f_U(u)\} + \Delta(u) - \int_0^{\Delta(u)} \frac{\Delta(u)-s}{(1+s)^2} ds \n\\
&=&\log |\det(\A)| + \log \{f_U(u)\} + \Delta(u) - \Delta(u)^2 \int_0^1 \frac{r}{\{(1-r)(\Delta(u)+1)+r\}^2} dr.
\ese
Therefore,
\be\label{eq:fu1}
&&n_2^{-1/2}\sumib\{\log f_{\wh U}(\wh u_i) - \log f_U(\wh u_i)\}\n\\
&=&n_2^{1/2}\log |\det(\A)|  + n_2^{-1/2}\sumib\Delta(\wh u_i) - n_2^{-1/2}\sumib R_i,
\ee
where
\be\label{eq:fua}
R_i=\Delta(\wh u_i)^2 \int_0^1 \frac{r}{\{(1-r)(\Delta(\wh u_i)+1)+r\}^2} dr.
\ee

\noindent\underline{\bf Expansion of $\sumib \Delta(\wh U_i)$}

In the following proof, we first give the expansion of the linear term
in the Taylor expansion, $\sumib\Delta(\wh U_i)$, by analyzing $B(u)$
and $D(u)$ separately. We start from expanding $\sumib B(\wh U_i)$.

We first analyze $\B_1(u)$,
\bse
&&n_2^{-1/2}\sumib \frac{1}{f_U(\wh U_i)}\tr[\{(\A-\I)\trans\otimes(\A-\I)\trans\}\B_1(\wh U_i)]\\
&=& \tr[\{(\A-\I)\trans\otimes(\A-\I)\trans\} n_2^{-1/2}\sumib \frac{1}{f_U(\wh U_i)}\int_{\|\y\|=\wh U_i}
  \int_0^1
 \vec\left\{\S_3(\y_t)\right\}(\y\trans\otimes\y\trans)f_\Y(\y_t)(1-t)dt
d\y]\\
&=& \tr\left(\{(\A-\I)\trans\otimes(\A-\I)\trans\} n_2^{-1/2}\sumib 
  E\left[\int_0^1
 \vec\left\{\S_3(\y_t)\right\}(\y\trans\otimes\y\trans)\frac{f_\Y(\y_t)}{f_\Y(\y)}(1-t)dt\Big| U=\wh U_i,\wh\bmu,\wh\bSigma\right]
\right)\\
&=& \tr\left\{\{(\A-\I)\trans\otimes(\A-\I)\trans\} n_2^{1/2} E_{\wh p}\left(
  E\left[\int_0^1
 \vec\left\{\S_3(\y_t)\right\}(\y\trans\otimes\y\trans)\frac{f_\Y(\y_t)}{f_\Y(\y)}(1-t)dt\Big|\wh U,\wh\bmu,\wh\bSigma\right]\Big|\wh\bmu,\wh\bSigma\right)
\right\}\\
&&+\tr\left[\{(\A-\I)\trans\otimes(\A-\I)\trans\} n_2^{1/2}o_{\wh p}(1)\right]\\ 
&=& \tr\left([(\A-\I)\trans\otimes(\A-\I)\trans] n_2^{1/2}
  E_{\wh p}\left[\int_0^1
 \vec\left\{\S_3(\y_t)\right\}(\y\trans\otimes\y\trans)\frac{f_\Y(\y_t)}{f_\Y(\y)}(1-t)dt\Big|\wh\bmu,\wh\bSigma\right]
\right) +o_p(n_1^{-1}n_2^{1/2})\\
&=& O_p(n_1^{-1}n_2^{1/2})
  \left\{E\left[\int_0^1
 \vec\left\{\S_3(\y_t)\right\}(\y\trans\otimes\y\trans)\frac{f_\Y(\y_t)}{f_\Y(\y)}(1-t)dt\right]+o_p(1)\right\}+o_p(n_1^{-1}n_2^{1/2})\\
&=& O_p(n_1^{-1}n_2^{1/2})
  E\left[\int_0^1
 \vec\left\{\S_3(\Y)\right\}(\Y\trans\otimes\Y\trans)(1-t)dt\right]+o_p(n_1^{-1}n_2^{1/2})\\
&=& O_p(n_1^{-1}n_2^{1/2})
  E\left[
 \vec\left\{\S_3(\Y)\right\}(\Y\trans\otimes\Y\trans)\right]+o_p(n_1^{-1}n_2^{1/2})\\
 &=& O_p(n_1^{-1}n_2^{1/2}),
\ese
where the fifth last equality  uses contiguity  under Assumption \ref{cond:D1}, the fourth last
equation uses Assumption \ref{cond:D2} with respect to
\be\label{eq:cont-4}
E_{\wt p}\left[\int_0^1
 \vec\left\{\S_3(\y_t)\right\}(\y\trans\otimes\y\trans)\frac{f_\Y(\y_t)}{f_\Y(\y)}(1-t)dt\Big|\wt\bmu,\wt\bSigma\right]
\ee
and the last
equation uses $ E[
\vec\{\S_3(\Y)\}(\Y\trans\otimes\Y\trans)]<\infty$,
which is equivalent to $E\{\Y\trans\S_3(\Y)\Y\}<\infty$, and is implied by Assumptions \ref{cond:A1}, \ref{cond:B1} and \ref{cond:B2}.

Similarly, for $\B_2(u)$, we have
\bse
n_2^{-1/2}\sumib \frac{1}{f_U(\wh U_i)}(\A\c)\trans\B_2(\wh U_i)(\A\c) = O_p(n_1^{-1}n_2^{1/2}) 
\ese
under Assumption \ref{cond:D2} with respect to
\be\label{eq:cont-5}
E_{\wt p} \left\{
  \int_0^1
 \S_3(\y_t) \frac{f_\Y(\y_t)}{f_\Y(\y)}(1-t)dt \Big|\wt\bmu,\wt\bSigma\right\}
\ee
and $E\{\S_3(\Y)\}<\infty$ implied by Assumptions \ref{cond:B1} and \ref{cond:B2}. 

Also similarly, for $\B_3(u)$, we have
\bse
n_2^{-1/2}\sumib \frac{1}{f_U(\wh U_i)}(\A\c)\trans\B_3(\wh U_i)\vec(\A-\I)  &=& O_p(n_1^{-1}n_2^{1/2}) 
\ese
under Assumption \ref{cond:D2} with respect to
\be\label{eq:cont-6}
E_{\wt p} \left[
  \int_0^1
 \left\{\y\trans \otimes \S_3(\y_t)\right\} \frac{f_\Y(\y_t)}{f_\Y(\y)}(1-t)dt \Big|\wt\bmu,\wt\bSigma \right]
\ee
and 
$E\{\Y\trans\otimes\S_3(\Y)\}<\infty$ implied by Assumptions \ref{cond:A1}, \ref{cond:B1} and \ref{cond:B2}.

Therefore, under Assumptions \ref{cond:D2} with respect to \eqref{eq:cont-4}, \eqref{eq:cont-5}, \eqref{eq:cont-6}, Assumptions  \ref{cond:A1}, \ref{cond:B1} and \ref{cond:B2}, we have
\be\label{eq:B}
n_2^{-1/2}\sumib B(\wh U_i) = O_p(n_1^{-1}n_2^{1/2}).
\ee

We finally consider $\sumib D(\wh U_i)$. Note that
\bse
D(u) &=&\frac{1}{f_U(u)} \int_{\|\y\|=u}\S_1(\Y)\trans \{(\A-\I)\y-\A\c\}f_\Y(\y) d\y \\
&=& E\left[\S_1(\Y)\trans \{(\A-\I)\Y-\A\c\}|U=u,\wh\bmu,\wh\bSigma\right] \\
&=&\tr\left[ (\A-\I)E\left\{\Y \S_1(\Y)\trans |U=u\right\}- (\A\c)
  E\left\{\S_1(\Y)\trans |U=u\right\}\right].
\ese
Thus, under Assumption \ref{cond:D2} with respect to
\be\label{eq:cont-7-8}
E_{\wt p}\left\{\Y \S_1(\Y)\trans \mid\wt\bmu,\wt\bSigma\right\} \quad \text{and} \quad E_{\wt p}\left\{\S_1(\Y)\trans\mid\wt\bmu,\wt\bSigma\right\},
\ee
contiguity implied by Assumption
\ref{cond:D1}, $E(\|\Y\|^4)<\infty$ and $E(\|\S_1(\Y)\|^2)<\infty$
implied by Assumptions \ref{cond:A1} and \ref{cond:B1}, using
$E\left\{\Y \S_1(\Y)\trans \right\}=-\I$ and $E\{\S_1(\Y)\}=0$ again,
we have
\be\label{eq:D}
&&n_2^{-1/2}\sumib D(\wh U_i) \n \\
&=&  \tr\left[ (\A-\I)n_2^{-1/2}\sumib E\left\{\Y \S_1(\Y)\trans |U=\wh U_i\right\}- (\A\c) n_2^{-1/2}\sumib E\left\{\S_1(\Y)\trans |U=\wh U_i\right\}\right]\n\\
&=&  n_2^{1/2}\tr\left\{ (\A-\I)\left(E_{\wh p}\left[ E\left\{\Y
        \S_1(\Y)\trans |\wh
        U\right\}{\mid\wh\bmu,\wh\bSigma}\right]+o_{\wh
      p}(1)\right)\right.\n\\
&&\left.- (\A\c) \left(E_{\wh p}\left[ E\left\{\S_1(\Y)\trans |\wh U\right\}{\mid\wh\bmu,\wh\bSigma}\right]+o_{\wh p}(1)\right)\right\}\n\\
&=&  n_2^{1/2}\tr\left[ (\A-\I)E_{\wh p}\left\{\Y \S_1(\Y)\trans \mid\wh\bmu,\wh\bSigma\right\} - (\A\c) E_{\wh p}\left\{\S_1(\Y)\trans\mid\wh\bmu,\wh\bSigma\right\}+o_{\wh p}(n_1^{-1/2})\right]\n\\
&=&  n_2^{1/2}\left\{\tr\left[ (\A-\I)E\left\{\Y \S_1(\Y)\trans \right\}\right] -  E\left\{\S_1(\Y)\trans\right\}(\A\c) +o_{p}(n_1^{-1/2})\right\}\n\\
&=& n_2^{1/2}\left\{-\tr (\A-\I)  +o_{p}(n_1^{-1/2})\right\}\n\\
&=& n_2^{1/2}\tr\left\{ \wh\bSigma^{1/2}n_1^{-1}\sumja\bpsi_{\bSigma^{-1/2}}(\X_j,\bmu,\bSigma)\right\}+O_p(n_2^{1/2}n_1^{-1}) +o_{p}(n_2^{1/2}n_1^{-1/2})\n\\
&=&n_2^{1/2}n_1^{-1}\sumja\tr\left\{ \bSigma^{1/2}\bpsi_{\bSigma^{-1/2}}(\X_j,\bmu,\bSigma)\right\}+o_{p}(n_2^{1/2}n_1^{-1/2}).
\ee 

Combining \eqref{eq:B} and \eqref{eq:D}, under Assumptions
\ref{cond:D1}, \ref{cond:D2} with respect to \eqref{eq:cont-4},
\eqref{eq:cont-5}, \eqref{eq:cont-6}, \eqref{eq:cont-7-8}, and
Assumptions \ref{cond:A1}, \ref{cond:B1}, \ref{cond:B2}, we have
\be\label{eq:Delta}
n_2^{-1/2}\sumib \Delta(\wh U_i) 
&=& n_2^{-1/2}\sumib B(\wh U_i) + n_2^{-1/2}\sumib D(\wh U_i) \n\\
&=& O_p(n_1^{-1}n_2^{1/2}) + n_2^{1/2}n_1^{-1}\sumja\tr\left\{ \bSigma^{1/2}\bpsi_{\bSigma^{-1/2}}(\X_j,\bmu,\bSigma)\right\}+o_{p}(n_2^{1/2}n_1^{-1/2}) \n\\
&=& n_2^{1/2}n_1^{-1}\sumja\tr\left\{ \bSigma^{1/2}\bpsi_{\bSigma^{-1/2}}(\X_j,\bmu,\bSigma)\right\}+o_{p}(n_2^{1/2}n_1^{-1/2}) \n\\
&=&O_{p}(n_1^{-1/2}n_2^{1/2}).
\ee

\noindent\underline{\bf Truncation for the remainder}

Similar to the difficulties in handling $\me_1$, here it is also
  hard to control the remainder term $R_i$ when $1+\Delta(\wh U_i)$ is
  too  
  close to 0 in the last term in \eqref{eq:fua}. We hence resort to a
  truncation treatment.
Let the event $\me_2$ be $\me_2=\{1+\Delta(\wh U_i)\ge \epsilon_n ,
\forall i\in\{n_1+1,\ldots,n\}\}$. We will follow the same logic as
the truncation procedure for $\me_1$: we first control the remainder
term under $\me_2$, and then further show that the events outside
$\me_2$ are rare enough to be ignored.

\noindent\underline{\bf Bounding the integral in the remainder term under $\me_2$}

Under $\me_2$, the integral part in $R_i$'s can be bounded as the
extreme values of $1+\Delta(\wh U_i)$ are ruled out. We have
  \bse
&&n_2^{-1/2}\sumib \Delta(\wh U_i)^2 \int_0^1 \frac{r}{\{(1-r)(\Delta(\wh U_i)+1)+r\}^2} dr I(\me_2)\\
&\le & n_2^{-1/2}\sumib \Delta(\wh U_i)^2 \int_0^1 \frac{r}{\{(1-r)\epsilon_n+r\}^2} dr \\
&=& n_2^{-1/2}\sumib \Delta(\wh U_i)^2 \int_0^1 \frac{r}{\{r(1-\epsilon_n)+\epsilon_n\}^2} dr \\
&=&n_2^{-1/2}\sumib \Delta(\wh U_i)^2\left\{\frac{1}{(1-\epsilon_n)^2}\log|r(1-\epsilon_n)+\epsilon_n|+\frac{\epsilon_n}{(1-\epsilon_n)^2}\frac{1}{r(1-\epsilon_n)+\epsilon_n}\right\}_0^1\\
&=&n_2^{-1/2}\sumib \Delta(\wh U_i)^2\left\{\frac{1}{(1-\epsilon_n)^2}\left(-\log|\epsilon_n|\right)+\frac{\epsilon_n}{(1-\epsilon_n)^2}\left(1-\frac{1}{\epsilon_n}\right)\right\}\\
&=&n_2^{-1/2}\sumib \Delta(\wh U_i)^2\left\{\frac{-\log(\epsilon_n)+\epsilon_n-1}{(1-\epsilon_n)^2}\right\}.
\ese
Let $\epsilon_n=n_1^{-k}$ where $k\ge 2$, then
\bse
\frac{-\log(\epsilon_n)+\epsilon_n-1}{(1-\epsilon_n)^2} = \frac{k\log(n_1)+n_1^{-k}-1}{(1-n_1^{-k})^2} \le 2k\log(n_1),
\ese
so
\be\label{eq:remainder-e2}
n_2^{-1/2}\sumib \Delta(\wh U_i)^2 \int_0^1 \frac{r}{\{(1-r)(\Delta(\wh U_i)+1)+r\}^2} dr I(\me_2) \le n_2^{-1/2}\sumib \Delta(\wh U_i)^2 2k\log(n_1).
\ee

\noindent\underline{\bf Expansion of $\sumib \Delta(\wh U_i)^2$}

Note that the integral term in the remainder term
    in \eqref{eq:remainder-e2}
  is uniformly bounded under $\me_2$, we now bound the rest  in the
  remainder, which is $n_2^{-1/2}\sumib \Delta(\wh U_i)^2$.
We divide $\Delta(u)$ into five terms and apply Lemma \ref{lem:a} as follows:
\bse
&&n_2^{-1/2}\sumib \Delta(\wh U_i)^2 
\le 5n_2^{-1/2}\sumib \left[\frac{1}{f_U(\wh U_i)}\tr\{[(\A-\I)\trans\otimes(\A-\I)\trans]\B_1(\wh U_i)\}\right]^2\\
&& \quad +5n_2^{-1/2}\sumib \left[\frac{1}{f_U(\wh U_i)}(\A\c)\trans\B_2(\wh U_i)(\A\c)\trans\right]^2 +5n_2^{-1/2}\sumib \left[\frac{2}{f_U(\wh U_i)}(\A\c)\trans\B_3(\wh U_i)\vec(\A-\I)\right]^2\\
&& \quad +5n_2^{-1/2}\sumib \tr\left[(\A-\I)E\left\{\Y\S_1(\Y)\trans |U=\wh U_i\right\}\right]^2 +5n_2^{-1/2}\sumib \left[(\A\c) E\left\{\S_1(\Y)\trans |U=\wh U_i\right\}\right]^2.
\ese 
We analyze the five terms one by one. For the first term, we have
\bse
&& n_2^{-1/2}\sumib \left[\frac{1}{f_U(\wh U_i)}\tr\{[(\A-\I)\trans\otimes(\A-\I)\trans]\B_1(\wh U_i)\}\right]^2 \\
&\le & \|(\A-\I)\trans\otimes(\A-\I)\trans\|_F^2 n_2^{-1/2}\sumib \frac{1}{f_U(\wh U_i)^2}\|\B_1(\wh U_i)\|_F^2.
\ese
We consider the two parts separately. Firstly,
\bse
&&\|(\A-\I)\trans\otimes(\A-\I)\trans\|_F^2 
= \tr [\{(\A-\I)\otimes(\A-\I)\} \{(\A-\I)\trans\otimes(\A-\I)\trans\}]\\
&=& \tr [ \{(\A-\I)(\A-\I)\trans\} \otimes \{(\A-\I)(\A-\I)\trans\} ] 
= [\tr  \{(\A-\I)(\A-\I)\trans\}]^2 
= \|\A-\I\|_F^4
= O_p(n_1^{-2}).
\ese
Then,
\bse
&&n_2^{-1/2}\sumib \frac{1}{f_U(\wh U_i)^2}\|\B_1(\wh U_i)\|_F^2\\
&=& n_2^{-1/2}\sumib \left\|\int_{\|\y\|=\wh U_i}
  \int_0^1
 \vec\left\{\S_3(\y_t)\right\}(\y\trans\otimes\y\trans)f_\Y(\y_t)(1-t)\frac{1}{f_U(\wh U_i)}dt
d\y\right\|_F^2\\
&=& n_2^{-1/2}\sumib \left\|E\left[
  \int_0^1
 \vec\left\{\S_3(\y_t)\right\}(\y\trans\otimes\y\trans)\frac{f_\Y(\y_t)}{f_\Y(\y)}(1-t)dt
\Big| U=\wh U_i,\wh\bmu,\wh\bSigma\right]\right\|_F^2\\
&\le& n_2^{-1/2}\sumib E\left[ \left\|
  \int_0^1
 \vec\left\{\S_3(\y_t)\right\}(\y\trans\otimes\y\trans)\frac{f_\Y(\y_t)}{f_\Y(\y)}(1-t)dt
\right\|_F^2\Big| U=\wh U_i,\wh\bmu,\wh\bSigma\right]\\
&=& n_2^{1/2}  E_{\wh p}\left\{ E\left[ \left\|
  \int_0^1
 \vec\left\{\S_3(\y_t)\right\}(\y\trans\otimes\y\trans)\frac{f_\Y(\y_t)}{f_\Y(\y)}(1-t)dt
\right\|_F^2\Big| \wh U,\wh\bmu,\wh\bSigma\right]\Big|\wh\bmu,\wh\bSigma\right\}+n_2^{1/2}o_{\wh p}(1)\\
&=& n_2^{1/2}   E_{\wh p}\left[ \left\|
  \int_0^1
 \vec\left\{\S_3(\y_t)\right\}(\y\trans\otimes\y\trans)\frac{f_\Y(\y_t)}{f_\Y(\y)}(1-t)dt
\right\|_F^2\Big|\wh\bmu,\wh\bSigma\right]+n_2^{1/2}o_p(1)\\
&=& n_2^{1/2}  E\left[ \left\|
  \int_0^1
 \vec\left\{\S_3(\y)\right\}(\y\trans\otimes\y\trans)(1-t)dt
\right\|_F^2\right]+n_2^{1/2}o_p(1)\\
&=& n_2^{1/2}  \frac{1}{4}E\left[ \left\|
 \vec\left\{\S_3(\Y)\right\}(\Y\trans\otimes\Y\trans)
\right\|_F^2\right]+n_2^{1/2}o_p(1)\\
&=& n_2^{1/2}  \frac{1}{4}E\left[ \tr \left\{
(\Y\otimes\Y) \vec\left\{\S_3(\Y)\right\}\trans\vec\left\{\S_3(\Y)\right\}(\Y\trans\otimes\Y\trans)
\right\}\right]+n_2^{1/2}o_p(1)\\
&=& n_2^{1/2}  \frac{1}{4}E\left\{\tr
(\Y\Y\trans\otimes\Y\Y\trans)
\|\S_3(\Y)\|_F^2\right\}+n_2^{1/2}o_p(1)\\
&=& n_2^{1/2}  \frac{1}{4}E\left\{ \|\Y\|^4\|\S_3(\Y)\|_F^2\right\}+n_2^{1/2}o_p(1),
\ese
 where in the above derivations, we use contiguity  due to Assumption \ref{cond:D1}
 in the sixth
 last equality
and also Assumption \ref{cond:D2} with respect to
\be\label{eq:cont-9}
E_{\wt p}\left[ \left\|
  \int_0^1
 \vec\left\{\S_3(\y_t)\right\}(\y\trans\otimes\y\trans)\frac{f_\Y(\y_t)}{f_\Y(\y)}(1-t)dt
\right\|_F^2\Big|\wt\bmu,\wt\bSigma\right]
\ee
 in the fifth last
equality, and use the boundedness of $E\left\{
  \|\Y\|^4\|\S_3(\Y)\|_F^2\right\}$
in Assumption implied by Assumptions \ref{cond:A1}, \ref{cond:B2}.
Thus, under all assumptions mentioned above, the first term has the order
\bse
n_2^{-1/2}\sumib \left[\frac{1}{f_U(\wh U_i)}\tr\{[(\A-\I)\trans\otimes(\A-\I)\trans]\B_1(\wh U_i)\}\right]^2 =O_p(n_1^{-2}n_2^{1/2}).
\ese 

Similarly,  for the second term, we have
\bse
n_2^{-1/2}\sumib \left[\frac{1}{f_U(\wh U_i)}(\A\c)\trans\B_2(\wh U_i)(\A\c)\right]^2 = O_p(n_1^{-2}n_2^{1/2}).
\ese
under Assumptions \ref{cond:D1}, \ref{cond:D2} with respect to
\be\label{eq:cont-10}
E_{\wt p}\left\{ \left\|
  \int_0^1
 \S_3(\y_t)  \frac{f_\Y(\y_t)}{f_\Y(\y)}(1-t)dt 
\right\|_F^2\Big|\wt\bmu,\wt\bSigma\right\},
\ee
and Assumptions  \ref{cond:A1},  \ref{cond:B2}.

Also similarly, for the third term, we have
\bse
n_2^{-1/2}\sumib \left[\frac{1}{f_U(\wh U_i)}(\A\c)\trans\B_3(\wh
  U_i)\vec(\A-\I)\right]^2 = O_p(n_1^{-2}n_2^{1/2})
\ese
under Assumptions \ref{cond:D1}, \ref{cond:D2} with respect to
\be\label{eq:cont-11}
E_{\wt p}\left[ \left\|
  \int_0^1
 \left\{\y\trans \otimes \S_3(\y_t)\right\} \frac{f_\Y(\y_t)}{f_\Y(\y)}(1-t)dt
\right\|_F^2\Big|\wt\bmu,\wt\bSigma\right],
\ee
and Assumptions  \ref{cond:A1},  \ref{cond:B2}.

Now we consider the fourth term, where the logic is also similar to
treating the first term,
\bse
&&n_2^{-1/2}\sumib \left(\tr\left[(\A-\I)E\left\{\Y\S_1(\Y)\trans |U=\wh U_i\right\}\right]\right)^2\\
&\le&n_2^{-1/2}\sumib \|\A-\I\|_F^2 \left\|E\left\{\Y\S_1(\Y)\trans |U=\wh U_i\right\}\right\|_F^2\\
&\le& \|\A-\I\|_F^2 n_2^{-1/2}\sumib  E\left\{\|\Y\S_1(\Y)\trans\|_F^2 |U=\wh U_i\right\}\\
&=& \|\A-\I\|_F^2 n_2^{-1/2}\sumib  E\left\{\|\Y\|^2\|\S_1(\Y)\|^2 |U=\wh U_i\right\}\\
&=& \|\A-\I\|_F^2 n_2^{1/2} \left\{E_{\wh p}\left[ E\left\{\|\Y\|^2\|\S_1(\Y)\|^2 |U=\wh U\right\}\mid\wh\bmu,\wh\bSigma\right]+o_{\wh p}(1)\right\}\\
&=& \|\A-\I\|_F^2 n_2^{1/2} \left[ E_{\wh p}\left\{\|\Y\|^2\|\S_1(\Y)\|^2\mid\wh\bmu,\wh\bSigma\right\}+o_{\wh p}(1)\right]\\
&=& O_p(n_1^{-1}) n_2^{1/2} \left[ E\left\{\|\Y\|^2\|\S_1(\Y)\|^2\right\}+o_p(1)\right]\\
&=& O_p(n_1^{-1} n_2^{1/2}),
\ese
under Assumptions \ref{cond:D1}, \ref{cond:D2} with respect to
\be\label{eq:cont-12}
E_{\wt p}\left\{\|\Y\|^2\|\S_1(\Y)\|^2\mid\wt\bmu,\wt\bSigma\right\},
\ee
and Assumptions  \ref{cond:A1},  \ref{cond:B1}.

Finally, similarly for the fifth term, we have
\bse
n_2^{-1/2}\sumib \left[ E\left\{\S_1(\Y)\trans |U=\wh U_i\right\}(\A\c)\right]^2 = O_p(n_1^{-1} n_2^{1/2})
\ese
under Assumptions \ref{cond:D1}, \ref{cond:D2} with respect to
\be\label{eq:cont-13}
E_{\wt p}\left\{\|\S_1(\Y)\|^2\mid\wt\bmu,\wt\bSigma\right\},
\ee
and Assumptions   \ref{cond:A1},  \ref{cond:B2}.

Therefore, under Assumptions \ref{cond:D1}, \ref{cond:D2} with respect
to \eqref{eq:cont-9}, \eqref{eq:cont-10}, \eqref{eq:cont-11},
\eqref{eq:cont-12}, \eqref{eq:cont-13},  and Assumptions
\ref{cond:A1},  \ref{cond:B1},  \ref{cond:B2}, we have 
\be\label{eq:sum-delta2}
n_2^{-1/2}\sumib \Delta(\wh U_i)^2 = O_p(n_1^{-1} n_2^{1/2}).
\ee

\noindent\underline{\bf Order of the remainder term under $\me_2$}

Inserting \eqref{eq:sum-delta2} into \eqref{eq:remainder-e2}, we have
\be\label{eq:delta2}
n_2^{-1/2}\sumib \Delta(\wh U_i)^2 \int_0^1 \frac{r}{\{(1-r)(\Delta(\wh U_i)+1)+r\}^2} dr I(\me_2) =O_p(n_1^{-1} n_2^{1/2}\log(n_1)).
\ee

\noindent\underline{\bf Analysis of $\me_2^C$}

Similar to the case for $\me_1$, we now show that the event $\me_2^C$
happens rare enough by estimating the rate of $I(\me_2^C)$ as follows.
\be\label{eq:comp}
I(\me_2^C)&=&I\left(\exists i\in\{n_1+1,\ldots,n\} \mbox{ s.t. } 1+\Delta(\wh U_i)<\epsilon_n \right)
\le\sumib I\left(1+\Delta(\wh U_i) \le \epsilon_n\right)\n\\
&=&\sumib I\left(\frac{\epsilon_n}{1+\Delta(\wh U_i)}  \ge1\right )
\le\sumib \frac{\epsilon_n}{1+\Delta(\wh U_i)}
= \epsilon_n \sumib \frac{1}{ 1+\Delta(\wh U_i)}\n\\
&=& n_2 \epsilon_n E_{\wh p}\left\{\frac{1}{ 1+\Delta(\wh U)}\Big|\wh\bmu,\wh\bSigma\right\} +  n_2 \epsilon_n o_{\wh p}(1)
=  O_p(n_2 \epsilon_n ),
\ee
under contiguity  due to Assumption \ref{cond:D1}, and Assumption \ref{cond:D2} with respect to
\be\label{eq:cont-14}
E_{\wt p}\left\{\frac{1}{ 1+\Delta(\wt U)}\Big|\wt\bmu,\wt\bSigma\right\}
\ee

When $\epsilon=n_1^{-k}$, then
\bse
I(\me_2^C)=  O_p(n_2 n_1^{-k} ),
\ese

\noindent\underline{\bf Summary of results under $\me_2$}

Following from \eqref{eq:fu1}, we have
\be\label{eq:fui1}
&&n_2^{-1/2}\sumib \{\log f_{\wh U}(\wh U_i)-\log f_{U}(\wh U_i)\}I(\me_2)\n\\
&=&n_2^{-1/2}\sumib \left[\log |\det(\A)|  + \Delta(\wh U_i) - \Delta(\wh U_i)^2 \int_0^1 \frac{r}{\{(1-r)(\Delta(\wh U_i)+1)+r\}^2} dr\right]I(\me_2)\n\\
&=&n_2^{1/2}\log |\det(\A)| I(\me_2) + n_2^{-1/2}\sumib\Delta(\wh U_i) I(\me_2) + O_p(n_1^{-1} n_2^{1/2}\log(n_1))\n\\
&=&n_2^{1/2}\log |\det(\A)|  + n_2^{-1/2}\sumib \Delta(\wh U_i)  + O_p(n_1^{-1} n_2^{1/2}\log(n_1))\n\\
&& -n_2^{1/2}\log |\det(\A)| I(\me_2^C) - n_2^{-1/2}\sumib\Delta(\wh U_i) I(\me_2^C).
\ee 
By continuous mapping theorem, we have $\log|\det(\A)|= O_p(n_1^{-1/2})$. Thus, 
\bse
n_2^{1/2}\log |\det(\A)| I(\me_2^C) = n_2^{1/2}  O_p(n_1^{-1/2}) O_p(n_2n_1^{-k}) =  O_p(n_2^{3/2}n_1^{-k-1/2})
\ese 

Also, by \eqref{eq:Delta}, we have
\bse
n_2^{-1/2}\sumib\Delta(\wh U_i) I(\me_2^C) =
 O_p(n_2^{1/2}n_1^{-1/2})O_p(n_2n_1^{-k})
=o_p(n_2^{3/2}n_1^{-k}).
\ese 

Therefore, under all above conditions, using
\eqref{eq:logdetA} and
\eqref{eq:Delta}, we have
\be\label{eq:fui}
&&n_2^{-1/2}\sumib \{\log f_{\wh U}(\wh U_i)-\log f_{U}(\wh U_i)\}I(\me_2)\n\\
&=&n_2^{1/2}\log |\det(\A)|  + n_2^{-1/2}\sumib \Delta(\wh U_i) + O_p(n_1^{-1} n_2^{1/2}\log(n_1))+ O_p(n_2^{3/2}n_1^{-k-1/2}) +o_p(n_2^{3/2}n_1^{-k})\n\\
&=&-n_2^{1/2}n_1^{-1} \sumja \tr\{\bpsi_{\bSigma^{-1/2}}(\X_j,\bmu,\bSigma)\bSigma^{1/2}\} + O_p(n_2^{1/2}n_1^{-1}) +n_2^{1/2}n_1^{-1}\sumja\tr\left\{ \bSigma^{1/2}\bpsi_{\bSigma^{-1/2}}(\X_j,\bmu,\bSigma)\right\}\n\\
&&+o_{p}(n_2^{1/2}n_1^{-1/2})+  O_p(n_1^{-1} n_2^{1/2}\log(n_1))+o_p(n_2^{3/2}n_1^{-k})\n\\
&=&O_p(n_1^{-1} n_2^{1/2}\log(n_1)) +o_p(n_2^{1/2}n_1^{-1/2}+n_2^{3/2}n_1^{-k}).
\ee

\noindent\underline{\bf Results under $\me_2^C$}

In contrast, under $\me_2^C$,
\be\label{eq:fuic}
&&n_2^{-1/2}\sumib \{\log f_{\wh U}(\wh U_i)- \log f_{U}(\wh U_i)\}I(\me_2^C)\n\\
&=&n_2^{1/2}[ E_{\wh p}\{\log f_{\wh U}(\wh
U)|\wh\bmu,\wh\bSigma\}-  E_{\wh p}\{\log f_{U}(\wh
  U)|\wh\bmu,\wh\bSigma\} +o_{\wh p}(1) ] O_p(n_2n_1^{-k})\n\\
&=& O_p(n_2^{3/2}n_1^{-k}),
\ee
under Assumption \ref{cond:D2} with respect to
\be\label{eq:cont-15-16}
E_{\wt p}\{\log f_{\wt U}(\wt U)|\wt\bmu,\wt\bSigma\} \quad \text{and} \quad E_{\wt p}\{\log f_{U}(\wh U)|\wt\bmu,\wt\bSigma\},
\ee
and contiguity  under Assumption \ref{cond:D1}. Here,
\bse
f_{\wt U} (u)= |\det(\wt \A)|\int_{\|\y\|=u}f_{\Y}\{\y+(\wt \A-\I)\y-\wt\A\wt\c\}d\y
\ese
where $\wt\A=\bSigma^{-1/2}\wt\bSigma^{1/2}$ and
$\wt\c=\wt\bSigma^{-1/2}(\bmu-\wt\bmu)$.

\noindent\underline{\bf Final expansion of $\sumib \{\log f_{\wh
    U}(\wh U_i)-\log f_{U}(\wh U_i)\}$}

 Combining \eqref{eq:fuic} and \eqref{eq:fui},
\be\label{eq:fub}
&&n_2^{-1/2}\sumib \{\log f_{\wh U}(\wh U_i)-\log f_{U}(\wh U_i)\}\n\\
&=&n_2^{-1/2}\sumib \{\log f_{\wh U}(\wh U_i)- \log f_{U}(\wh U_i)\}I(\me_2)+n_2^{-1/2}\sumib \{\log f_{\wh U}(\wh U_i)-\log f_{U}(\wh U_i)\}I(\me_2^C)\n\\
&=&O_p(n_1^{-1} n_2^{1/2}\log(n_1)) +o_p(n_2^{1/2}n_1^{-1/2}+n_2^{3/2}n_1^{-k}) + O_p(n_2^{3/2}n_1^{-k})\n\\
&=&
O_p(n_2^{3/2}n_1^{-k}+n_1^{-1} n_2^{1/2}\log(n_1))
+o_{p}(n_2^{1/2}n_1^{-1/2}) .
\ee

\subsubsection{Asymptotic expansion of $\sumib\{\log f_U(\wh U_i)-\log f_U(U_i)\}$}

\noindent\underline{\bf Main expansion of $\sumib\{\log f_U(\wh U_i)-\log f_U(U_i)\}$}

By Taylor expansion, we have
\be\label{eq:fu3}
&&n_2^{-1/2}\sumib\{\log f_U(\wh u_i)-\log f_U(u_i)\}\n\\
&=&n_2^{-1/2}\sumib Q_1(u_i)(\wh u_i-u_i)+
\frac{1}{2}n_2^{-1/2}\sumib Q_2(u_i^*)(\wh
u_i-u_i)^2,
\ee
where $u_i^*$ is between $u_i$ and $\wh u_i$. Note that
\bse
&&n_2^{-1/2}\sumib Q_2(u_i^*)(\wh u_i-u_i)^2\\
&=&n_2^{-1/2}\sumib Q_2(u_i)(\wh u_i-u_i)^2 + n_2^{-1/2}\sumib \{Q_2(u_i^*)-Q_2(u_i)\}(\wh u_i-u_i)^2\\
\ese
By Cauchy-Schwarz inequality, we have
\bse
n_2^{-1/2}\sumib Q_2(u_i)(\wh u_i-u_i)^2
&\le& n_2^{1/2}\left\{n_2^{-1}\sumib Q_2(u_i)^2\right\}^{1/2}\left\{n_2^{-1}\sumib(\wh u_i-u_i)^4\right\}^{1/2} \\
&=& n_2^{1/2}  \left[ E \{ Q_2(U)^2 \} + o_p (1) \right]^{1/2} \left\{ O_p(n_1^{-2}) \right\}^{1/2} \\
&=& O_p(n_1^{-1}n_2^{1/2}),
\ese
where the second last equality uses Assumptions \ref{cond:C2} and
$E(\|\Y\|^4)<\infty$ implied by Assumption \ref{cond:A1}, and
Lemma \ref{lem:b} is also applied here. The other term can be bounded
under Assumptions \ref{cond:F2} and $E(\|\Y\|^3)<\infty$ implied by
Assumption \ref{cond:A1} using  Lemma \ref{lem:b} again, as 
\bse
&&\left| n_2^{-1/2}\sumib \{Q_2(u_i^*)-Q_2(u_i)\}(\wh u_i-u_i)^2 \right| 
\le n_2^{-1/2}\sumib |Q_2(u_i^*)-Q_2(u_i)|(\wh u_i-u_i)^2 \\
&\le& n_2^{-1/2}\sumib L|u_i^*-u_i|(\wh u_i-u_i)^2 
\le n_2^{-1/2}L \sumib |\wh u_i-u_i|^3 
= O_p(n_1^{-3/2}n_2^{1/2}).
\ese
Plugging back into \eqref{eq:fu3}, and using \eqref{eq:uihat-ui}, we have
\be\label{eq:fu4}
&&n_2^{-1/2}\sumib\{\log f_U(\wh u_i)-\log f_U(u_i)\}\n\\
&=&n_2^{-1/2}\sumib Q_1(u_i) (\wh u_i-u_i)+ O_p(n_1^{-1}n_2^{1/2})\n\\
&=&n_2^{-1/2}\sumib Q_1(u_i)
n_1^{-1}\sumja\y_i\trans
\{\bpsi_{\bSigma^{-1/2}}(\x_j,\bmu,\bSigma)\bSigma^{1/2}\y_i-\y_j\}/u_i
\n\\
&&+ n_2^{-1/2}\sumib Q_1(u_i) (S_{i4}+S_{i3} 
+ S_{i1}+S_{i2})+O_p(n_1^{-1}n_2^{1/2}).
\ee

\noindent\underline{\bf Ignorable terms in \eqref{eq:fu4}}

Note that
$|S_{i4}|\le \|\Y_i\| \|\r_4\|$ , $|S_{i3}|\le \|\r_3\|$, so
\bse
&&\left|n_2^{-1/2}\sumib Q_1(u_i) S_{i4} \right| 
\le n_2^{-1/2}\sumib |Q_1(u_i)| |S_{i4} | 
\le n_2^{-1/2}\sumib |Q_1(u_i)| \|\Y_i\| \|\r_4\|\\
&&= n_2^{1/2}\{E(\|\Y\| |Q_1(U)|)+o_p(1)\}O_p(n_1^{-1}) = O_p(n_1^{-1}n_2^{1/2})
\ese
under $E(\|\Y\| |Q_1(U)|)<\infty$ implied by Assumptions \ref{cond:A1} and \ref{cond:C1}, and
\bse
\left|n_2^{-1/2}\sumib Q_1(u_i) S_{i3} \right| 
\le n_2^{-1/2}\sumib |Q_1(u_i)| |S_{i3} | 
\le n_2^{1/2}\{E(|Q_1(U)|)+o_p(1)\}\|\r_3\| = O_p(n_1^{-1}n_2^{1/2})
\ese
under $E(|Q_1(U)|)<\infty$ implied by Assumption \ref{cond:C1}. 
We then consider the terms regarding $S_{i1}$ and $S_{i2}$.
For the term related to $S_{i1}$, using Lemma \ref{lem:c} and \eqref{eq:si1}, we have
\bse
\left|n_2^{-1/2}\sumib Q_1(u_i) S_{i1}\right| \le n_2^{-1/2}\sumib |Q_1(u_i)||S_{i1}| \le n_2^{-1/2}\sumib \frac{5}{2}|Q_1(u_i)| \frac{\|\wh\Y_i-\Y_i\|^2}{\|\Y_i\|} =O_p(n_1^{-1}n_2^{1/2})
\ese
under $E(|Q_1(U)|\|\Y\|)<\infty$ and  $E(|Q_1(U)|/\|\Y\|)<\infty$, which are implied by Assumptions \ref{cond:A1}, \ref{cond:A2} and \ref{cond:C1}.
Also, for the term related to $S_{i2}$, using Lemma \ref{lem:c} again and \eqref{eq:si2}, we have
\bse
\left|n_2^{-1/2}\sumib Q_1(u_i) S_{i2}\right| \le n_2^{-1/2}\sumib |Q_1(u_i)| |S_{i2}| \le n_2^{-1/2}\sumib |Q_1(u_i)|\frac{\|\wh\Y_i-\Y_i\|^3}{\|\Y_i\|^2} =O_p(n_1^{-3/2}n_2^{1/2})
\ese
under $E(|Q_1(U)|\|\Y\|)<\infty$ and  $E(|Q_1(U)|/\|\Y\|^2)<\infty$.
Here, $E(|Q_1(U)|\|\Y\|)<\infty$ is ensured by Assumptions
  \ref{cond:A1} and \ref{cond:C1}, while 
  $E(|Q_1(U)|/\|\Y\|^2) \le \{E(|Q_1(U)|^3)\}^{1/3}
  \{E(1/\|\Y\|^3)\}^{2/3}<\infty$ is
  ensured by H\"older's inequality, 
  Assumptions
  \ref{cond:A2} and
\ref{cond:C1}.

Therefore, we have
\bse
n_2^{-1/2}\sumib Q_1(u_i) (S_{i4}+S_{i3} 
+ S_{i1}+S_{i2}) =O_p(n_1^{-1}n_2^{1/2}).
\ese

\noindent\underline{\bf Final expansion of $\sumib\{\log f_U(\wh U_i)-\log f_U(U_i)\}$}

Continuing from \eqref{eq:fu4}, we have
\be\label{eq:fu5}
&&n_2^{-1/2}\sumib\{\log f_U(\wh u_i)-\log f_U(u_i)\}\n\\
&=&n_2^{-1/2}\sumib Q_1(u_i)
n_1^{-1}\sumja\y_i\trans
\{\bpsi_{\bSigma^{-1/2}}(\x_j,\bmu,\bSigma)\bSigma^{1/2}\y_i-\y_j\}/u_i+O_p(n_1^{-1}n_2^{1/2})\n\\
&=&n_2^{1/2}n_1^{-1}\sumja \tr\{n_2^{-1}\sumib Q_1(u_i)
\bSigma^{1/2}\y_i \y_i\trans/u_i
\bpsi_{\bSigma^{-1/2}}(\x_j,\bmu,\bSigma)\}\n\\
&&-n_2^{1/2} n_1^{-1}\sumja  n_2^{-1}\sumib Q_1(u_i)\v_i\trans
\y_j+O_p(n_1^{-1}n_2^{1/2})\n\\
&=&n_2^{1/2}n_1^{-1}\sumja \tr[\bSigma^{1/2}
E\{Q_1(U)
U\V\V\trans\}\bpsi_{\bSigma^{-1/2}}
(\x_j,\bmu,\bSigma)]-n_2^{1/2}
n_1^{-1}\sumja
E\{Q_1(U)\V\trans\}
\y_j\n\\
&&+O_p(n_1^{-1}n_2^{1/2}+n_1^{-1/2}).
\ee

\subsubsection{Asymptotic expansion of $\sumib\{\log f_{\wh U}(\wh u_i)-\log f_U(u_i)\}$}

Combining \eqref{eq:fub} and \eqref{eq:fu5},
\be\label{eq:fu-1}
&&n_2^{-1/2}\sumib\{\log f_{\wh U}(\wh U_i)-\log f_U(U_i)\}\n\\
&=&n_2^{-1/2}\sumib\{\log f_{\wh U}(\wh U_i)-\log f_U(\wh U_i)\} + n_2^{-1/2}\sumib\{\log f_U(\wh U_i)-\log f_U(U_i)\}\n\\
&=& O_p(n_2^{3/2}n_1^{-k}+n_1^{-1} n_2^{1/2}\log(n_1))
+o_{p}(n_2^{1/2}n_1^{-1/2}) \n\\
&&+n_2^{1/2}n_1^{-1}\sumja \tr[\bSigma^{1/2}
E\{Q_1(U)
U\V\V\trans\}\bpsi_{\bSigma^{-1/2}}
(\x_j,\bmu,\bSigma)]-n_2^{1/2}
n_1^{-1}\sumja
E\{Q_1(U)\V\trans\}
\y_j\n\\
&&+O_p(n_1^{-1}n_2^{1/2}+n_1^{-1/2})\n\\
&=& n_2^{1/2}n_1^{-1}\sumja \tr[\bSigma^{1/2}
E\{Q_1(U)
U\V\V\trans\}\bpsi_{\bSigma^{-1/2}}
(\x_j,\bmu,\bSigma)]-n_2^{1/2}
n_1^{-1}\sumja
E\{Q_1(U)\V\trans\}
\y_j\n\\
&&+o_{p}(n_2^{1/2}n_1^{-1/2}) +
O_p(n_1^{-1} n_2^{1/2}\log(n_1)+n_1^{-1/2}+n_2^{3/2}n_1^{-k}).
\ee

\subsubsection{Final result for $-\wh H_{n_2}(\wh U) + H (U)$}

By \eqref{eq:fu-1} and Section 4 of \cite{bsy2019}, we give the final
asymptotic expansion of $-\wh H_{n_2}(\wh U) + H (U)$ as follows:
\be\label{eq:fu}
&&n_2^{1/2} \{ \wh H_{n_2}(\wh U)- H (U) \}\n \\
&=&  n_2^{-1/2}\sumib \left\{\wh H_{n_2}(\wh U) +\log f_{\wh U}(\wh u_i) \right\} - n_2^{-1/2}\sumib \{\log f_{\wh U}(\wh u_i)-\log f_U(u_i)\} \n\\
&&-n_2^{-1/2}\sumib \left\{\log f_U(u_i) +  H (U)\right\} \n\\
&=& o_{\wh p}(1)  - n_2^{1/2}n_1^{-1}\sumja \tr[\bSigma^{1/2}
E\{Q_1(U)
U\V\V\trans\}\bpsi_{\bSigma^{-1/2}}
(\x_j,\bmu,\bSigma)]+n_2^{1/2}
n_1^{-1}\sumja
E\{Q_1(U)\V\trans\}
\y_j\n\\
&&+o_{p}(n_2^{1/2}n_1^{-1/2}) +
O_p(n_1^{-1} n_2^{1/2}\log(n_1)+n_1^{-1/2}+n_2^{3/2}n_1^{-k})-n_2^{-1/2}\sumib \{\log f_U(u_i) +  H (U)\} \n\\
&=& - n_2^{1/2}n_1^{-1}\sumja \tr[\bSigma^{1/2}
E\{Q_1(U)
U\V\V\trans\}\bpsi_{\bSigma^{-1/2}}
(\x_j,\bmu,\bSigma)]+n_2^{1/2}
n_1^{-1}\sumja
E\{Q_1(U)\V\trans\}
\y_j\n\\
&&-n_2^{-1/2}\sumib \{\log f_U(u_i) +  H (U)\}+o_{p}(1+n_2^{1/2}n_1^{-1/2}) \n\\
&&+
O_p(n_1^{-1} n_2^{1/2}\log(n_1)+n_1^{-1/2}+n_2^{3/2}n_1^{-k}),
\ee 
where the last equality also uses contiguity implied by Assumption \ref{cond:D1}.

\subsection{Summary}

Combining \eqref{eq:uvy},
  \eqref{eq:logu}, \eqref{eq:fy}, \eqref{eq:fu}, as long as $k\ge 2$, we obtain
\bse
&&n_2^{1/2}\{T_1 -d(f_{U,\V}\|f_Uf_0) \} \\
&=&n_2^{1/2} \{ -\wh H_{n_2}(\wh \Y) + H(\Y) \}  + n_2^{1/2} (p-1) [\wh E_{n_2} \{ \log ( \wh U) \} - E\{\log(U)\}] + n_2^{1/2} \{\wh H_{n_2}(\wh U)- H(U)\} \\
&=&-n_2^{1/2} n_1^{-1}\sumja \tr\{\bpsi_{\bSigma^{-1/2}}(\X_j,\bmu,\bSigma)\bSigma^{1/2}\}   + n_2^{-1/2}\sumib \{\log f_\Y(\y_i) +  H (\Y)\}\n\\
&& +n_1^{-1} n_2^{1/2}  (p-1)\sumja  \left[\tr\left\{ E\left(\V\V\trans\right)
\bpsi_{\bSigma^{-1/2}}(\X_j,\bmu,\bSigma) \bSigma^{1/2}\right\}
-E\left(\V\trans/U\right)\y_j \right]\n\\
&& + n_2^{-1/2} (p-1)\sumib\{\log(U_i) -E\{\log(U)\}\} 
- n_2^{1/2}n_1^{-1}\sumja \tr[\bSigma^{1/2}
E\{Q_1(U)
U\V\V\trans\}\bpsi_{\bSigma^{-1/2}}
(\x_j,\bmu,\bSigma)]\n\\
&&+n_2^{1/2}
n_1^{-1}\sumja
E\{Q_1(U)\V\trans\}
\y_j-n_2^{-1/2}\sumib \{\log f_U(u_i) +  H (U)\}+o_{p}(1+n_2^{1/2}n_1^{-1/2}) \n\\
&&+
O_p(n_2^{3/2}n_1^{-k}+n_1^{-1/2}+n_2^{1/2}n_1^{-1}\log(n_1))\n\\
&=&n_2^{1/2}n_1^{-1} \sumja  \tr\left(
    [(p-1) E\left(\V\V\trans\right)-\I-E\{Q_1(U)
U\V\V\trans\}]
\bpsi_{\bSigma^{-1/2}}(\X_j,\bmu,\bSigma) \bSigma^{1/2}
\right)\n\\
&&- n_2^{1/2}n_1^{-1} \sumja \left[ (p-1) E(\V\trans/U)
-E\{Q_1(U)\V\trans\}\right]
\y_j  \n\\
&& + n_2^{-1/2}\sumib \{ \log f_\Y(\y_i) + (p-1)\log(u_i) 
-\log f_U(u_i)\} +n_2^{1/2}\{H(\Y) -(p-1)E\{\log(U)\}-H(U)\}\n\\
&&+o_{p}(1+n_2^{1/2}n_1^{-1/2}) +
O_p(n_2^{3/2}n_1^{-k}+n_1^{-1/2}+n_2^{1/2}n_1^{-1}\log(n_1))\n\\
&=& n_2^{1/2}n_1^{-1} \sumja \psi_1(\X_j,\bmu,\bSigma) + n_2^{-1/2}\sumib \psi_2(\X_i,\bmu,\bSigma)\n\\
&&+o_{p}(1+n_2^{1/2}n_1^{-1/2}) +
O_p(n_2^{3/2}n_1^{-k}+n_1^{-1/2}+n_2^{1/2}n_1^{-1}\log(n_1))
\ese
where
\be
\psi_1(\x,\bmu,\bSigma) 
&=& \tr\left(
    \left[(p-1) E\left(\V\V\trans\right)-\I-E\{Q_1(U)
U\V\V\trans\}\right]
\bpsi_{\bSigma^{-1/2}}(\x,\bmu,\bSigma) \bSigma^{1/2}
\right)\n\\
&&-  \left[ (p-1) E(\V\trans/U)
-E\{Q_1(U)\V\trans\}\right]
\y, \label{eq:psi1}\\
\psi_2(\x,\bmu,\bSigma) &=&  \log f_\Y(\y)+(p-1)\log(u) 
-\log f_U(u)+H(\Y) -(p-1)E\{\log(U)\} -H(U).\label{eq:psi2}
\ee
\qed

\section{Proof of Theorem \ref{th:unknown}}\label{sec:proofth3}
We consider the test statistics
\bse
 T_1\equiv -\wh H_{n_2}(\wh U,\wh\V)+\wh H_{n_2}(\wh U)-
  \log c_p = -\wh H_{n_2}(\wh \Y) + (p-1)\wh E_{n_2} \{ \log ( \wh U) \}+\wh H_{n_2}(\wh U)-
  \log c_p
  \ese
based on the data $(\wh U_i, \wh\Y_i), i=n_1+1, \dots, n$. 
 Similarly, we define
\bse
 T_2\equiv -\wh H_{n_1}(\wh U,\wh\V)+\wh H_{n_1}(\wh U)-
  \log c_p = -\wh H_{n_1}(\wh \Y) + (p-1)\wh E_{n_1} \{ \log ( \wh U) \}+\wh H_{n_1}(\wh U)-
  \log c_p
  \ese
based on the data $(\wh U_i, \wh\Y_i), i=1, \dots, n_1$. Recall the
notations in \eqref{eq:T-def} and \eqref{eq:hbar-ebar}.
We set $n_1=n_2=\lfloor n/2 \rfloor$. We first consider the $\wb H_{n}(\wh \Y)$ term.
\bse
&&n^{1/2} \{ -\wb H_{n}(\wh \Y) + H (\Y) \} \\
&=&2^{-1/2}\left[n_2^{1/2} \{ -\wh H_{n_2}(\wh \Y)+ H (\Y) \} + n_1^{1/2} \{ -\wh H_{n_1}(\wh \Y)+ H (\Y) \}\right]+o_p(1)\\
&=&-2^{-1/2}\left(n_2^{1/2} n_1^{-1}\sumja + n_1^{1/2} n_2^{-1}\sumjb \right) \tr\{\bpsi_{\bSigma^{-1/2}}(\X_j,\bmu,\bSigma)\bSigma^{1/2}\}  \\
&& + 2^{-1/2}\left(n_2^{-1/2}\sumib + n_1^{-1/2}\sumia \right) \{\log f_\Y(\y_i) +  H (\Y)\}+O_p(n^{-1/2}) +o_p(1)\\
&=&- n^{-1/2}\sumj \tr\{\bpsi_{\bSigma^{-1/2}}(\X_j,\bmu,\bSigma)\bSigma^{1/2}\}   + n^{-1/2}\sumi \{\log f_\Y(\y_i) +  H (\Y)\}
+O_p(n^{-1/2}) +o_p(1)\\
&=&n^{-1/2}\sumi \psi_{T1}(\x_i,\bmu,\bSigma)+O_p(n^{-1/2}) +o_p(1)
\ese
where
\bse
\psi_{T1}(\x,\bmu,\bSigma)= -  \tr\{\bpsi_{\bSigma^{-1/2}}(\x,\bmu,\bSigma)\bSigma^{1/2}\} + \log f_\Y(\y) +  H (\Y).
\ese

We then consider the $\wb E_n \{ \log ( \wh U) \}$ term.
\bse
&&n^{1/2}[\wb E_n \{ \log ( \wh U) \} - E\{\log(U)\}]\\
&=&2^{-1/2}n_2^{-1/2}\sumib\{\log(\wh U_i) -E\{\log(U)\}\} + 2^{-1/2}n_1^{-1/2}\sumia\{\log(\wh U_i) -E\{\log(U)\}\} \\ 
&=& 2^{-1/2} \left( n_1^{-1} n_2^{1/2} \sumja + n_2^{-1} n_1^{1/2} \sumjb \right) \left[\tr\left\{ E\left(\V\V\trans\right)
\bpsi_{\bSigma^{-1/2}}(\X_j,\bmu,\bSigma) \bSigma^{1/2}\right\}
-E\left(\V\trans/U\right)\Y_j \right]\n\\
&& + 2^{-1/2}\left(n_2^{-1/2}\sumib+n_1^{-1/2}\sumia\right)\{\log(U_i) -E\{\log(U)\}\} +O_p( n^{-1/2}\log(n)+ n^{3/2-k} )\\
&=& n^{-1/2} \sumj  \left[\tr\left\{ E\left(\V\V\trans\right)
\bpsi_{\bSigma^{-1/2}}(\X_j,\bmu,\bSigma) \bSigma^{1/2}\right\}
-E\left(\V\trans/U\right)\Y_j \right]\n\\
&& + n^{-1/2}\sumi\{\log(U_i) -E\{\log(U)\}\} +O_p( n^{-1/2}\log(n)+ n^{3/2-k} )\\
&=& n^{-1/2} \sumi \psi_{T2}(\x_i,\bmu,\bSigma) +O_p( n^{-1/2}\log(n)+ n^{3/2-k} )
\ese
where
\bse
 \psi_{T2}(\x,\bmu,\bSigma) 
&=& \tr\left\{ E\left(\V\V\trans\right)
\bpsi_{\bSigma^{-1/2}}(\x,\bmu,\bSigma) \bSigma^{1/2}\right\}
-E\left(\V\trans/U\right)\y +\log(u) -E\{\log(U)\}.
\ese

We finally consider the $\wb H_n(\wh U)$ term.
\bse
&&n^{1/2} \{ \wb H_{n}(\wh U)- H (U) \} \\
&=&2^{-1/2} \left[n_2^{1/2} \{ \wh H_{n_2}(\wh U)- H (U) \} +n_1^{1/2}
  \{ \wh H_{n_1}(\wh U)- H (U) \} \right]  +o_p(1)\\
&=& - 2^{-1/2} \left(n_2^{1/2}n_1^{-1}\sumja + n_1^{1/2}n_2^{-1}\sumjb \right) \tr[\bSigma^{1/2}
E\{Q_1(U)
U\V\V\trans\}\bpsi_{\bSigma^{-1/2}}
(\x_j,\bmu,\bSigma)]\n\\
&&+2^{-1/2} \left(n_2^{1/2}
n_1^{-1}\sumja + n_1^{1/2}n_2^{-1}\sumjb \right)
E\{Q_1(U)\V\trans\}
\y_j\\
&&-2^{-1/2} \left(n_2^{-1/2}\sumib +n_1^{-1/2}\sumia\right)\{\log f_U(u_i) +  H (U)\}
+o_{p}(1) +
O_p(n^{3/2-k} +n^{-1/2} \log(n) )\\
&=& - n^{-1/2}\sumj \tr[\bSigma^{1/2}
E\{Q_1(U)
U\V\V\trans\}\bpsi_{\bSigma^{-1/2}}
(\x_j,\bmu,\bSigma)]+n^{-1/2}\sumj
E\{Q_1(U)\V\trans\}
\y_j\\
&&- n^{-1/2}\sumi \{\log f_U(u_i) +  H (U)\}+o_{p}(1) +
O_p(n^{3/2-k} +n^{-1/2} \log(n) )\\
&=&  n^{-1/2}\sumi \psi_{T3}(\x_i,\bmu,\bSigma)+o_{p}(1) +
O_p(n^{3/2-k} +n^{-1/2} \log(n) )
\ese
where
\bse
\psi_{T3}(\x,\bmu,\bSigma)&=&
 -\tr[\bSigma^{1/2}
E\{Q_1(U)
U\V\V\trans\}\bpsi_{\bSigma^{-1/2}}
(\x,\bmu,\bSigma)]+
E\{Q_1(U)\V\trans\}
\y  - \log f_U(u) -  H (U).
\ese

Therefore, as long as $k \ge 2$, we have
\bse
&&n^{1/2}\{T -d(f_{U,\V}\|f_Uf_0) \} \\
&=& n^{1/2}\{- \wb H_n(\wh \Y) +H(\Y) + (p-1)\wb E_n \{ \log ( \wh U) \} - (p-1)E\{\log(U)\}+ \wb H_n(\wh U) - H(U)\} \\
&=& n^{-1/2} \sumi \{\psi_{T1}(\X_i,\bmu,\bSigma) + (p-1)\psi_{T2}(\X_i,\bmu,\bSigma) + \psi_{T3}(\X_i,\bmu,\bSigma)\}+o_p(1)\\
&=& n^{-1/2} \sumi \Big[-  \tr\{\bpsi_{\bSigma^{-1/2}}(\X_i,\bmu,\bSigma)\bSigma^{1/2}\} + \log f_\Y(\Y_i) +  H (\Y)\\
&&  + (p-1) \tr\left\{ E\left(\V\V\trans\right)
\bpsi_{\bSigma^{-1/2}}(\X_i,\bmu,\bSigma) \bSigma^{1/2}\right\}
- (p-1)E\left(\V\trans/U\right)\Y_i  \n\\
&&+ (p-1)[\log(U_i) -E\{\log(U)\}]  -\tr[\bSigma^{1/2}
E\{Q_1(U)
U\V\V\trans\}\bpsi_{\bSigma^{-1/2}}
(\X_i,\bmu,\bSigma)]\\
&&+
E\{Q_1(U)\V\trans\}
\Y_i  - \log f_U(U_i) -  H (U)\Big]+o_p(1)\\
&=&n^{-1/2} \sumi  \tr\left(
    \left[(p-1) E\left(\V\V\trans\right)-\I-E\{Q_1(U)
U\V\V\trans\}\right]
\bpsi_{\bSigma^{-1/2}}(\X_i,\bmu,\bSigma) \bSigma^{1/2}
\right)\n\\
&&- n^{-1/2} \sumi [ (p-1) E(\V\trans/U)
-E\{Q_1(U)\V\trans\}]
\y_i 
+ n^{-1/2}\sumi \{ \log f_\Y(\y_i) +(p-1)\log(u_i) 
-\log f_U(u_i)\}\n\\
&& +n^{1/2}\{H(\Y) -(p-1)E\{\log(U)\}-H(U)\}+o_p(1)\\
&=&n^{-1/2}\sumi\psi(\x_i,\bmu,\bSigma)+o_p(1),
\ese
where $\psi(\x,\bmu,\bSigma)=\psi_1(\x,\bmu,\bSigma)+\psi_2(\x,\bmu,\bSigma)$, as defined in \eqref{eq:psi1} and \eqref{eq:psi2}.

Under $H_0$, $E(\V)=\0, E(\V\V\trans)=\I/p$, $E(\V/U) = E(\V) E(1/U) =\0$, and
\bse
E\left\{Q_1(U)\V\right\}&=&
E\left\{Q_1(U)\right\}E(\V)
=\0,\\
E\left\{
Q_1(U)
U\V\V\trans \right\}
&=&E\left\{
Q_1(U)
U \right\}E\left(\V\V\trans \right)
=-p^{-1}\I.
\ese
This leads to under $H_0$, $n^{1/2}T = o_p(1)$, which has the same
as in the known $\mu, \bSigma$ case.
Under $H_a$,
\bse
n^{1/2}\{T -d(f_{U,\V}\|f_Uf_0) \}
\sim N[0, E\{\psi(\X,\bmu,\bSigma)^2\}].
\ese
\qed

\newpage

\section{Values of empirical sizes and powers in Section \ref{sec:comp}} \label{sec:values}


\begin{table}[!h]
\centering
\begin{tabular}{lrrr}
\multicolumn{4}{c}{$n=500,\quad p=2$} \\
\hline
$s$ & 0 & 1 & 2 \\
\hline
KL & 0.014 & 0.968 & 0.999\\
KE & 0.024 & 1.000 & 1.000\\
HP & 0.046 & 1.000 & 1.000\\
MPQ & 0.052 & 0.995 & 1.000\\
PG & 0.047 & 1.000 & 1.000\\
SW & 0.051 & 0.892 & 0.698\\
SO & 0.040 & 1.000 & 1.000\\
\hline
\end{tabular}
\quad\quad
\begin{tabular}{lrrrrrr}
\multicolumn{7}{c}{$n=500,\quad p=5$} \\
\hline
$s$ & 0 & 1 & 2 & 3 & 4 & 5\\
\hline
KL & 0.036 & 0.837 & 1.000 & 1.000 & 1.000 & 1.000\\
KE & 0.024 & 0.516 & 0.998 & 0.998 & 1.000 & 1.000\\
HP & 0.034 & 0.925 & 1.000 & 1.000 & 1.000 & 1.000\\
MPQ & 0.065 & 1.000 & 1.000 & 1.000 & 1.000 & 1.000\\
PG & 0.054 & 1.000 & 1.000 & 1.000 & 1.000 & 1.000\\
SW & 0.043 & 0.977 & 0.982 & 0.991 & 0.995 & 0.995\\
SO & 0.049 & 1.000 & 1.000 & 1.000 & 1.000 & 1.000\\
\hline
\end{tabular}
\begin{tabular}{lrrrrrrrrrrr}
\multicolumn{12}{c}{$n=500,\quad p=10$} \\
\hline
$s$ & 0 & 1 & 2 & 3 & 4 & 5 & 6 & 7 & 8 & 9 & 10\\
\hline
KL & 0.027 & 0.126 & 0.337 & 0.597 & 0.825 & 0.938 & 0.986 & 0.999 & 1.000 & 1.000 & 1.000\\
KE & 0.012 & 0.018 & 0.081 & 0.270 & 0.751 & 0.986 & 0.995 & 0.999 & 0.999 & 1.000 & 0.999\\
HP & 0.057 & 0.110 & 0.281 & 0.483 & 0.713 & 0.888 & 0.967 & 0.994 & 0.996 & 1.000 & 1.000\\
MPQ & 0.068 & 0.918 & 0.999 & 1.000 & 1.000 & 1.000 & 1.000 & 1.000 & 1.000 & 1.000 & 1.000\\
PG & 0.057 & 1.000 & 1.000 & 1.000 & 1.000 & 1.000 & 1.000 & 1.000 & 1.000 & 1.000 & 1.000\\
SW & 0.055 & 0.953 & 0.992 & 1.000 & 1.000 & 1.000 & 1.000 & 1.000 & 1.000 & 1.000 & 1.000\\
SO & 0.048 & 1.000 & 1.000 & 1.000 & 1.000 & 1.000 & 1.000 & 1.000 & 1.000 & 1.000 & 1.000\\
\hline
\end{tabular}
\vspace{20pt}

\begin{tabular}{lrrr}
\multicolumn{4}{c}{$n=1000,\quad p=2$} \\
\hline
$s$ & 0 & 1 & 2 \\
\hline
KL & 0.006 & 0.999 & 1.000\\
KE & 0.03* & 1.00* & 1.00*\\
HP & 0.065 & 1.000 & 1.000\\
MPQ & 0.049 & 1.000 & 1.000\\
PG & 0.039 & 1.000 & 1.000\\
SW & 0.035 & 0.970 & 0.919\\
SO & 0.043 & 1.000 & 1.000\\
\hline
\end{tabular}
\quad\quad
\begin{tabular}{lrrrrrr}
\multicolumn{7}{c}{$n=1000,\quad p=5$} \\
\hline
$s$ & 0 & 1 & 2 & 3 & 4 & 5\\
\hline
KL & 0.044 & 0.996 & 1.000 & 1.000 & 1.000 & 1.000\\
KE & 0.02* & 1.00* & 1.00* & 1.00* & 1.00* & 1.00*\\
HP & 0.050 & 1.000 & 1.000 & 1.000 & 1.000 & 1.000\\
MPQ & 0.048 & 1.000 & 1.000 & 1.000 & 1.000 & 1.000\\
PG & 0.056 & 1.000 & 1.000 & 1.000 & 1.000 & 1.000\\
SW & 0.051 & 0.999 & 0.997 & 1.000 & 1.000 & 0.999\\
SO & 0.055 & 1.000 & 1.000 & 1.000 & 1.000 & 1.000\\
\hline
\end{tabular}
\begin{tabular}{lrrrrrrrrrrr}
\multicolumn{12}{c}{$n=1000,\quad p=10$} \\
\hline
$s$ & 0 & 1 & 2 & 3 & 4 & 5 & 6 & 7 & 8 & 9 & 10\\
\hline
KL & 0.029 & 0.249 & 0.688 & 0.940 & 0.994 & 1.000 & 1.000 & 1.000 & 1.000 & 1.000 & 1.000\\
KE & 0.01* & 0.04* & 0.58* & 1.00* & 1.00* & 1.00* & 1.00* & 1.00* & 1.00* & 1.00* & 1.00*\\
HP & 0.059 & 0.253 & 0.695 & 0.916 & 0.997 & 1.000 & 1.000 & 1.000 & 1.000 & 1.000 & 1.000\\
MPQ & 0.043 & 1.000 & 1.000 & 1.000 & 1.000 & 1.000 & 1.000 & 1.000 & 1.000 & 1.000 & 1.000\\
PG & 0.058 & 1.000 & 1.000 & 1.000 & 1.000 & 1.000 & 1.000 & 1.000 & 1.000 & 1.000 & 1.000\\
SW & 0.060 & 1.000 & 1.000 & 1.000 & 1.000 & 1.000 & 1.000 & 1.000 & 1.000 & 1.000 & 1.000\\
SO & 0.050 & 1.000 & 1.000 & 1.000 & 1.000 & 1.000 & 1.000 & 1.000 & 1.000 & 1.000 & 1.000\\
\hline
\end{tabular}
\caption{Values of empirical sizes and powers for Setting 1 in Figure \ref{fig:res-comparison-1}. The entries with * are based on 100 experiments, while others are based on 1000 experiments.}
\label{tab:res-comparison-1}
\end{table}


\begin{table}[!h]
\centering
\begin{tabular}{lrrr}
\multicolumn{4}{c}{$n=500,\quad p=2$} \\
\hline
$s$ & 0 & 1 & 2 \\
\hline
KL & 0.016 & 0.986 & 0.682\\
KE & 0.033 & 0.998 & 0.731\\
HP & 0.063 & 0.425 & 0.434\\
MPQ & 0.044 & 1.000 & 1.000\\
PG & 0.048 & 0.133 & 0.136\\
SW & 0.036 & 0.958 & 0.869\\
SO & 0.038 & 0.073 & 0.073\\
\hline
\end{tabular}
\quad\quad
\begin{tabular}{lrrrrrr}
\multicolumn{7}{c}{$n=500,\quad p=5$} \\
\hline
$s$ & 0 & 1 & 2 & 3 & 4 & 5\\
\hline
KL & 0.007 & 0.915 & 0.877 & 0.995 & 0.992 & 0.999\\
KE & 0.035 & 0.119 & 0.169 & 0.711 & 0.456 & 0.177\\
HP & 0.082 & 0.370 & 0.399 & 0.638 & 0.663 & 0.984\\
MPQ & 0.046 & 1.000 & 1.000 & 1.000 & 1.000 & 1.000\\
PG & 0.041 & 0.389 & 0.390 & 0.350 & 0.412 & 0.386\\
SW & 0.044 & 0.970 & 0.946 & 0.965 & 0.966 & 0.972\\
SO & 0.048 & 0.120 & 0.141 & 0.079 & 0.091 & 0.055\\
\hline
\end{tabular}
\begin{tabular}{lrrrrrrrrrrr}
\multicolumn{12}{c}{$n=500,\quad p=10$} \\
\hline
$s$ & 0 & 1 & 2 & 3 & 4 & 5 & 6 & 7 & 8 & 9 & 10\\
\hline
KL & 0.014 & 0.295 & 0.305 & 0.743 & 0.744 & 0.991 & 0.991 & 0.996 & 1.000 & 1.000 & 0.999\\
KE & 0.074 & 0.078 & 0.062 & 0.081 & 0.089 & 0.121 & 0.127 & 0.142 & 0.163 & 0.540 & 0.536\\
HP & 0.091 & 0.151 & 0.149 & 0.208 & 0.191 & 0.463 & 0.454 & 0.527 & 0.492 & 0.638 & 0.635\\
MPQ & 0.062 & 1.000 & 1.000 & 1.000 & 1.000 & 1.000 & 1.000 & 1.000 & 1.000 & 1.000 & 1.000\\
PG & 0.047 & 0.538 & 0.569 & 0.652 & 0.617 & 0.689 & 0.719 & 0.683 & 0.694 & 0.715 & 0.687\\
SW & 0.050 & 0.992 & 0.982 & 0.990 & 0.977 & 0.994 & 0.989 & 0.993 & 0.994 & 0.994 & 0.994\\
SO & 0.047 & 0.106 & 0.114 & 0.098 & 0.074 & 0.069 & 0.066 & 0.045 & 0.076 & 0.070 & 0.065\\
\hline
\end{tabular}
\vspace{20pt}

\begin{tabular}{lrrr}
\multicolumn{4}{c}{$n=1000,\quad p=2$} \\
\hline
$s$ & 0 & 1 & 2 \\
\hline
KL & 0.005 & 1.000 & 0.940\\
KE & 0.00* & 1.00* & 1.00*\\
HP & 0.072 & 0.454 & 0.457\\
MPQ & 0.052 & 1.000 & 1.000\\
PG & 0.045 & 0.138 & 0.126\\
SW & 0.042 & 0.987 & 0.911\\
SO & 0.045 & 0.079 & 0.075\\
\hline
\end{tabular}
\quad\quad
\begin{tabular}{lrrrrrr}
\multicolumn{7}{c}{$n=1000,\quad p=5$} \\
\hline
$s$ & 0 & 1 & 2 & 3 & 4 & 5\\
\hline
KL & 0.003 & 0.999 & 0.990 & 1.000 & 0.999 & 1.000\\
KE & 0.03* & 0.34* & 0.54* & 1.00* & 0.98* & 0.35*\\
HP & 0.076 & 0.420 & 0.410 & 0.646 & 0.675 & 0.991\\
MPQ & 0.046 & 1.000 & 1.000 & 1.000 & 1.000 & 1.000\\
PG & 0.041 & 0.385 & 0.346 & 0.348 & 0.366 & 0.411\\
SW & 0.039 & 0.997 & 0.976 & 0.983 & 0.977 & 0.985\\
SO & 0.036 & 0.139 & 0.126 & 0.088 & 0.072 & 0.071\\
\hline
\end{tabular}
\begin{tabular}{lrrrrrrrrrrr}
\multicolumn{12}{c}{$n=1000,\quad p=10$} \\
\hline
$s$ & 0 & 1 & 2 & 3 & 4 & 5 & 6 & 7 & 8 & 9 & 10\\
\hline
KL & 0.010 & 0.547 & 0.644 & 0.980 & 0.979 & 1.000 & 1.000 & 1.000 & 1.000 & 1.000 & 1.000\\
KE & 0.06* & 0.09* & 0.07* & 0.18* & 0.06* & 0.20* & 0.34* & 0.54* & 0.41* & 0.73* & 0.65*\\
HP & 0.098 & 0.141 & 0.143 & 0.219 & 0.222 & 0.600 & 0.583 & 0.637 & 0.651 & 0.798 & 0.767\\
MPQ & 0.051 & 1.000 & 1.000 & 1.000 & 1.000 & 1.000 & 1.000 & 1.000 & 1.000 & 1.000 & 1.000\\
PG & 0.046 & 0.521 & 0.532 & 0.605 & 0.602 & 0.680 & 0.717 & 0.719 & 0.699 & 0.693 & 0.714\\
SW & 0.052 & 0.999 & 0.989 & 0.990 & 0.987 & 0.994 & 0.991 & 0.994 & 0.995 & 0.994 & 0.996\\
SO & 0.044 & 0.112 & 0.084 & 0.095 & 0.099 & 0.085 & 0.090 & 0.079 & 0.085 & 0.088 & 0.072\\
\hline
\end{tabular}
\caption{Values of empirical sizes and powers for Setting 2 in Figure
  \ref{fig:res-comparison-2}. The entries with * are based on 100
  experiments, while others are based on 1000 experiments.}
\label{tab:res-comparison-2}
\end{table}


\begin{table}[!h]
\centering
\begin{tabular}{lrrr}
\multicolumn{4}{c}{$n=500,\quad p=2$} \\
\hline
$s$ & 0 & 1 & 2 \\
\hline
KL & 0.004 & 0.984 & 1.000\\
KE & 0.059 & 0.213 & 0.273\\
HP & 0.242 & 0.251 & 0.665\\
MPQ & 0.053 & 0.111 & 0.123\\
PG & 0.039 & 0.060 & 0.058\\
SW & 0.043 & 1.000 & 1.000\\
SO & 0.040 & 0.053 & 0.055\\
\hline
\end{tabular}
\quad\quad
\begin{tabular}{lrrrrrr}
\multicolumn{7}{c}{$n=500,\quad p=5$} \\
\hline
$s$ & 0 & 1 & 2 & 3 & 4 & 5\\
\hline
KL & 0.001 & 0.126 & 1.000 & 1.000 & 1.000 & 1.000\\
KE & 0.134 & 0.083 & 0.345 & 0.978 & 0.890 & 0.884\\
HP & 0.610 & 0.227 & 0.332 & 0.386 & 0.298 & 0.204\\
MPQ & 0.054 & 0.093 & 0.363 & 0.437 & 0.312 & 0.204\\
PG & 0.032 & 0.111 & 0.146 & 0.143 & 0.125 & 0.110\\
SW & 0.044 & 1.000 & 1.000 & 1.000 & 1.000 & 1.000\\
SO & 0.017 & 0.060 & 0.083 & 0.075 & 0.066 & 0.057\\
\hline
\end{tabular}
\begin{tabular}{lrrrrrrrrrrr}
\multicolumn{12}{c}{$n=500,\quad p=10$} \\
\hline
$s$ & 0 & 1 & 2 & 3 & 4 & 5 & 6 & 7 & 8 & 9 & 10\\
\hline
KL & 0.017 & 0.025 & 0.213 & 0.537 & 0.711 & 0.801 & 0.838 & 0.833 & 0.822 & 0.822 & 0.737\\
KE & 0.148 & 0.082 & 0.091 & 0.098 & 0.146 & 0.118 & 0.154 & 0.214 & 0.305 & 0.394 & 0.551\\
HP & 0.650 & 0.178 & 0.135 & 0.176 & 0.169 & 0.146 & 0.105 & 0.121 & 0.099 & 0.091 & 0.084\\
MPQ & 0.055 & 0.061 & 0.246 & 0.396 & 0.437 & 0.363 & 0.293 & 0.222 & 0.171 & 0.165 & 0.120\\
PG & 0.016 & 0.080 & 0.164 & 0.202 & 0.178 & 0.202 & 0.179 & 0.195 & 0.174 & 0.170 & 0.156\\
SW & 0.061 & 0.348 & 0.779 & 0.829 & 0.911 & 0.965 & 0.981 & 0.999 & 1.000 & 1.000 & 1.000\\
SO & 0.000 & 0.020 & 0.070 & 0.063 & 0.069 & 0.056 & 0.059 & 0.079 & 0.060 & 0.063 & 0.057\\
\hline
\end{tabular}
\vspace{20pt}

\begin{tabular}{lrrr}
\multicolumn{4}{c}{$n=1000,\quad p=2$} \\
\hline
$s$ & 0 & 1 & 2 \\
\hline
KL & 0.004 & 1.000 & 1.000\\
KE & 0.10* & 0.86* & 0.94*\\
HP & 0.317 & 0.285 & 0.629\\
MPQ & 0.042 & 0.108 & 0.136\\
PG & 0.049 & 0.055 & 0.063\\
SW & 0.059 & 1.000 & 1.000\\
SO & 0.049 & 0.045 & 0.052\\
\hline
\end{tabular}
\quad\quad
\begin{tabular}{lrrrrrr}
\multicolumn{7}{c}{$n=1000,\quad p=5$} \\
\hline
$s$ & 0 & 1 & 2 & 3 & 4 & 5\\
\hline
KL & 0.000 & 0.476 & 1.000 & 1.000 & 1.000 & 1.000\\
KE & 0.13* & 0.10* & 1.00* & 0.99* & 1.00* & 1.00*\\
HP & 0.686 & 0.250 & 0.397 & 0.446 & 0.349 & 0.254\\
MPQ & 0.052 & 0.090 & 0.670 & 0.732 & 0.533 & 0.318\\
PG & 0.048 & 0.111 & 0.146 & 0.137 & 0.134 & 0.132\\
SW & 0.049 & 1.000 & 1.000 & 1.000 & 1.000 & 1.000\\
SO & 0.037 & 0.054 & 0.084 & 0.096 & 0.059 & 0.073\\
\hline
\end{tabular}
\begin{tabular}{lrrrrrrrrrrr}
\multicolumn{12}{c}{$n=1000,\quad p=10$} \\
\hline
$s$ & 0 & 1 & 2 & 3 & 4 & 5 & 6 & 7 & 8 & 9 & 10\\
\hline
KL & 0.005 & 0.026 & 0.535 & 0.935 & 0.985 & 0.994 & 0.996 & 0.998 & 0.995 & 0.990 & 0.985\\
KE & 0.12* & 0.10* & 0.12* & 0.20* & 0.23* & 0.30* & 0.67* & 0.94* & 0.99* & 0.84* & 1.00*\\
HP & 0.711 & 0.157 & 0.126 & 0.149 & 0.149 & 0.157 & 0.132 & 0.085 & 0.071 & 0.066 & 0.078\\
MPQ & 0.060 & 0.076 & 0.309 & 0.602 & 0.623 & 0.523 & 0.425 & 0.323 & 0.211 & 0.161 & 0.137\\
PG & 0.018 & 0.084 & 0.187 & 0.205 & 0.177 & 0.190 & 0.201 & 0.181 & 0.161 & 0.169 & 0.162\\
SW & 0.042 & 0.828 & 1.000 & 0.998 & 0.999 & 1.000 & 1.000 & 1.000 & 1.000 & 1.000 & 1.000\\
SO & 0.008 & 0.032 & 0.075 & 0.072 & 0.069 & 0.072 & 0.070 & 0.070 & 0.063 & 0.066 & 0.076\\
\hline
\end{tabular}
\caption{Values of empirical sizes and powers for Setting 3 in Figure \ref{fig:res-comparison-3}. The entries with * are based on 100 experiments, while others are based on 1000 experiments.}
\label{tab:res-comparison-3}
\end{table}


\begin{table}[!h]
\centering
\begin{tabular}{lrrr}
\multicolumn{4}{c}{$n=500,\quad p=2$} \\
\hline
$s$ & 0 & 1 & 2 \\
\hline
KL & 0.004 & 1.000 & 1.000\\
KE & 0.040 & 0.308 & 1.000\\
HP & 0.068 & 0.490 & 0.770\\
MPQ & 0.057 & 0.999 & 1.000\\
PG & 0.042 & 0.106 & 0.008\\
SW & 0.029 & 0.919 & 0.929\\
SO & 0.049 & 0.044 & 0.051\\
\hline\end{tabular}
\quad\quad
\begin{tabular}{lrrrrrr}
\multicolumn{7}{c}{$n=500,\quad p=5$} \\
\hline
$s$ & 0 & 1 & 2 & 3 & 4 & 5\\
\hline
KL & 0.009 & 1.000 & 1.000 & 1.000 & 1.000 & 1.000\\
KE & 0.051 & 0.234 & 0.382 & 0.654 & 0.996 & 0.995\\
HP & 0.097 & 0.089 & 0.161 & 0.179 & 0.168 & 0.200\\
MPQ & 0.049 & 1.000 & 1.000 & 1.000 & 1.000 & 1.000\\
PG & 0.046 & 0.089 & 0.051 & 0.040 & 0.032 & 0.032\\
SW & 0.031 & 0.411 & 0.407 & 0.407 & 0.383 & 0.375\\
SO & 0.051 & 0.051 & 0.051 & 0.059 & 0.058 & 0.039\\
\hline
\end{tabular}
\begin{tabular}{lrrrrrrrrrrr}
\multicolumn{12}{c}{$n=500,\quad p=10$} \\
\hline
$s$ & 0 & 1 & 2 & 3 & 4 & 5 & 6 & 7 & 8 & 9 & 10\\
\hline
KL & 0.019 & 0.683 & 0.722 & 0.710 & 0.705 & 0.746 & 0.707 & 0.737 & 0.731 & 0.718 & 0.705\\
KE & 0.042 & 0.068 & 0.067 & 0.054 & 0.069 & 0.053 & 0.078 & 0.078 & 0.067 & 0.179 & 0.227\\
HP & 0.085 & 0.066 & 0.087 & 0.087 & 0.086 & 0.085 & 0.099 & 0.089 & 0.078 & 0.085 & 0.101\\
MPQ & 0.055 & 1.000 & 1.000 & 1.000 & 1.000 & 1.000 & 1.000 & 1.000 & 1.000 & 1.000 & 1.000\\
PG & 0.064 & 0.095 & 0.066 & 0.044 & 0.056 & 0.055 & 0.045 & 0.052 & 0.039 & 0.045 & 0.051\\
SW & 0.043 & 0.085 & 0.077 & 0.080 & 0.086 & 0.092 & 0.079 & 0.080 & 0.093 & 0.098 & 0.076\\
SO & 0.044 & 0.042 & 0.049 & 0.055 & 0.047 & 0.047 & 0.044 & 0.041 & 0.034 & 0.041 & 0.054\\
\hline
\end{tabular}
\vspace{20pt}

\begin{tabular}{lrrr}
\multicolumn{4}{c}{$n=1000,\quad p=2$} \\
\hline
$s$ & 0 & 1 & 2 \\
\hline
KL & 0.001 & 1.000 & 1.000\\
KE & 0.07* & 1.00* & 1.00*\\
HP & 0.081 & 0.497 & 0.939\\
MPQ & 0.049 & 1.000 & 1.000\\
PG & 0.041 & 0.086 & 0.009\\
SW & 0.021 & 0.936 & 0.945\\
SO & 0.053 & 0.046 & 0.056\\
\hline
\end{tabular}
\quad\quad
\begin{tabular}{lrrrrrr}
\multicolumn{7}{c}{$n=1000,\quad p=5$} \\
\hline
$s$ & 0 & 1 & 2 & 3 & 4 & 5\\
\hline
KL & 0.002 & 1.000 & 1.000 & 1.000 & 1.000 & 1.000\\
KE & 0.02* & 0.91* & 1.00* & 1.00* & 1.00* & 1.00*\\
HP & 0.108 & 0.105 & 0.237 & 0.286 & 0.329 & 0.324\\
MPQ & 0.052 & 1.000 & 1.000 & 1.000 & 1.000 & 1.000\\
PG & 0.027 & 0.086 & 0.058 & 0.053 & 0.043 & 0.040\\
SW & 0.039 & 0.536 & 0.526 & 0.519 & 0.541 & 0.535\\
SO & 0.047 & 0.052 & 0.050 & 0.045 & 0.044 & 0.051\\
\hline
\end{tabular}
\begin{tabular}{lrrrrrrrrrrr}
\multicolumn{12}{c}{$n=1000,\quad p=10$} \\
\hline
$s$ & 0 & 1 & 2 & 3 & 4 & 5 & 6 & 7 & 8 & 9 & 10\\
\hline
KL & 0.024 & 0.966 & 0.979 & 0.978 & 0.982 & 0.985 & 0.991 & 0.985 & 0.987 & 0.989 & 0.983\\
KE & 0.00* & 0.06* & 0.06* & 0.10* & 0.08* & 0.13* & 0.17* & 0.10* & 0.22* & 0.72* & 0.86*\\
HP & 0.073 & 0.068 & 0.077 & 0.094 & 0.110 & 0.089 & 0.094 & 0.106 & 0.095 & 0.097 & 0.105\\
MPQ & 0.047 & 1.000 & 1.000 & 1.000 & 1.000 & 1.000 & 1.000 & 1.000 & 1.000 & 1.000 & 1.000\\
PG & 0.046 & 0.081 & 0.063 & 0.053 & 0.058 & 0.047 & 0.049 & 0.044 & 0.041 & 0.050 & 0.039\\
SW & 0.030 & 0.095 & 0.092 & 0.121 & 0.112 & 0.109 & 0.103 & 0.131 & 0.127 & 0.095 & 0.107\\
SO & 0.033 & 0.048 & 0.045 & 0.042 & 0.058 & 0.046 & 0.046 & 0.048 & 0.051 & 0.042 & 0.057\\
\hline
\end{tabular}
\caption{Values of empirical sizes and powers for Setting 4 in Figure
  \ref{fig:res-comparison-4}. The entries with * are based on 100
  experiments, while others are based on 1000 experiments.} 
\label{tab:res-comparison-4}
\end{table}

\clearpage

\section{Additional plots in data analysis}\label{sec:plots}
\begin{figure}[!htbp]
\centering
\includegraphics[width=0.9\textwidth]{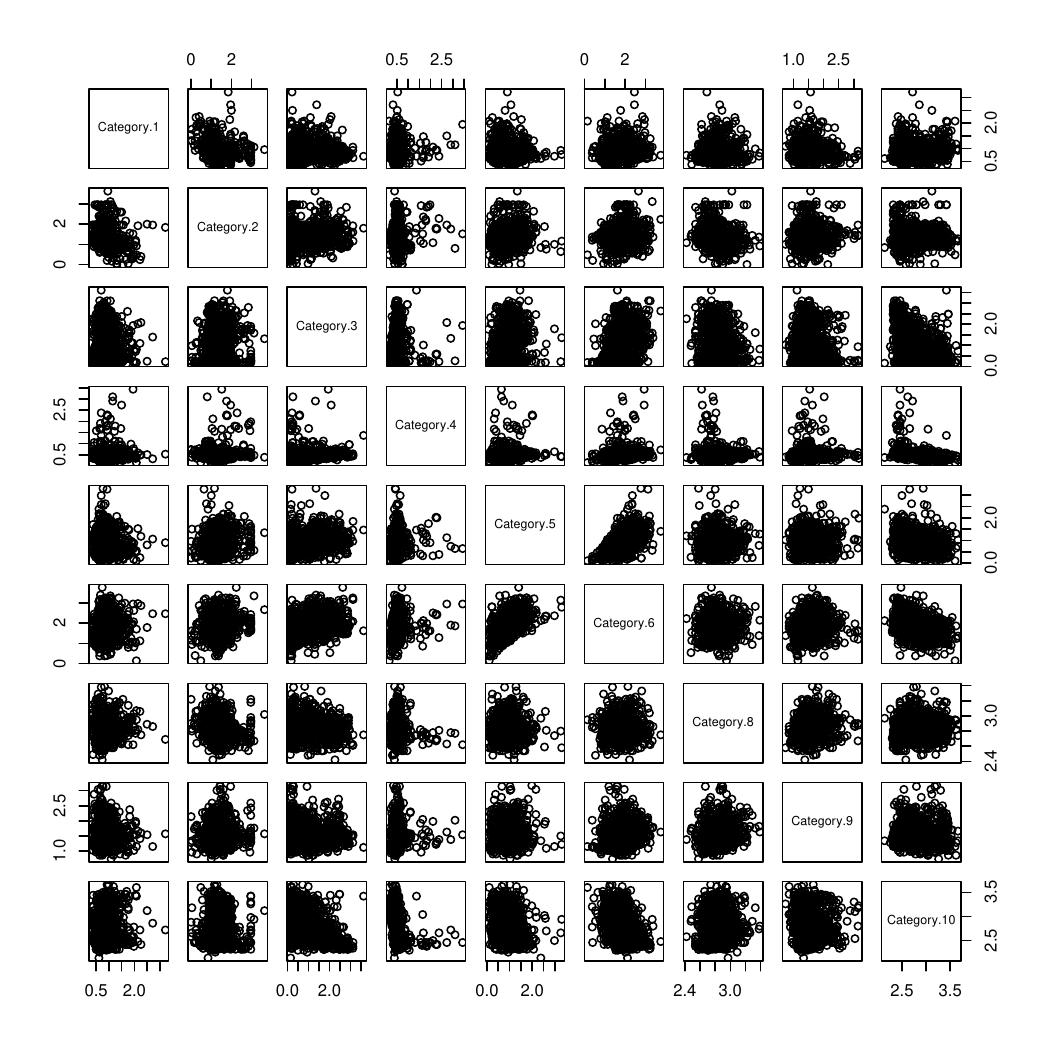}
\caption{Scatterplot of the original dataset on Travel Reviews.}
\label{fig:original-dataset}
\end{figure}

\begin{figure}[!htbp]
\centering
\includegraphics[width=0.9\textwidth]{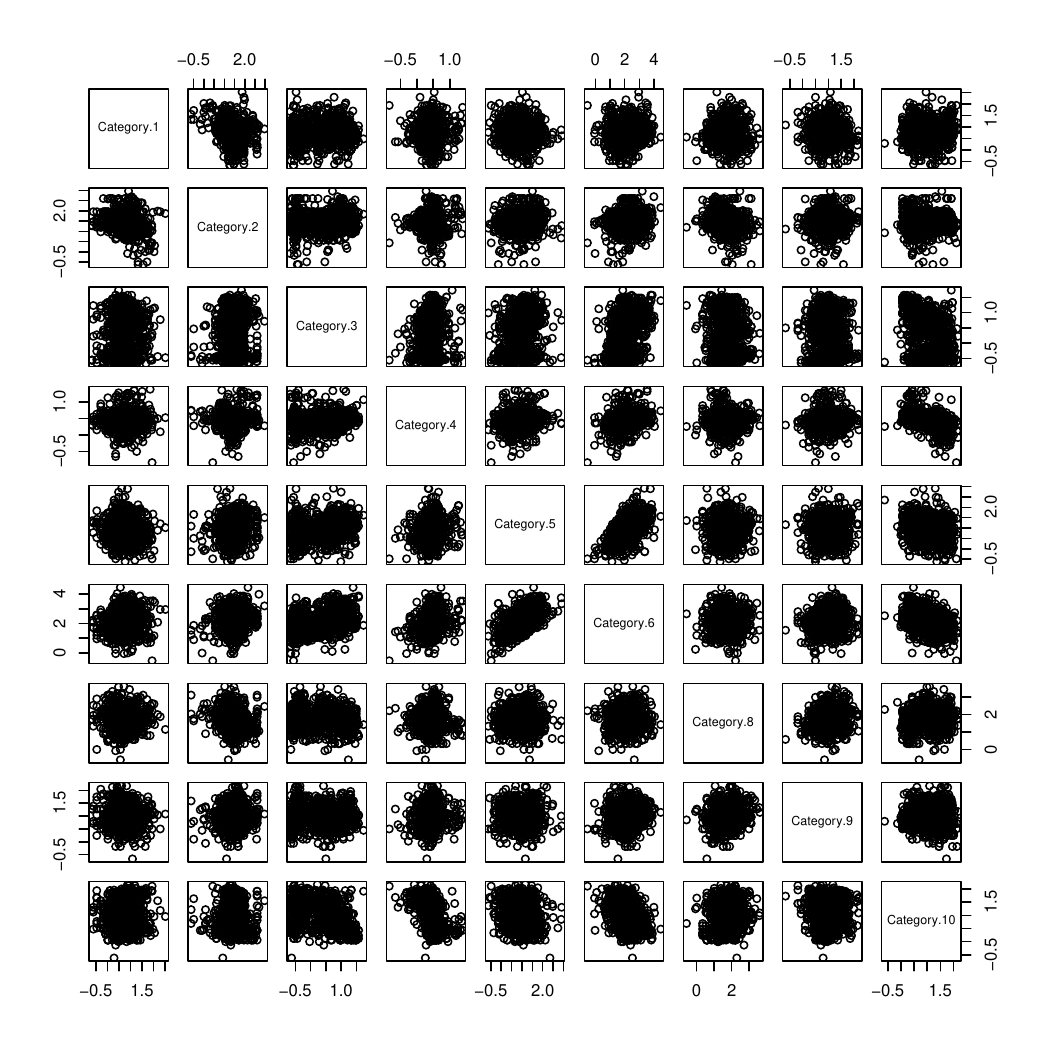}
\caption{Scatterplot of the Box-Cox transformed dataset on Travel Reviews.}
\label{fig:bctrans-dataset}
\end{figure}

\clearpage

\bibliographystyle{agsm}
\bibliography{ellipref}

\end{document}